\documentclass[sigconf]{acmart}
\renewcommand\footnotetextcopyrightpermission[1]{} 
\setcopyright{none}

\usepackage[normalem]{ulem}
\usepackage{algorithmic}
\usepackage{array}
\usepackage{url}

\usepackage{soul}

\usepackage{tikz}
\usetikzlibrary{calc,spy,shapes}
\usepackage{graphicx}

\usepackage{subfig}
\usepackage{caption}

\usepackage{amsmath}
\usepackage{courier}
\usepackage{listings}

\usepackage{mathtools}

\usepackage{pgfplots}
\usetikzlibrary{pgfplots.groupplots}

\usepackage{enumitem}
\usepackage{booktabs}
\usepackage{multirow}
\usepackage{makecell}

\newcommand{\multirowbt}[3]{%
\multirow{#1}{#2}{#3}%
}

\usepackage{color}
\usepackage{listings}
\usepackage{rotating}

\newcommand{\goal}[1]{\textcolor{blue}{<***GOAL: #1>}}
\renewcommand{\goal}[1]{}

\DeclareMathSizes{9}{8}{6}{6}

\usepackage{caption}

\usepackage[utf8]{inputenc}
\usepackage{cleveref}
\crefname{section}{§}{§§}
\Crefname{section}{§}{§§}

\begin{document}

\setlength{\pdfpageheight}{\paperheight}
\setlength{\pdfpagewidth}{\paperwidth}

\title{Evaluating the Cost of Atomic Operations\\ on Modern Architectures}

\author{Hermann Schweizer$^*$, Maciej Besta$^{*\dagger}$, Torsten Hoefler}
       \affiliation{\vspace{0.3em}Department of Computer Science, ETH Zurich\\
       {$^*$}Both authors contributed equally to this work\\
        {$^\dagger$}Corresponding author\\
}

\begin{abstract}
Atomic operations (atomics) such as Compare-and-Swap (CAS) or Fetch-and-Add
(FAA) are ubiquitous in parallel programming. Yet, performance tradeoffs
between these operations and various characteristics of such systems, such as
the structure of caches, are unclear and have not been thoroughly analyzed.
In this paper we establish an evaluation methodology, develop a performance
model, and present a set of detailed benchmarks for latency and bandwidth of
different atomics. We consider various state-of-the-art x86 architectures:
Intel Haswell, Xeon Phi, Ivy Bridge, and AMD Bulldozer. The results unveil
surprising performance relationships between the considered atomics and
architectural properties such as the coherence state of the accessed cache
lines.
One key finding is that all the tested atomics have comparable latency and
bandwidth even if they are characterized by different consensus numbers.
Another insight is that the hardware implementation of atomics prevents any
instruction-level parallelism even if there are no dependencies between the
issued operations.
Finally, we discuss solutions to the discovered performance issues in the
analyzed architectures.  Our analysis enables simpler and more effective
parallel programming and accelerates data processing on various architectures
deployed in both off-the-shelf machines and large compute systems.
\end{abstract}


\maketitle
\pagestyle{plain}

{\vspace{-0.5em}\noindent \textbf{This is an extended version of a paper published at\\ PACT'15 under the same title}}

{\vspace{1em}\small\noindent\textbf{Project website:}\\\url{https://spcl.inf.ethz.ch/Research/Parallel\_Programming/Atomics}}

\section{Introduction}
\label{sec:introduction}

Multi- and manycore architectures are established in both commodity
off-the-shelf desktop and server computers, as well as large-scale datacenters
and supercomputers.
Example designs include Intel Xeon Phi with 61 cores on a chip installed in
Tianhe-2~\cite{liao2014milkyway}, or AMD Bulldozer with 32 cores per node
deployed in Cray XE6~\cite{vaughan2011application}.
Moreover, the number of cores on a chip is growing steadily and CPUs with
hundreds of cores are predicted to be manufactured in the
foreseeable future~\cite{Esmaeilzadeh:2011:DSE:2000064.2000108}.
The common feature of all these architectures is the increasing complexity of
the memory subsystems characterized by multiple cache levels with different
inclusion policies, various cache coherence protocols, and different on-chip
network topologies connecting the cores and the
caches~\cite{Hackenberg:2009:CCA:1669112.1669165}.

Virtually all such architectures provide atomic operations that have numerous
applications in parallel codes.  Many of them (e.g., \textsf{Test-and-Set}) can
be used to implement locks and other synchronization
mechanisms~\cite{herlihy2012art}.  Others, e.g., \textsf{Fetch-and-Add} and
\textsf{Compare-and-Swap}, enable constructing miscellaneous lock-free and
wait-free algorithms and data structures that have stronger progress guarantees
than lock-based codes~\cite{herlihy2012art}.

Despite their importance and widespread utilization, the performance of
atomic operations has not been thoroughly analyzed so far.  For example,
according to the common view, \textsf{Compare-and-Swap} is slower than
\textsf{Fetch-and-Add}~\cite{Morrison:2013:FCQ:2442516.2442527}.  However,
it was only shown that the semantics of \textsf{Compare-and-Swap} introduce
the notion of ``wasted work'' resulting in lower performance of some
codes~\cite{Morrison:2013:FCQ:2442516.2442527,Harris:2001:PIN:645958.676105}.
Other works provide basic insights and illustrate that the performance of
atomics is similar on multi-socket systems due to the overheads from
socket-to-socket hops~\cite{David:2013:EYA:2517349.2522714}.  Yet, to the
best of our knowledge, no model and benchmarks analyze in detail the
latency or bandwidth of the execution of the actual operations in the
context of complex multilevel cache and memory hierarchies.
Even more importantly, the performance tradeoffs between atomics and
various characteristics of multi- and manycore systems (cache coherency
protocol, number of memory hierarchy levels, etc.) have also not been
thoroughly studied so far.
For example, a single node in popular Cray XE6 cabinets provides two AMD
Bulldozer CPUs connected with a HyperTransport (HT) link, each CPU consists
of two dies also connected with HT, and each die provides one L3 cache and
four L2 caches shared by eight cores~\cite{vaughan2011application}.
It is unclear what the performance of different atomics is on such a
system, what is the influence of the cache coherency protocol, what is the
performance impact of mechanisms such as adjacent cache line prefetchers,
and whether optimizations such as instruction-level parallelism are
available for atomics.

In this paper, we introduce a performance model and establish a methodology
for benchmarking atomics. Then, we use it to analyze the latency and
bandwidth of the most popular atomic operations (\textsf{Compare-and-Swap},
\textsf{Fetch-and-Add}, \textsf{Swap}). Our results unveil undocumented
architectural properties of the tested systems and identify several
performance issues of the evaluated operations. We discuss solutions to
these problems and we illustrate how our model and analysis simplify
parallel programming in areas such as graph analytics.
The key contributions of this work are:

\begin{figure*}[!t]
\centering
 \subfloat[The evaluated Haswell.]{
  \includegraphics[width=0.175\textwidth]{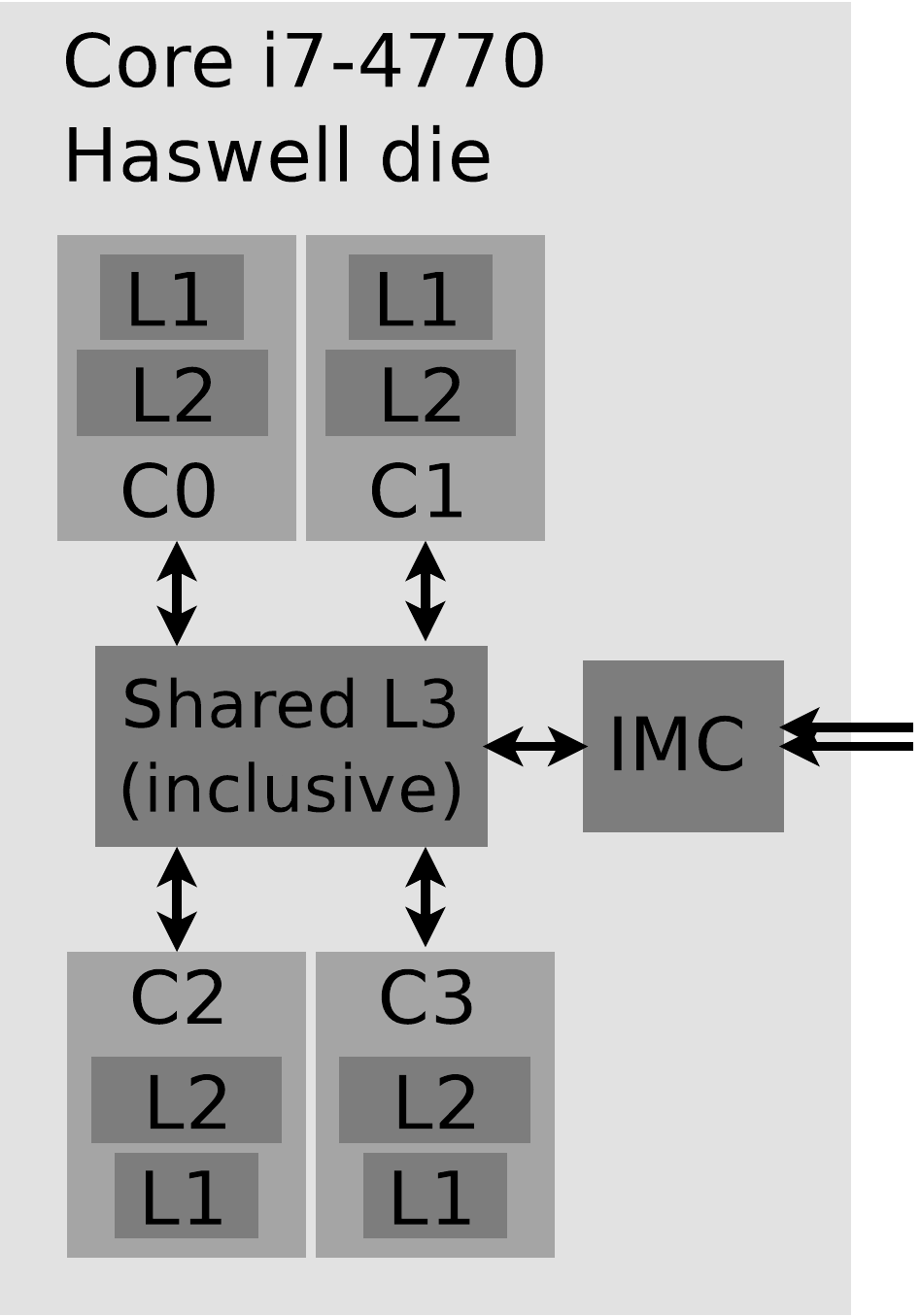}
  \label{fig:galilei_arch}
 }
 \subfloat[The evaluated Bulldozer.]{
  \includegraphics[width=0.72\textwidth]{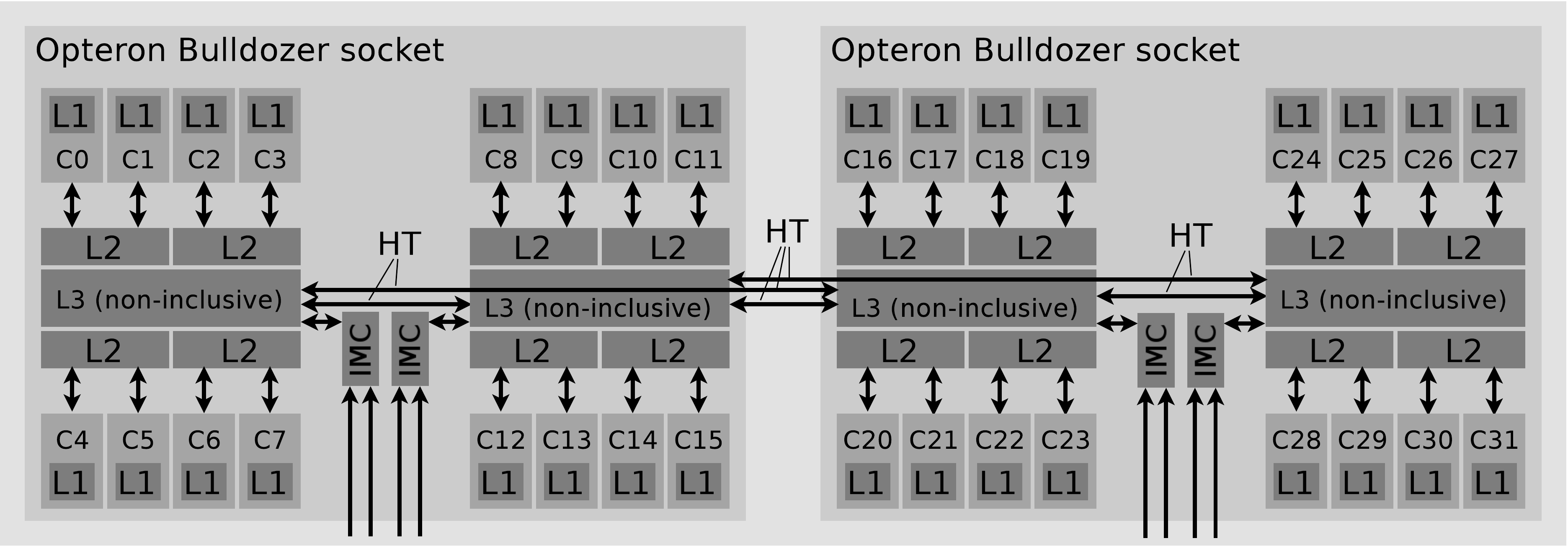}
  \label{fig:amd_arch}
 }
 \\
  \subfloat[The evaluated Ivy Bridge.]{
  \includegraphics[width=0.8\textwidth]{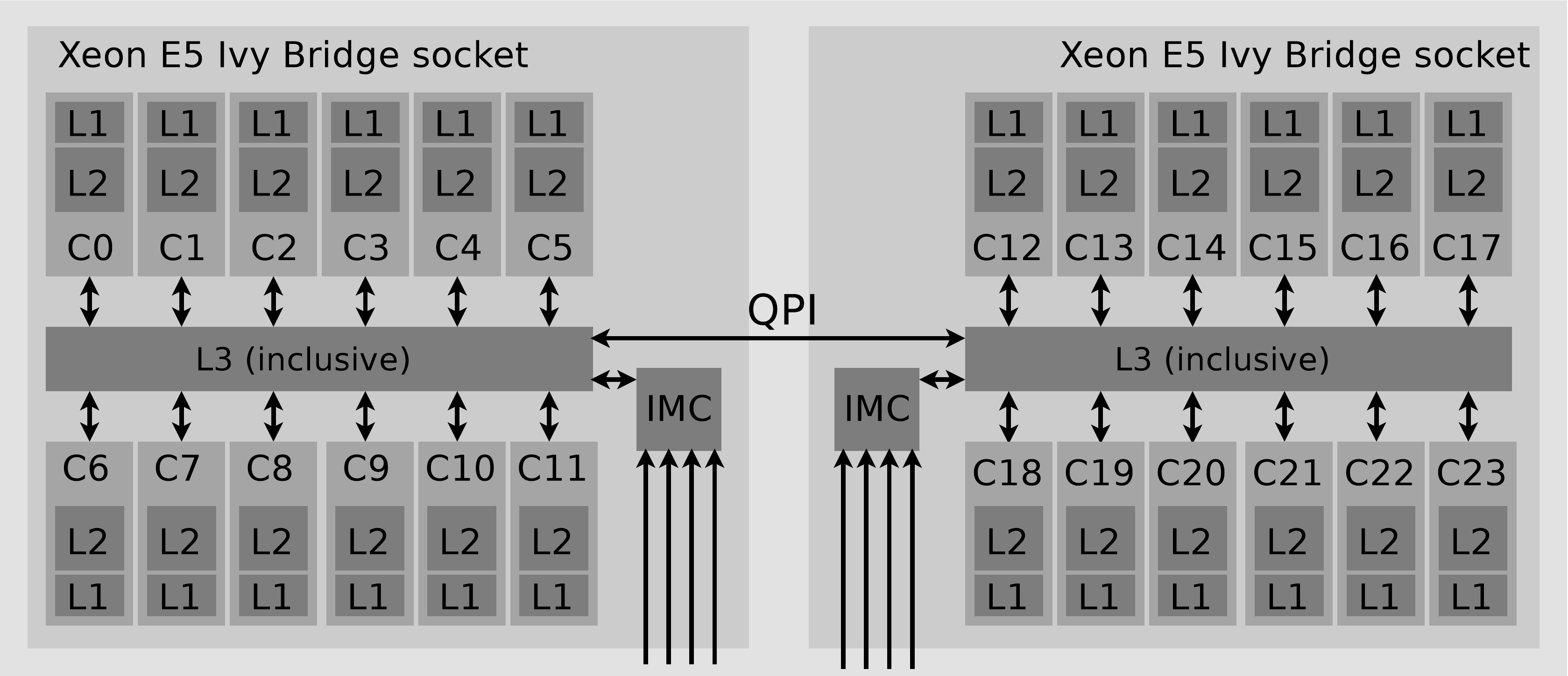}
  \label{fig:euler_arch}
 }
\caption{(\cref{sec:tested_systems}) The illustration of three of the
analyzed architectures: Intel Haswell, AMD Bulldozer, and
the Intel Ivy Bridge. The first one 
combines the features of the Haswell and Bulldozer testbeds as it hosts
private L1s and L2s and offers 24 cores grouped in two CPUs. The second is
a manycore design where each core has a private L1 and L2; the cores form
a ring.}
\label{fig:tested_architectures}
\end{figure*}

\begin{itemize}[leftmargin=0.5em]
\item We introduce a performance model for the latency and bandwidth of
atomics. The model takes into account different cache coherency states and
the structure of the caching hierarchy.
\item We establish a methodology for benchmarking atomic operations
targeting state-of-the-art multi- and manycore architectures with deep
memory hierarchies.
\item We conduct a detailed performance analysis of
\textsf{Compare-and-Swap}, \textsf{Fetch-and-Add}, and \textsf{Swap}. We
use the analysis to validate the model, to illustrate undocumented
architectural properties of the tested systems, and to suggest several
improvements in the hardware implementation of respective atomics.
\item We discuss how our analysis simplifies parallel programming using the
example of graph traversal algorithms.
\end{itemize}

\section{Background}
\label{sec:background}

We now present a general approach for benchmarking memory accesses
(Sec.~\ref{sec:background_benchmarking_accesses}) that we will later use and
extend.  Then, we discuss the evaluated architectures and atomics
(Sections~\ref{sec:tested_systems} and~\ref{sec:tested_mechanisms}).

\subsection{Benchmarking Memory Accesses}
\label{sec:background_benchmarking_accesses}

In our analysis we use and extend the X86membench infrastructure for
benchmarking memory accesses~\cite{Hackenberg:2009:CCA:1669112.1669165}
that utilizes the high resolution \texttt{RDTSC} time stamp counter.  Each
benchmark consists of the following phases:

\begin{description}[leftmargin=0.5em]
\item[\textsf{Preparation}:] A buffer of the selected size is allocated and
filled with the data specific to each benchmark (see more details
in Section~\ref{benchmarkDesign}). The TLB is warmed up and the data is placed in
caches in the selected coherency state.  
\item[\textsf{Synchronization}:] This phase makes sure that all threads
finished the preparation phase and it defines a future moment in time when
all the threads will start the measurement phase.  
\item[\textsf{Measurement}:] All participating threads: take a time stamp
\textsf{t\_start}, do a measurement, and take the other time stamp
\textsf{t\_end}. 
\item[\textsf{Result collection}:] The time stamps of all participating
cores are communicated and the total time of execution is calculated as
\textsf{max(t\_end) - min(t\_start)}.
\end{description}

\subsection{Evaluated Architectures and Systems}
\label{sec:tested_systems}

Next, we present the targeted architectures and systems. We illustrate more
details in Figure~\ref{fig:tested_architectures} and
Table~\ref{tab:tested_systems}.

\begin{description}[leftmargin=1em]
\item[\textsf{Haswell}] is an Intel state-of-the-art microarchitecture that
offers sophisticated mechanisms such as hardware transactional memory
(HTM)~\cite{Yoo:2013:PEI:2503210.2503232, besta2015accelerating}. In our benchmarks we use a
quadcore Haswell chip included in a commodity off-the-shelf server machine;
see Figure~\ref{fig:galilei_arch}. The L1 and L2 caches are private to each
core and the L3 inclusive cache is shared by all the cores. We select this
configuration to analyze a simple commodity multicore system.
\item[\textsf{Ivy Bridge}] is an Intel microarchitecture used in various
supercomputers such as Tianhe-2~\cite{liao2014milkyway} or NASA
Pleiades~\cite{pleiades}. Here, we evaluate an Ivy Bridge configuration
installed in a cluster Euler from ETH Zurich that contains two
12-core CPUs connected with Quick Path Interconnect (QPI).  The L1 and L2
are private to each core and the L3 inclusive cache is shared by all
the cores. We use this configuration to analyze the performance
characteristics of deep memory hierarchies with three cache levels.
\item[\textsf{Bulldozer}] is an AMD microarchitecture designed to improve
power efficiency for HPC applications~\cite{butler2010amd}. Here, we
evaluate a configuration included in the Cray XE6 Monte Rosa
supercomputer~\cite{vaughan2011application}; see
Figure~\ref{fig:amd_arch}. A compute node deploys two 16-core AMD Bulldozer
Interlagos CPUs. Each CPU hosts two 8-core dies that are connected with
HyperTransport (HT)~\cite{Slogsnat:2007:VLL:1216919.1216926}. We selected
this system to unveil differences between Intel and AMD systems and to
analyze the effects coming from a particularly complex design with 
three cache levels, multiple CPUs, shared L2
caches, and multiple dies per CPU.
\item[\textsf{MIC}] is an Intel state-of-the-art manycore architecture
deployed in Xeon Phi processors that targets massively parallel systems. We
evaluate a configuration with 61 cores. Each core has a private L1 and L2;
there is no L3. The cores are connected with a ring topology.  We use MIC
to analyze a highly parallel coprocessor installed in supercomputing
machines such as Tianhe-2~\cite{liao2014milkyway}.
\end{description}

\begin{table*}[!t]
\centering
\begin{tabular}{l|l||l|l|l|l}
\toprule
&\textsf{\textbf{Architecture:}} &\textsf{\textbf{Haswell}}& \textsf{\textbf{Ivy Bridge}} & \textsf{\textbf{Bulldozer}} & \textsf{\textbf{MIC}}  \\
\midrule
\multirowbt{7}{*}{\begin{turn}{90}\scriptsize \textsf{Processor}\end{turn}}
& Manufacturer &Intel &Intel &AMD & Intel \\
& CPU model& Core i7-4770& Xeon E5-2697v2 & Opteron 6272 & Xeon Phi 7120\\
&Cores/CPU &4 &12 &16(2x8)  & 61\\
&CPUs &1 &2 &2 & 1\\
&Core frequency &3400 MHz &2700 MHz &2100 MHz 1238 MHz &\\
&Interconnect & - & 2x QPI (8.0 GT/s) &4x HT 3.1 (6.4 GT/s) & - \\
\midrule
\multirowbt{10}{*}{\begin{turn}{90}\scriptsize \textsf{Caches}\end{turn}} & Cache line size &64B &64B &64B & 64B \\
&L1 cache &32KB per core &32KB per core&16KB per core & 32KB per core \\
&L1 Update policy& write back & write back& write through & write back \\
&L2 cache &256KB per core&256KB per core &2MB per 2 cores & 512KB per core \\
&L2 Update policy& write back & write back& write back & write back \\
&L2 incl/excl: & neither& neither & neither  & inclusive\\
&L3 cache &8MB fully shared& 30MB fully shared&8MB per 8 cores & - \\
&L3 Update policy& write back & write back& write back & - \\
&L3 incl/excl: &inclusive* & inclusive* & non-inclusive  & - \\
&CC protocol &MESIF &MESIF &MOESI  & MESI-GOLS \\
\midrule
\multirowbt{4}{*}{\scriptsize \textsf{Memory}} & Main memory &8GB &64GB &32GB & 8GB \\
&memory channels/CPU&1x dual channel &2x dual channel &2x dual channel  & 8x dual channel\\
&Huge page size &2MB & 2MB &2MB  & 2MB\\
\midrule
\multirowbt{4}{*}{\scriptsize \textsf{Others}} & Linux kernel used & 3.14-1 & 2.6.32 & 2.6.32 & 2.6.38.8\\

& \textsf{CAS} assembly instruction & \texttt{Cmpxchg} & \texttt{Cmpxchg} & \texttt{Cmpxchg} & \texttt{Cmpxchg} \\
 & \textsf{FAA} assembly instruction&\texttt{Xadd} & \texttt{Xadd} & \texttt{Xadd}  & \texttt{Xadd}\\
  & \textsf{SWP} assembly instruction&\texttt{Xchg} & \texttt{Xchg} & \texttt{Xchg} & \texttt{Xchg} \\

\bottomrule
\end{tabular}
\caption{The comparison of the tested systems. We denote the cache
coherency protocol as \texttt{CC protocol}. ``*'' indicates that the shared
inclusive L3 cache in Intel Haswell and Ivy Bridge contains a \emph{core
valid bit} for each core on the CPU that indicates whether a respective
core may contain a given cache line in its private higher level cache (the
bit is set) or whether it certainly does not contain this cache line (the
bit is zeroed).} \label{tab:tested_systems}
\end{table*}

The considered systems represent both multicore commodity off-the-shelf
machines (Haswell) and high-end manycore HPC systems (MIC, Ivy Bridge,
Bulldozer). They all use the same cache line size (64B) and use extensions
of the well known MESI~\cite{Molka:2009:MPC:1636712.1637764} cache
coherency protocol.
Haswell and Ivy Bridge utilize MESIF; it avoids redundant data
transfers from other cores or memory by adding the \textbf{F}orward state
to designate a cache to respond to any requests for the given shared
line~\cite{Hackenberg:2009:CCA:1669112.1669165}. 
AMD Bulldozer deploys the MOESI protocol that prevents write-backs to
memory by introducing the \textbf{O}wned state which allows a dirty cache
line to be shared~\cite{Hackenberg:2009:CCA:1669112.1669165}.
%
%
Finally, Xeon Phi deploys a protocol based on MESI that extends
the Shared state with a directory-based cache coherency protocol
GOLS (Globally Owned Locally Shared) to simulate the Owned state to enable
sharing of a modified line.

Another difference between the tested systems is the structure of L3; we
will later show that it impacts the performance of atomics.
Xeon Phi hosts no L3.  Ivy Bridge and Haswell deploy the inclusive L3 cache
where each cache entry contains a \emph{core valid bit} for each core on
the CPU. If this bit is set then the related core \emph{may} have the
respective cache line in its L1 or L2, possibly in a dirty state. If none
of the core valid bits is set (or if the cache line is not present in L3)
then the respective cache line is also not present in L1 and L2.
On the contrary, L3 in AMD Bulldozer is neither exclusive nor inclusive:
the presence of a cache line in L2 does not determine its presence in
higher level caches.  This will have a detrimental effect on the
performance of atomics as we will illustrate in
Section~\ref{sec:performanceAnalysis}.

\subsection{Evaluated Atomics Operations}
\label{sec:tested_mechanisms}

Finally, we discuss the selection of the evaluated atomics.

\begin{description}[leftmargin=0.5em]
\item[\textsf{Compare-and-Swap(*mem, reg1, reg2) (CAS):}] it loads the
value stored in \textsf{*mem} into \textsf{reg1}. If the original value in
\textsf{reg1} is equal to \textsf{*mem} then it writes \textsf{reg2} into
\textsf{*mem}. We select \textsf{CAS} because it is utilized in numerous
lock-free and wait-free data structures and
algorithms~\cite{herlihy2012art} as well as various graph processing codes
such Graph500~\cite{murphy2010introducing}. 
\item[\textsf{Fetch-and-Add(*mem, reg) (FAA):}] it fetches the value from a
memory location \textsf{*mem} into a register \textsf{reg} and adds the
previous value from \textsf{reg} to \textsf{*mem}. We selected \textsf{FAA}
because of its importance for implementing shared counters and various data
structures~\cite{Morrison:2013:FCQ:2442516.2442527}, and to analyze the
performance differences between \textsf{FAA} and \textsf{CAS}.
\item[\textsf{Swap(*mem, reg) (SWP):}] it swaps the values in a memory
location \textsf{*mem} and a register \textsf{reg}. We choose \textsf{SWP}
due to its significance in, e.g., implementing locks~\cite{herlihy2012art}.
\end{description}

Here, we focus on benchmarking the atomic assembly operations
and we thus assume that each operation loads only one operand from the memory subsystem.
The remaining operands are precomputed and stored in the respective registers.
For \textsf{CAS} we also evaluate the variant with two operands fetched.
Our strategy reflects many parallel codes and data structures where the 
arguments of the atomic function calls are constants or precomputed values;
for example BFS traversals~\cite{murphy2010introducing} or
distributed hashtables~\cite{fompi-paper, besta2014fault, besta2015active}.

The analyzed atomics have different \emph{consensus numbers}, where
\emph{consensus} is the problem of agreeing on one value in the presence of
many parties~\cite{herlihy2012art}. The consensus number of an operation
\texttt{op}, denoted as \texttt{CN(op)}, is the maximum number of threads
that can reach consensus with a wait-free algorithm that only uses reads,
writes, and \texttt{op}. In this evaluation, we select both the operations
that have smaller consensus numbers (\texttt{CN(SWP) = \texttt{CN(FAA) =
2}}) and the operation with a high consensus number (\texttt{CN(CAS) =
$\infty$}) to analyze whether it has any performance implications.

\section{Design of Benchmarks}
\label{benchmarkDesign}

Measuring the performance of atomics is non-trivial due to the complexity
of deep memory hierarchies, various types of workloads with different
caching patterns, and the richness of hardware mechanisms such as cache
prefetchers that influence the performance
results~\cite{Hackenberg:2009:CCA:1669112.1669165}.
We now present the methodology that overcomes these challenges.
%
%
%
We conduct:

\begin{description}[leftmargin=0.5em]
\item[\textsf{Latency benchmarks:} ] Here, pointer chasing is used to
obtain the average latency of an atomic. This benchmark targets
latency-constrained codes such as shared counters or synchronization
variables used in parallel data structures.
\item[\textsf{Bandwidth benchmarks:} ] Here, all the memory cells of a
given buffer are accessed sequentially and the bandwidth is measured.
While this part targets some bandwidth-intensive codes such as graph
traversals~\cite{murphy2010introducing}, it also shows that the tested
atomics do not enable any instruction-level parallelism (ILP) even if there
are no dependencies between issued operations.
\end{description}

\subsection{Relevant Parameters}

We focus on the following parameters that impact the performance of
atomics:

\begin{description}[leftmargin=0.5em]
\item[\textsf{Type of atomic:}] we evaluate different atomic operations to cover
a broad range of data structures and workloads (e.g., different types of concurrent
queues may use either \textsf{CAS}~\cite{shann2000practical} or \textsf{FAA}~\cite{wilson1988operating}). In addition, we
illustrate the tradeoffs between the consensus number of atomics~\cite{coulouris2005distributed} 
and their performance. We evaluate \textsf{CAS}, \textsf{SWP}, and \textsf{FAA}.
\item[\textsf{Cache coherency state:}] we use cache lines in various CC
states (\textsf{M},\textsf{E},\textsf{S},\textsf{O},\textsf{I}) to analyze
the impact of the CC protocol on the performance of atomics.
\item[\textsf{Cache proximity:}] we place the accessed cache line in
different caches to evaluate the impact of state-of-the-art deep cache
hierarchies. The data accessed by a core can be in its local cache or in
another core's cache located: on the same die, on a different die
but on the same CPU, or on a different CPU.
\item[\textsf{Memory proximity:}] we use memories with different
proximities to cover today's NUMA systems. We will refer to a memory that
can be accessed by a core without using a die-die interconnection as the
\emph{local memory} and anything else as the \emph{remote memory}~\cite{tate2014programming}.
\item[\textsf{Thread count:}] we vary the number of threads accessing the
same cache line to analyze the overheads due to contention.
\item[\textsf{Operand size:}] we evaluate operations that modify operands
of various sizes to discover the most advantageous size to be used for
shared counters or synchronization variables.
\end{description}

\subsection{Structure of Benchmarks}
\label{sec:benchmark_structure}

The general structure of the benchmarks is similar to the structure
described in Section~\ref{sec:background_benchmarking_accesses} with the
difference of measuring atomic instructions instead of reads or writes.
\textsf{CAS} however is a special case which needs further adjustments.
When the old value in the register (\textsf{reg1}) does not correspond to
the value in memory (\textsf{*mem}), \textsf{CAS} fails and no memory
location will be modified. However, when the old value \textsf{reg1} is
equal to \textsf{*mem}, \textsf{CAS} succeeds and there will be a write to
memory. We investigate these cases separately.

\textbf{\textsf{CAS: Bandwidth Benchmarks }}
In the bandwidth benchmarks for successful \textsf{CAS}, we fill the buffer
with zeros and use zero as the old value (\textsf{reg1}). For unsuccessful
\textsf{CAS}, we fill the buffer with the increasing byte values. When a
\textsf{CAS} fails, \textsf{reg1} is updated to \textsf{*mem}. This value
will differ from the next one in the buffer, ensuring that all the issued
\textsf{CAS} operations will fail. 

\textbf{\textsf{CAS: Latency Benchmarks }}
We measure the latency of the unsuccessful \textsf{CAS} by filling the
buffer with the increasing values and comparing each new fetched value with
the previous one.  This ensures that each \textsf{CAS} fails.
For \textsf{CAS} to be successful we need to know \textsf{*mem} in advance.
With the pseudo random addresses this cannot be achieved without additional
memory accesses that would interfere with the benchmarks. Instead, we use
another approach: We fill the buffer with zeros, split it into equally
sized chunks, and perform a predefined access pattern using the beginning of
each chunk as the base address. If we benchmarked reads with this approach
they would be executed in parallel because there is no data dependency
between the reads, preventing the exact latency measurement. However,
\textsf{CAS} affects the register containing the old value and that value
also affects the outcome of the next operation so there is a data
dependency. Thus, the instructions are serialized and their latency
can be correctly measured.


\textbf{\textsf{CAS vs FAA vs SWP: Instruction Level Parallelism }}
On all the tested systems the \textsf{CAS} assembly operation always
modifies the same predefined register. Thus, the CPU cannot execute
multiple \textsf{CAS}es simultaneously because the result of one
\textsf{CAS} affects the outcome of the next \textsf{CAS}. \textsf{FAA} and
\textsf{SWP} however have only one explicit argument. Our bandwidth
benchmarks avoid data dependencies between the instructions to allow
parallel execution of \textsf{FAA} and \textsf{SWP}.  We will later
illustrate that the hardware implementation of each atomic still enforces
fully serialized execution.

\subsection{Interference from Hardware Mechanisms}

There are several mechanisms that could introduce significant
noise in the benchmarks; we turn them off where possible.
First, we avoid TLB misses by using hugepages and filling
the TLB with proper entries prior to the measurements.
Second, we disable the respective mechanisms that affect the clock frequency; these
are Turbo Boost, Enhanced Intel SpeedStep (EIST), and CPU C-states.  Thus,
the frequency of each core is always as specified in
Table~\ref{tab:tested_systems}.
Third, we turned off prefetchers (Hardware Prefetcher, Adjacent Cache Line
Prefetch) to prevent false speedups in the latency benchmarks.
In some of the systems (Ivy Bridge, Bulldozer) we could not influence the
hardware configuration and we avoided prefetching by applying sparser
access patterns.
Finally, by switching off HyperThreading we make sure that any two cores
visible to the programmer are also two physical cores.

\section{Performance Model}
\label{sec:performanceModel}

We now introduce our performance model.
We concretize the model by assuming that we model caching architectures
that match the considered Intel and AMD systems
(cf.~Section~\ref{sec:tested_systems} and~Table~\ref{tab:tested_systems}).
We will later (Section~\ref{sec:performanceAnalysis}) validate the model
and explain several differences between the predictions and the data that
illustrate interesting architectural properties of the considered systems.

\subsection{Latency}
\label{sec:latency_model}

Each atomic fetches and modifies a given cache line
(``read-modify-write'').
We predict that an atomic first issues a read for ownership in order to
fetch the respective cache line and invalidate the cache line copies in
other caches.
Then the operation is executed and the result is written in a modified
state to the local L1 cache.
We thus model the latency $\mathcal{L}$ of an atomic operation $A$
executing with an operand from a cache line
in a coherency state $S$ as:

\begin{alignat}{2}
\mathcal{L}(A,S) &= \mathcal{R}_{O}(S) + \mathcal{E}(A) + \mathcal{O}
\end{alignat}

$A$ denotes the analyzed atomic; $A \in \{\textsf{CAS}, \textsf{FAA},
\textsf{SWP}\}$.
$S$ denotes the coherency state; $S \in
\{\textsf{E},\textsf{M},\textsf{S},\textsf{O}\}$.
$\mathcal{R}_{O}(S)$ is the latency of the read for ownership (reading
a cache line in a coherency state $S$ and invalidating other caches).
$\mathcal{E}(A)$ is the latency of: locking a cache line, executing
$A$ by the CPU, and writing the operation result into a cache line in
the coherency state \textsf{M}.  As all other copies of the cache line are
invalidated, this will be a write into L1 local to the core executing the
instruction.
Finally, $\mathcal{O}$ denotes additional overheads related to various
proprietary optimizations of the coherence protocols that we describe
in Section~\ref{sec:performanceAnalysis}.
%
%
%
%
We conjecture that the most dominant element of $\mathcal{L}(A,S)$ is
$\mathcal{R}_{O}(S)$; a prediction supported by several studies
illustrating high latencies of reads for
ownership~\cite{Hackenberg:2009:CCA:1669112.1669165,Molka:2009:MPC:1636712.1637764,Molka:2014:MMC:2618128.2618129}.

$\mathcal{R}_{O}(S)$ strongly depends on $S$ and the location of the cache line.
We start with modeling operations that access cache lines located on the
same die as the requesting core. 

\subsubsection{On-die Accesses: E/M states}
\label{sec:on-die-model}

If $S$ is \textsf{E} or \textsf{M} then there is a single copy of the related cache line and
no invalidations will be issued. Thus, $\mathcal{R}_{O}(\textsf{E})$ and
$\mathcal{R}_{O}(\textsf{M})$ will be equal to the latency of a simple
read denoted as $\mathcal{R}$:

\begin{alignat}{2}
\mathcal{R}_{O}(\textsf{E/M}) &= \mathcal{R}(\textsf{E/M})
\end{alignat}

\textbf{\textsf{Private L1 and L2, shared L3 }}
We first assume that each core has private L1 and L2 caches and there is a
shared L3 across all the cores.
Examples of such systems are the considered Intel Ivy Bridge and Intel
Haswell configurations.
%
%
%
We first denote the latency of reading a cache line by a core from a local
L1, L2, and L3 cache as $\mathcal{R}_{L1,l}$, $\mathcal{R}_{L2,l}$, and
$\mathcal{R}_{L3,l}$, respectively.
Then, we have:

\begin{alignat}{2}
\mathcal{R}(\textsf{E/M}) &= \mathcal{R}_{L,l}  \mbox{ iff the cache line is in L}
\end{alignat}

where $L \in \{L1, L2, L3\}$. 
%
%
%
%
%
We now model the latency of accessing a cache line in L1 or L2 of a different core.
Here, we assume that the latency of transferring a cache line between L1 and L3 
can be estimated as $\mathcal{R}_{L3,l} - \mathcal{R}_{L1,l}$.
%
%
The total latency is increased by an additional cache line transfer from L3 to the requesting core:


\begin{alignat}{1}
\mathcal{R}(\textsf{E/M}) = \mathcal{R}_{L3,l} + \mathcal{R}_{L3,l} - \mathcal{R}_{L1,l}
\end{alignat}


\textbf{\textsf{Private L1, shared L2 and L3 }}
In some architectures (e.g., Bulldozer) L2 is shared.
%
%
For such systems, if the cache line is in the L1 owned by a core that shares L2 with the requesting core, then:

\begin{alignat}{1}
\mathcal{R}(\textsf{E/M}) = \mathcal{R}_{L2,l} + \mathcal{R}_{L2,l} - \mathcal{R}_{L1,l}
\end{alignat}


\begin{figure*}[!t]
\centering
 \subfloat[\hspace{-0.3em}\textsf{SWP/CAS/FAA}, Exclusive state]{
  \includegraphics[width=0.3\textwidth]{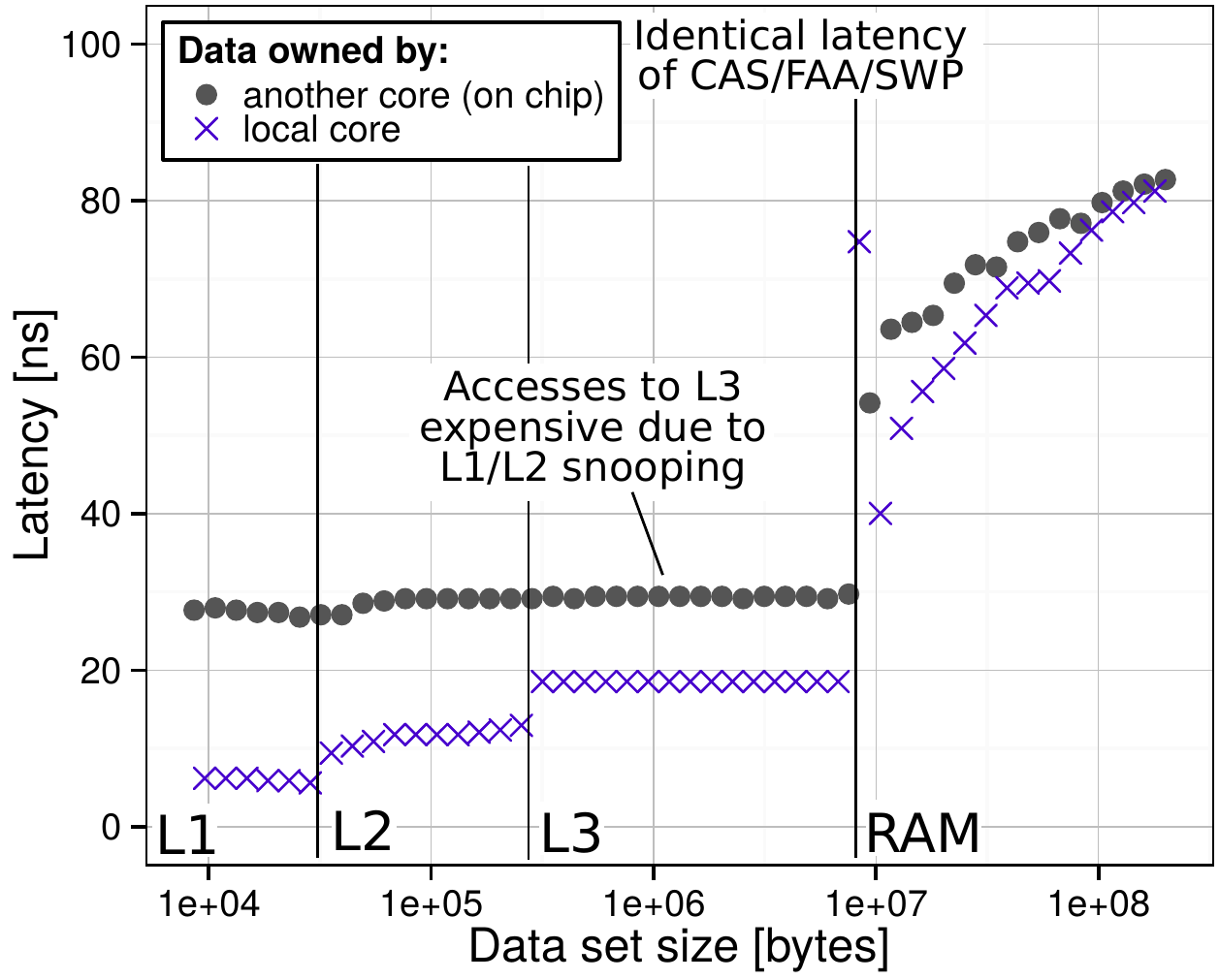}
  \label{fig:Haswell_cas_64_E}
 }
 \subfloat[\textsf{SWP/CAS/FAA}, Modified state]{
  \includegraphics[width=0.3\textwidth]{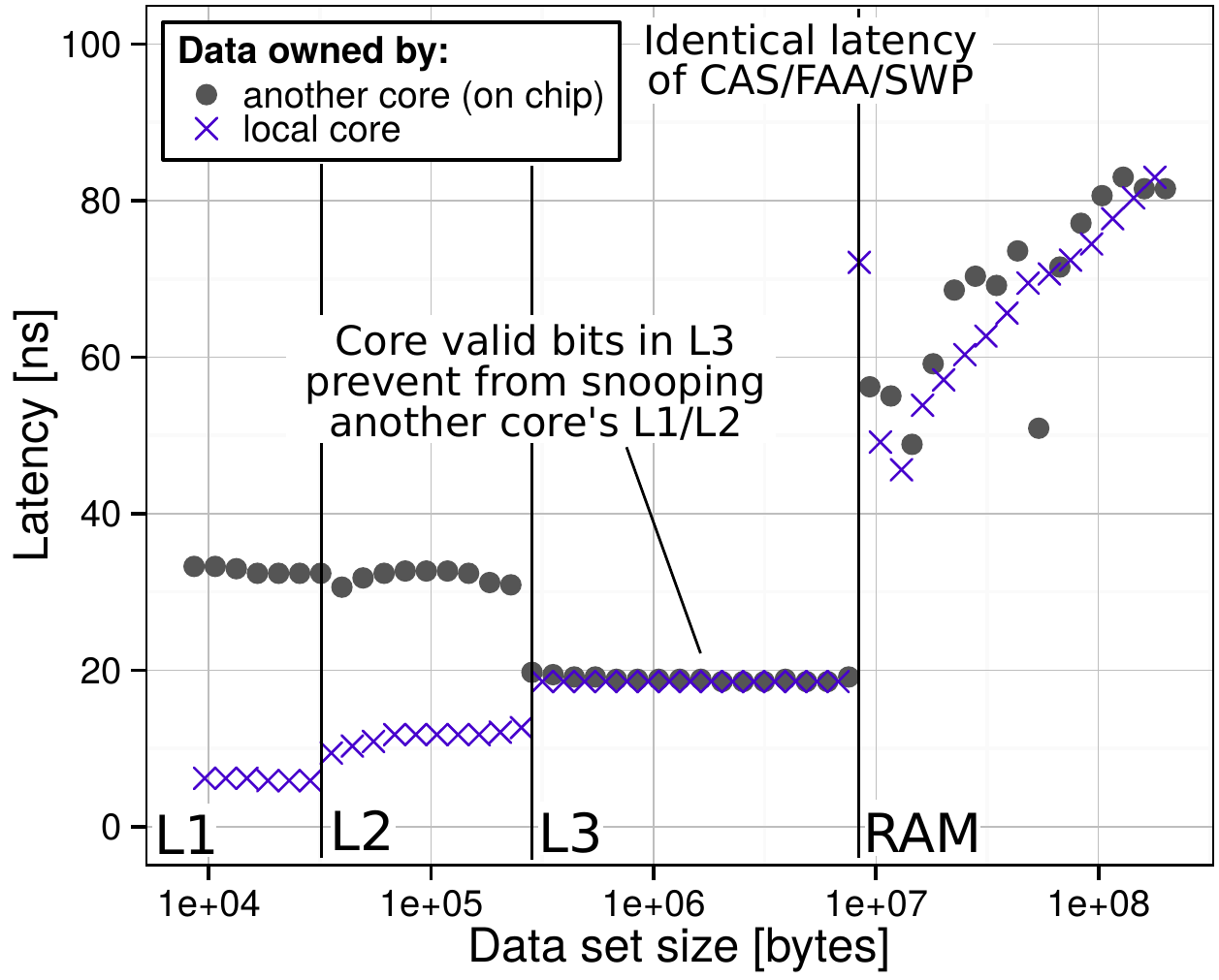}
  \label{fig:Haswell_cas_64_M}
 }
 \subfloat[\textsf{SWP/CAS/FAA}, Shared state]{
  \includegraphics[width=0.3\textwidth]{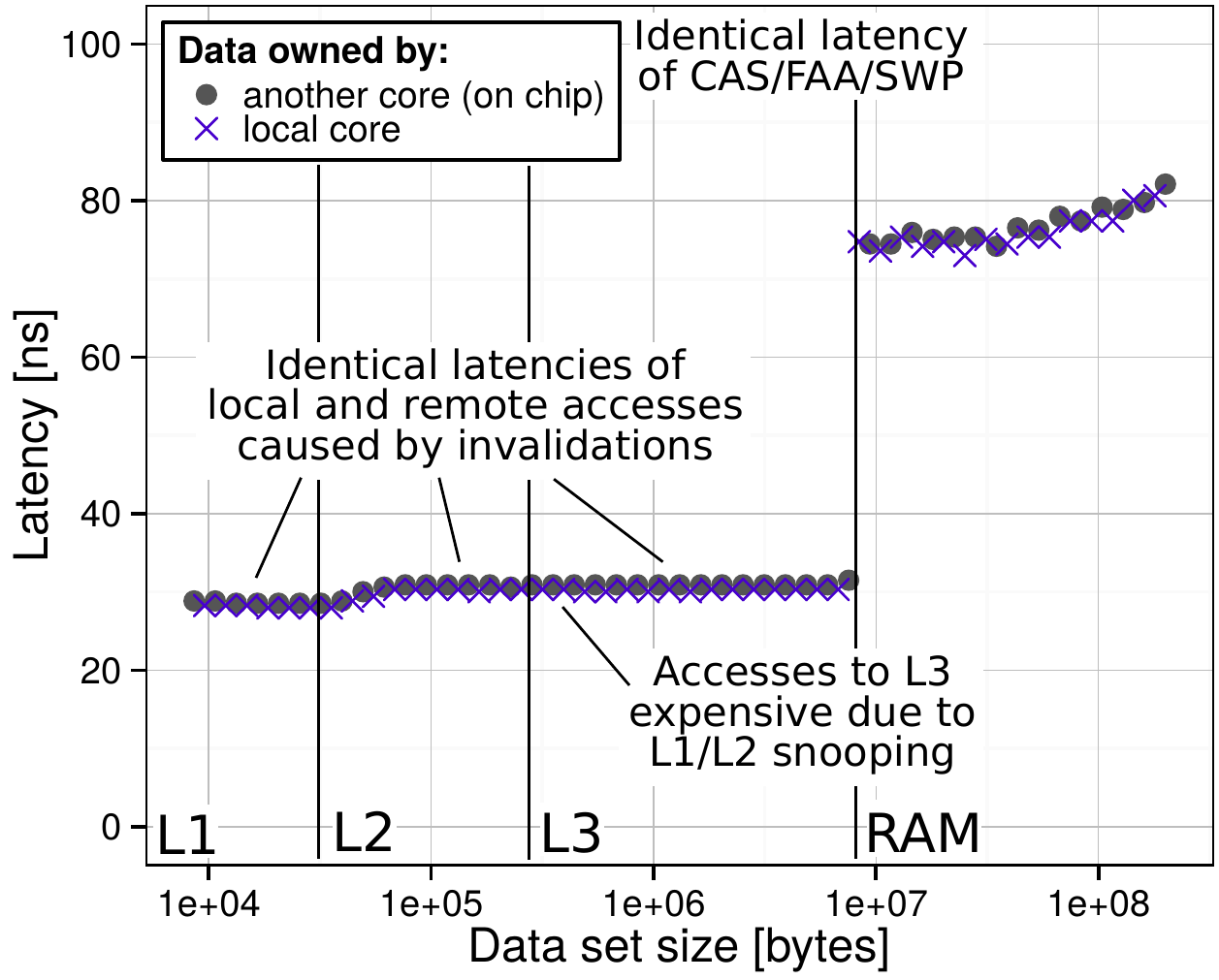}
  \label{fig:Haswell_cas_64_S}
 }\\
   \subfloat[\textsf{read}, Exclusive state]{
  \includegraphics[width=0.3\textwidth]{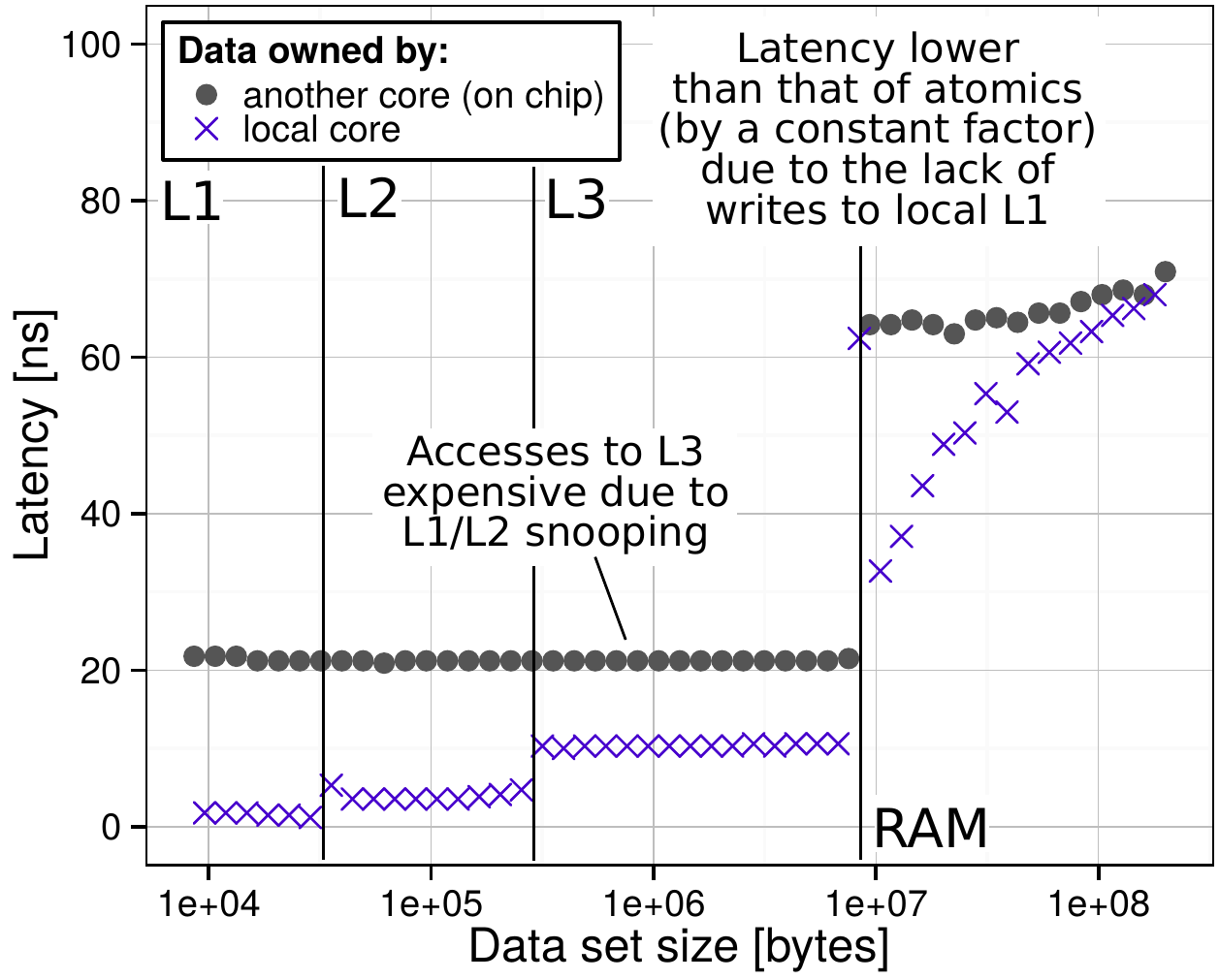}
  \label{fig:Haswell_read_64_E}
 }
  \subfloat[\textsf{read}, Modified state]{
  \includegraphics[width=0.3\textwidth]{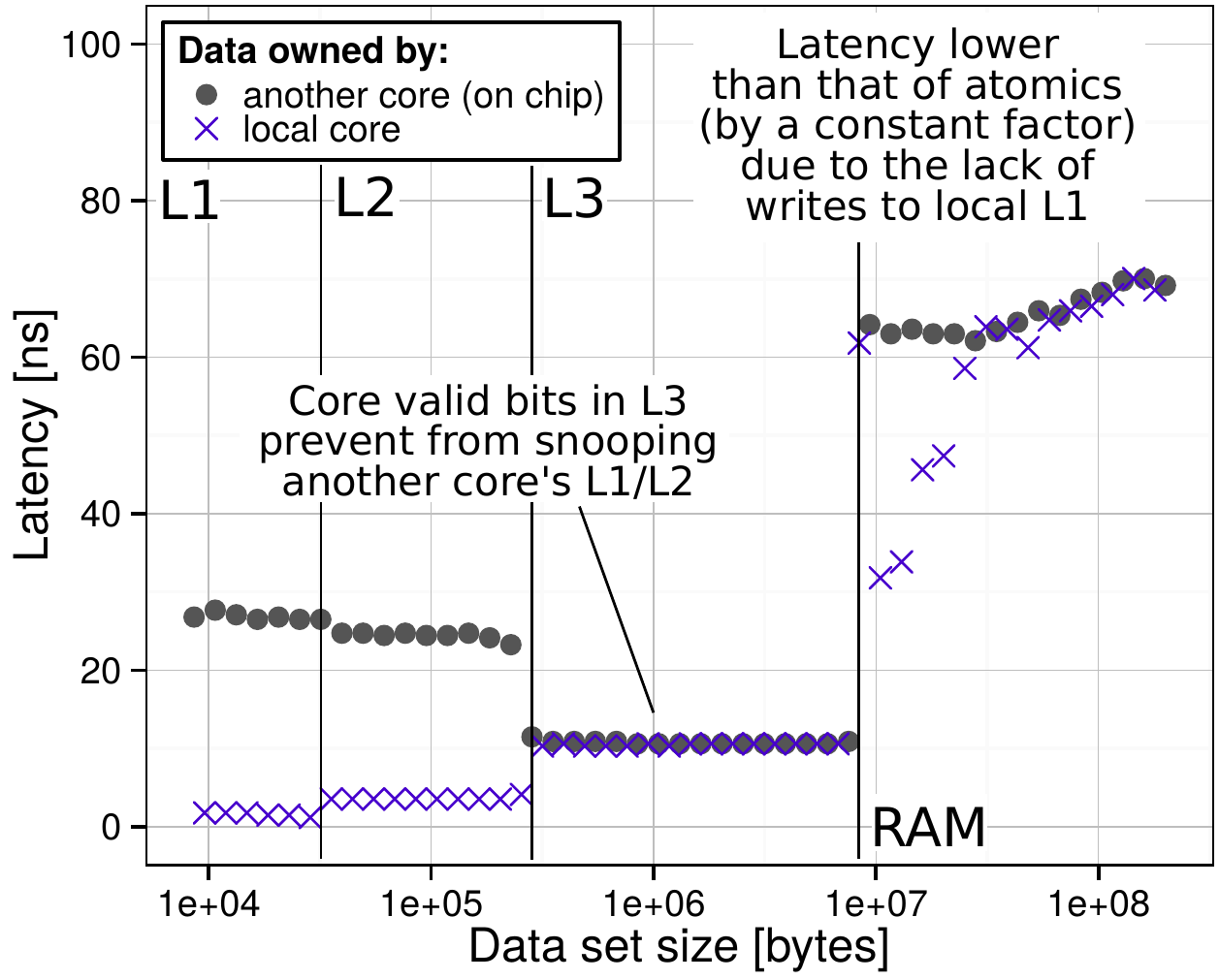}
  \label{fig:test2}
 }
  \subfloat[\textsf{read}, Shared state]{
  \includegraphics[width=0.3\textwidth]{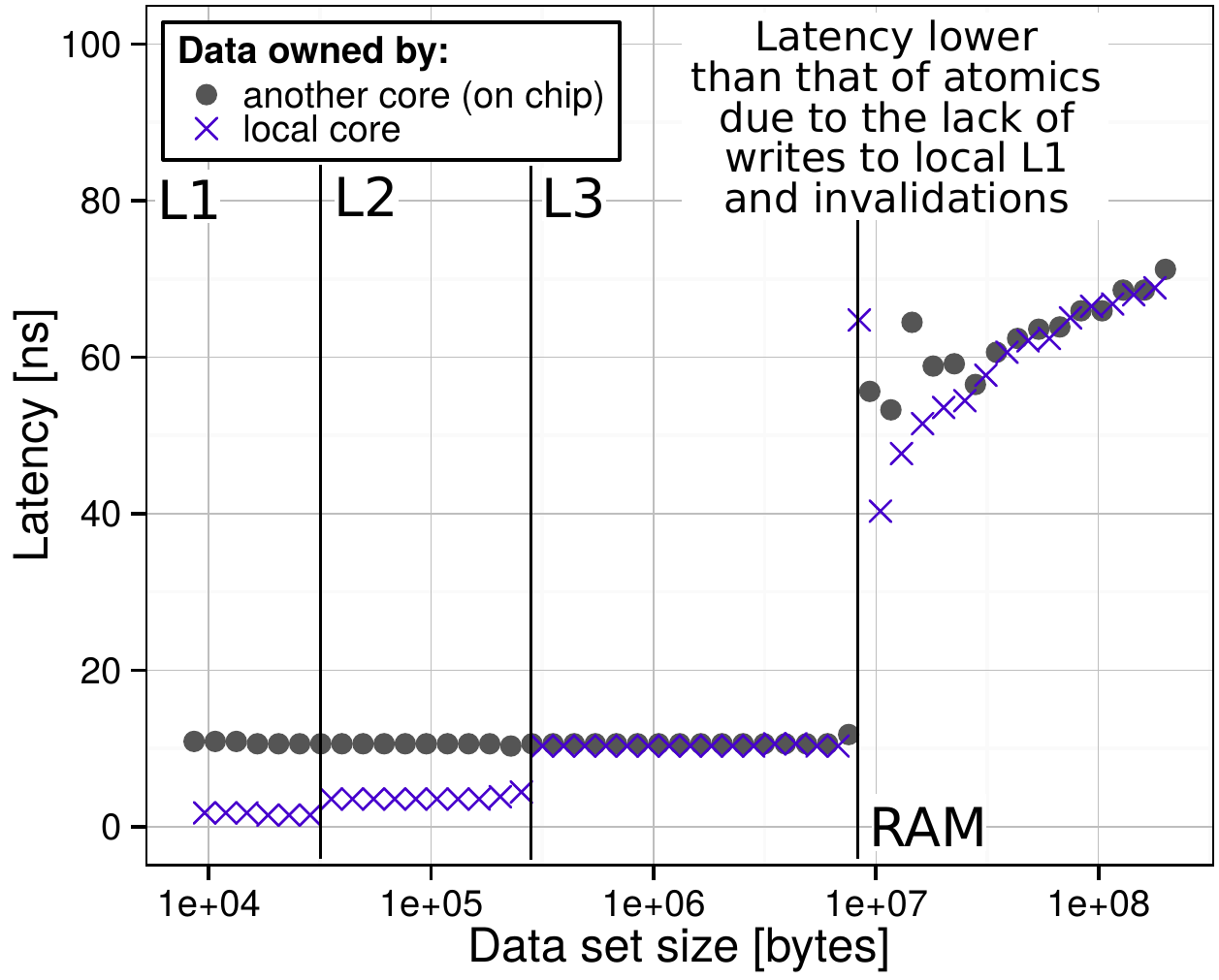}
  \label{fig:test2}
 }
\caption{The comparison of the latency of \textsf{CAS}, \textsf{FAA},
\textsf{SWP}, and \textsf{read} on Haswell. The requesting core accesses
its own cache lines (local) and  cache lines of a different core from the
same chip (on chip).} \label{fig:Latency_results_Haswell}
\end{figure*}

\textbf{\textsf{Private L1 and L2, no L3 }}
Finally, Xeon Phi represents systems with private L1/L2 and no L3.  Recent
research~\cite{Ramos:2013:MCC:2493123.2462916} illustrated that the
latency of a cache line transfer between any two cores on a Xeon Phi chip
can be assumed identical. 
Thus, we have:

\begin{alignat}{1}
\mathcal{R}(\textsf{E/M}) = \mathcal{R}_{L2,l} + \mathcal{R}_{L2,l} - \mathcal{R}_{L1,l} + \mathcal{H}
\end{alignat}

where $\mathcal{H}$ is the latency of the hop from the local L2 to the
remote L2, including the overhead from accessing the directories maintained
by the cache coherency protocol.

\subsubsection{On-die Accesses: S/O states}
\label{sec:on-die-model}

If the cache line is in \textsf{S} or \textsf{O}, then the read for ownership
invalidates its copies in other caches.  Assuming there are $N$
copies, we have:

\begin{alignat}{2}
\mathcal{R}_{O}(\textsf{S/O}) &= \mathcal{R}(\textsf{S/O}) + \max_{i \in \{1,...,N\}} \mathcal{L}_{inv,i}
\end{alignat}

where $\mathcal{L}_{inv,i}$ is the latency of invalidating the $i$th cache line.
Here, we assume that multiple invalidations are executed in parallel, thus
we take the maximum of the latencies.
We also predict that $\mathcal{L}_{inv,i}$ should not significantly
differ from $\mathcal{R}(\textsf{E})$ (of the $i$th cache line), because both require
invalidating private caches independent of data being cached there.
Similarly, we approximate reads of \textsf{S/O} cache lines with
reads of \textsf{E} cache lines:

\begin{alignat}{2}
\mathcal{R}_{O}(\textsf{S/O}) &= \mathcal{R}(\textsf{E}) + \max_{i \in \{1,...,N\}} \mathcal{R}_{i}(\textsf{E})
\end{alignat}


\subsubsection{Off-die Accesses}
\label{sec:off-die-model}

The operations accessing cache lines located on a different die include an
additional penalty from the underlying network (QPI on Intel and HT on AMD
systems). Here, we assume a constant overhead $\mathcal{H}$ per one
cache-to-cache hop (spanning two dies) that we add to the respective
latency expressions from Section~\ref{sec:on-die-model}.
The latency of accesses to the main memory $\mathcal{M}$ is modeled as a
sum of the L3 miss and the overhead introduced by processing the request by
the memory controller. For NUMA systems we also add $\mathcal{H}$ if necessary
for an additional die-to-die hop.
Finally, on Intel systems we also add $\mathcal{M}$ to each
$\mathcal{R}(\textsf{M})$ because such accesses require writebacks to
memory; AMD prevents it with the \textsf{O} state.

\subsection{Bandwidth}
\label{sec:bandwidth_model}

Here, we assume that the tested atomics always flush the write
buffers and thus do not allow for
ILP~\cite{intel_arch_opt_manual,intel_soft_opt_manual}.
Thus, the bandwidth $\mathcal{B}$ of an atomic $A$ executing with an
operand from a cache line in a coherency state $S$ can be simply
modeled as:

\begin{alignat}{2}
\mathcal{B}(A,S) &= \mathcal{C}_{size}/\mathcal{L}(A,S) 
\end{alignat}

where $\mathcal{C}_{size}$ is the cache line size.
This model assumes that each atomic modifies a different cache line. In
the case where the continuous memory block is accessed sequentially and
thus each cache line is hit multiple times, we have:
 
\begin{alignat}{2}
\mathcal{B}(A,S) &= \frac{ \mathcal{N} }{\mathcal{L}(A,S) + (\mathcal{N} - 1)\cdot (\mathcal{R}_{L1,l} + \mathcal{E}(A))} \cdot \mathcal{C}_{size}
\end{alignat}
 
where $O_{size}$ is the operand size and $\mathcal{N} = \mathcal{C}_{size}
/ \mathcal{O}_{size}$ is the number of operands that fit into a
cache line. Eq.~(10) is valid for Intel systems. L1 on AMD is
write-through, thus $\mathcal{R}_{L2,l}$ would replace
$\mathcal{R}_{L1,l}$:
 
\begin{alignat}{2}
\mathcal{B}(A,C,S) &=  \frac{ \mathcal{N} }{\mathcal{L}(A,C,S) + (\mathcal{N} - 1)\cdot \mathcal{L}_{R,L2,l}} \cdot \mathcal{C}_{size}
\end{alignat}

%
%
%
%
%
%

\section{Performance Analysis}
\label{sec:performanceAnalysis}


We now illustrate our main results and provide several surprising insights into
the performance of the tested atomics.
We exclude the results that show similar performance trends
and provide them in the Appendix.
We use the results to validate the model. Here, we first calculate the
median values of the parameters from Section~\ref{sec:performanceModel}.
The obtained numbers can be found in Table~\ref{tab:model_params}.
We omit the model lines for clarity of plots and we discuss each case where
the differences between the model and the data exceed 10\% of the
normalized root mean square error (NRMSE) defined as:
%

\begin{alignat}{2}
\text{NRMSE} &= \frac{1}{\bar{x}} \sqrt{\frac{1}{n} \sum_{i=1}^{n}(\hat{x}_i-x_i)^2}
\end{alignat}

where $\hat{x}_i$ are predictions, $x_i$ are the observed data points,
and $\bar{x}$ is the mean of the observed values.

\begin{table}[!h]
\centering
\begin{tabular}{l||l|l|l|l}
\toprule
\textsf{\textbf{Param.:}} &\textsf{\textbf{Haswell}}& \textsf{\textbf{Ivy Bridge}} & \textsf{\textbf{Bulldozer}} & \textsf{\textbf{Xeon Phi}} \\
\midrule
$\mathcal{R}_{L1,l}$ &1.17 &1.8 & 5.2 & 2.4\\
$\mathcal{R}_{L2,l}$ &3.5  &3.7  & 8.8 & 19.4\\
$\mathcal{R}_{L3,l}$ &10.3  &14.5  & 30 & -\\
$\mathcal{H}$ &	- &66 & 62& 161.2\\ 
$\mathcal{M}$ & 65  &80  & 75& 340\\ 
$\mathcal{E}(\textsf{CAS},.)$ &4.7  &4.8 & 25 & 12.4\\ 
$\mathcal{E}(\textsf{FAA},.)$ &5.6  &5.9  & 25 & 2.4\\ 
$\mathcal{E}(\textsf{SWP},.)$ &5.6  &5.9 &  25 &  3.1\\ 
\bottomrule
\end{tabular}
\caption{The model parameters (all numbers are in nanoseconds).}


\label{tab:model_params}
\end{table}

The overhead term $\mathcal{O}$ (see Eq.~(1)) depends on the
operation type, the coherency state, the accessed cache
line, and the architecture; we present a selection of $\mathcal{O}$ values in Table~\ref{tab:O}.
The available vendor documentations and manuals prevent the
definite explanation of the reasons behind the variability of
$\mathcal{O}$~\cite{amd_opt_manual,intel_soft_opt_manual,intel_arch_opt_manual}.
We conjecture that the reasons may include: proprietary undocumented
optimizations, variable locking overheads, or different snooping techniques
(e.g., whether or not the snoop request bypasses the targeted core).


\begin{table}[h]
\centering
\begin{tabular}{@{}l|llllll@{}}
\toprule
              & \multicolumn{3}{c}{{\bf Local}} & \multicolumn{3}{c}{{\bf Remote}} \\
                            & {\bf L1}  & {\bf L2} & {\bf L3} & {\bf L1}  & {\bf L2}  & {\bf L3} \\ \midrule
                            {\bf E state} & 0         & 3.8      & 3.5      & 3         & 5         & 5        \\
                            {\bf M state} & 0         & 3.8      & 3.5      & 9         & 8         & 5        \\
                            {\bf S state} & 3         & 1.4      & -4       & -15       & -14       & -12      \\ \bottomrule
                            \end{tabular}
\caption{The O term for Haswell.}
                            \label{tab:O}
                            \end{table}

\subsection{Latency}

\begin{figure}[!h]
\centering
  \includegraphics[width=0.33\textwidth]{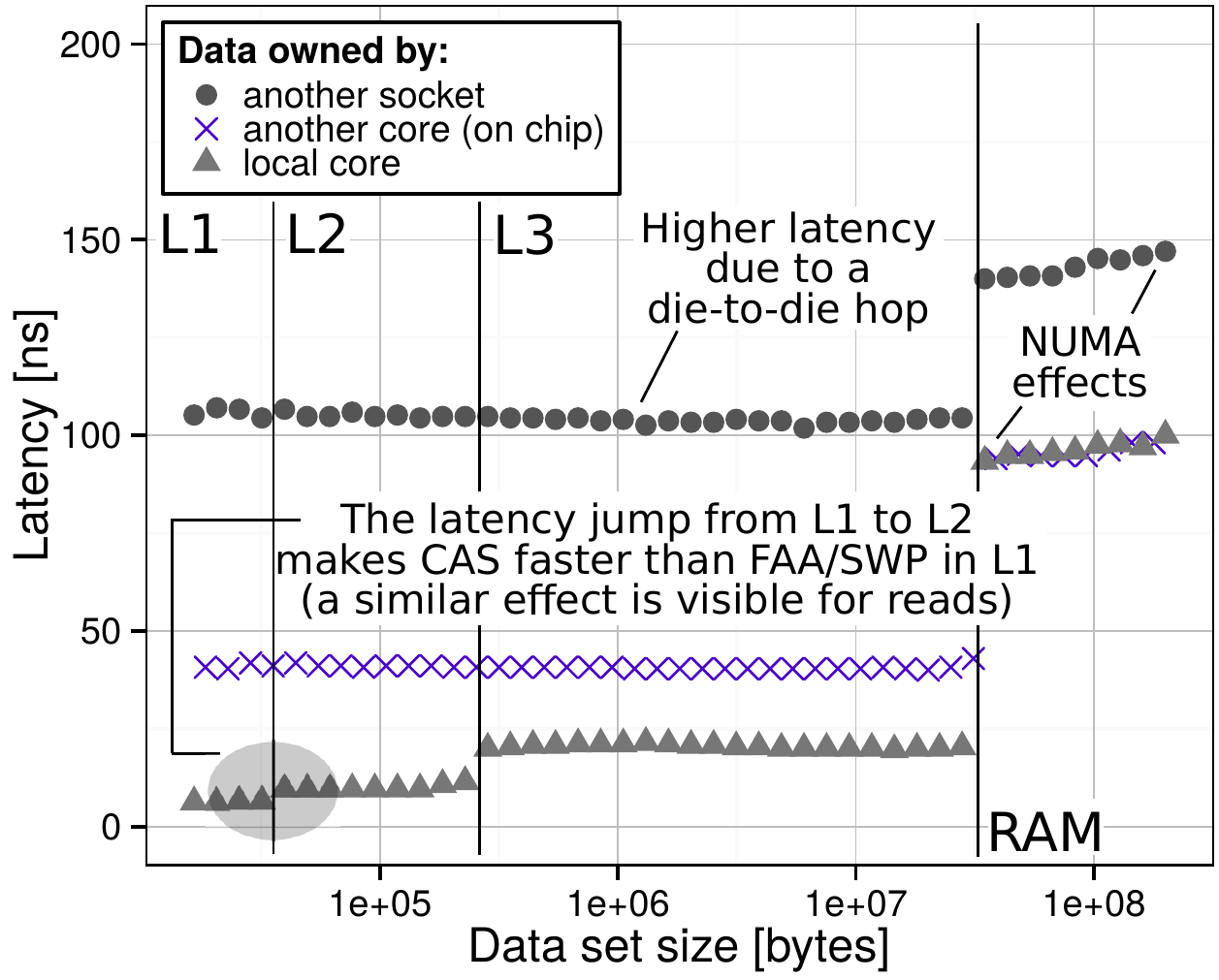}
\caption{The analysis of the CAS latency (Exclusive state) on Ivy Bridge
and the comparison to \textsf{FAA}/\textsf{SWP}. The
requesting core accesses its own cache lines (local), cache lines of a
different core from the same chip (on chip), and cache lines of a
core from a different socket (another socket).}
\label{fig:Latency_results_Ivy}
\end{figure}

\begin{figure*}[!t]
\centering
 \subfloat[\textsf{CAS/SWP}, Exclusive state]{
  \includegraphics[width=0.3\textwidth]{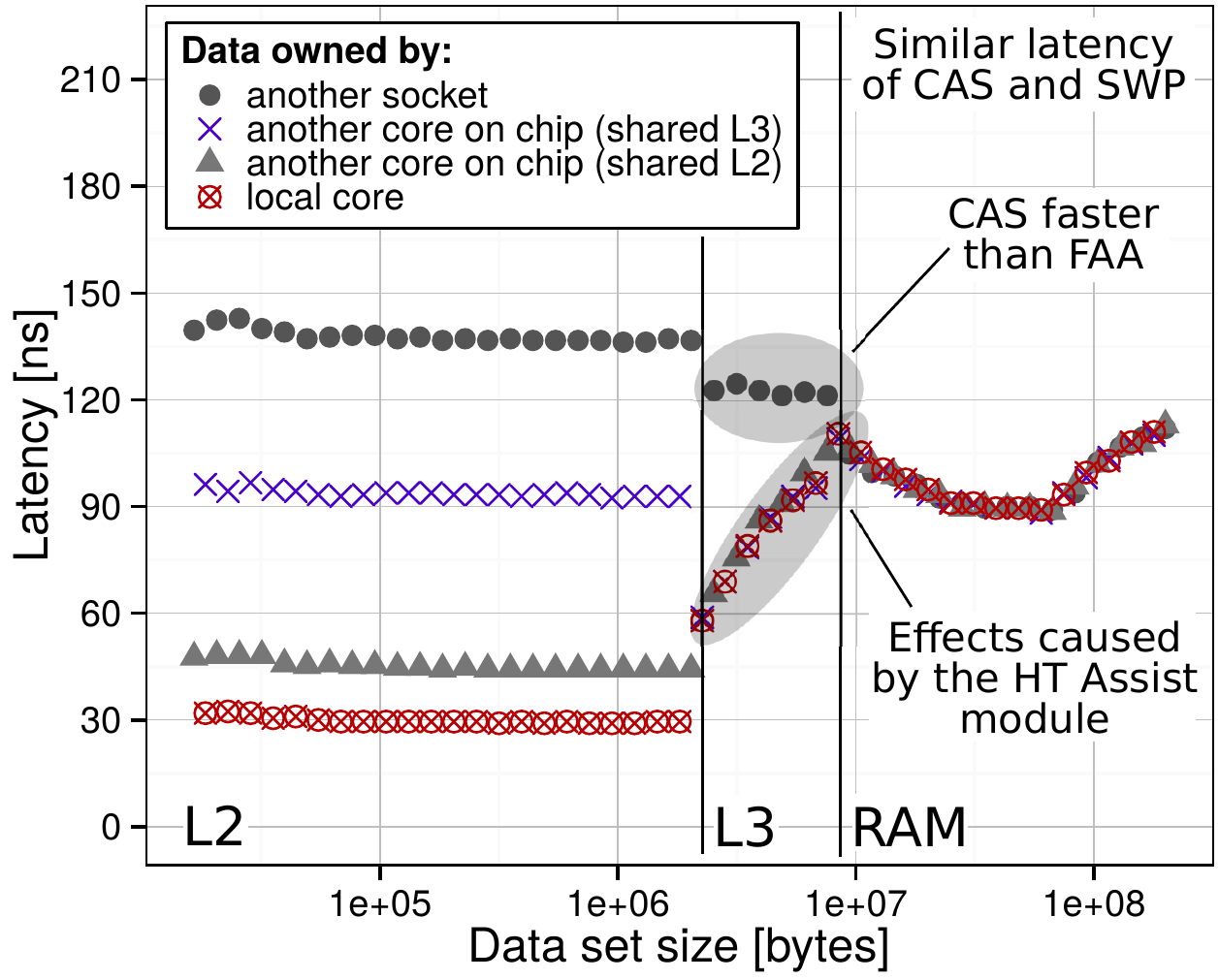}
  \label{fig:Bull_cas_E}
 }
 \subfloat[\textsf{FAA}, Exclusive state]{
  \includegraphics[width=0.3\textwidth]{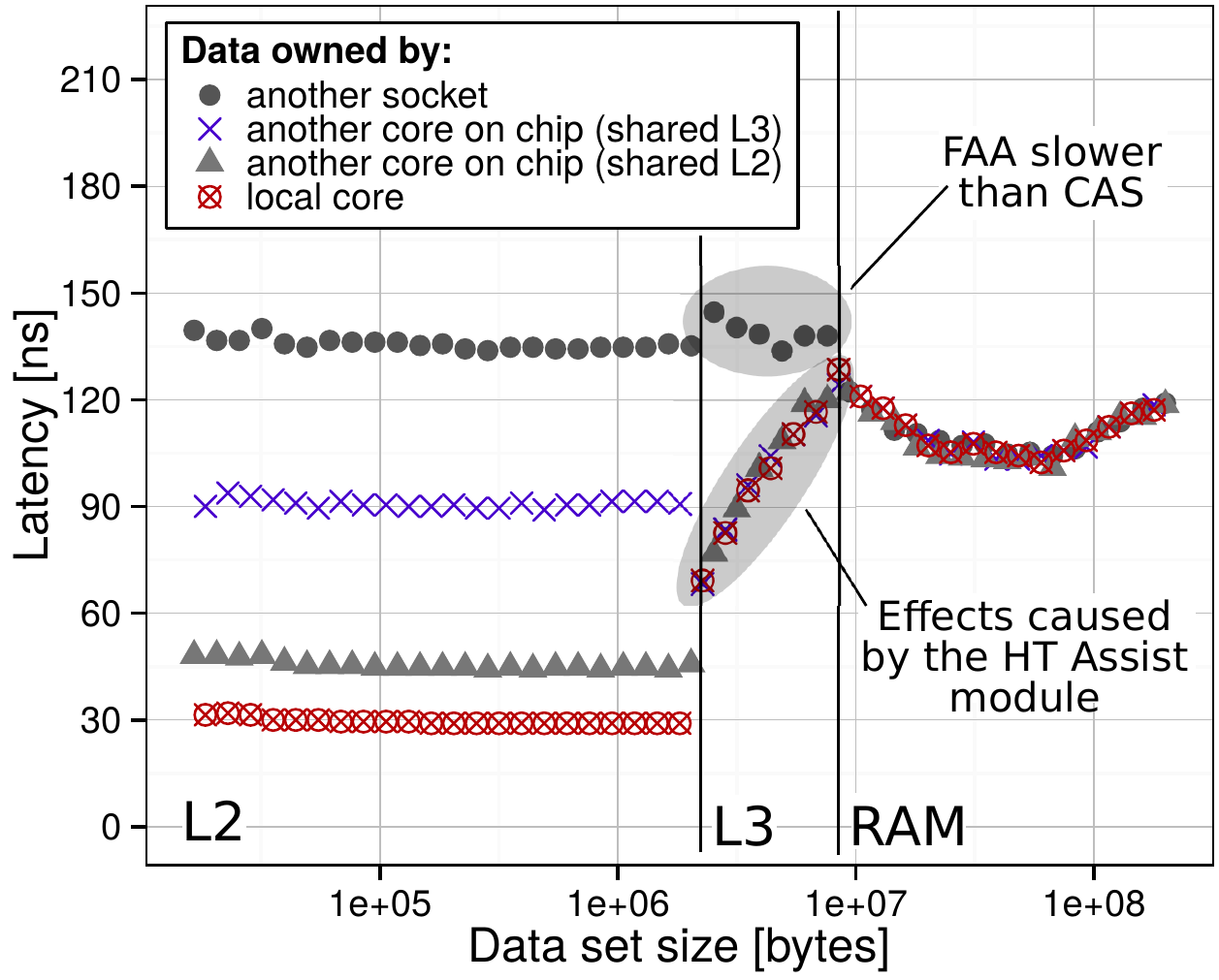}
  \label{fig:Bull_fad_E}
 }
  \subfloat[\textsf{read}, Exclusive state]{
  \includegraphics[width=0.3\textwidth]{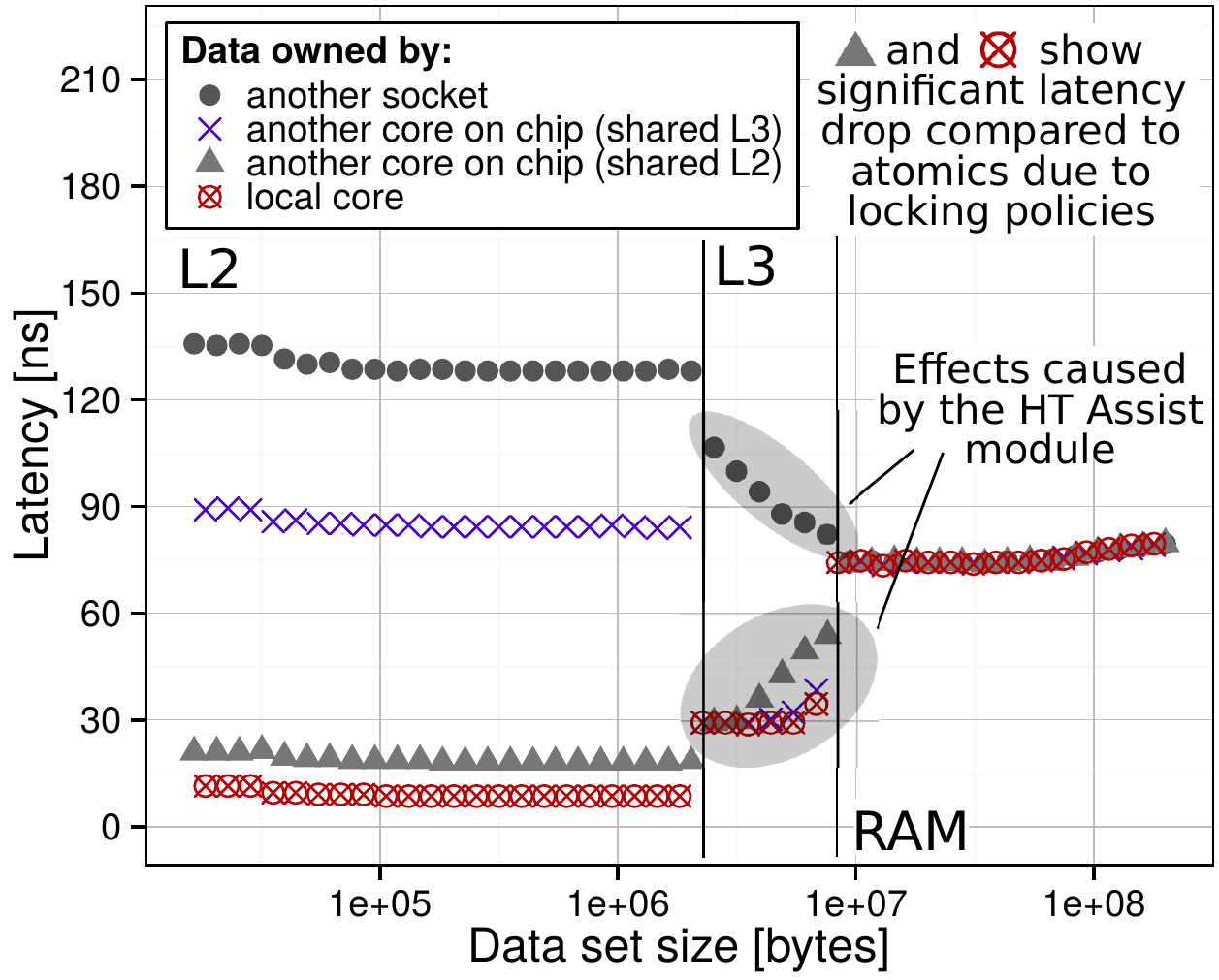}
  \label{fig:Bull_read_E}
 }
\caption{The latency comparison of
\textsf{CAS}/\textsf{FAA}/\textsf{SWP}/\textsf{read} on Bulldozer. The
requesting core accesses its own cache lines (local),  cache lines of
different cores that share L2 and L3 with the requesting core (on chip,
shared L2 and L3, respectively), and cache lines of a core from a
different socket (other socket).} \label{fig:Latency_results_Bull}
\end{figure*}

First, we present a selection of the latency results.
%
We compare \textsf{CAS}, \textsf{FAA}, and \textsf{SWP} that access cache
lines in the \textsf{E}, \textsf{M}, \textsf{S} and \textsf{O} states.
We exclude the \textsf{F} state
from the Intel analysis as it starts to affect the performance when more
than two CPUs are used~\cite{Molka:2014:MMC:2618128.2618129} while our
Intel testbeds host at most two CPUs.
We illustrate the results for unsuccessful \textsf{CAS}; successful
\textsf{CAS} follows similar performance patterns.
Finally, the latency of reads (\textsf{read}) is also plotted for a
baseline comparison with a simple memory access.
%

\subsubsection{Intel Haswell and Ivy Bridge}

We illustrate the latency results of the Intel systems in
Figures~\ref{fig:Latency_results_Haswell}
and~\ref{fig:Latency_results_Ivy}.
%
%
Atomics are consistently slower than reads by $\approx$5-10ns on both
systems for the \textsf{E}/\textsf{M} states (cf.
Figures~\ref{fig:Haswell_cas_64_E} and~\ref{fig:Haswell_read_64_E}). From
this we conjecture that atomics trigger a read for ownership and the
latency difference between atomics and simple reads stems from
$\mathcal{E}$, as predicted by Eq.~(1).
The desired cache line is read into the private cache of the core and all
its copies are invalidated. For cache lines in the \textsf{E} and
\textsf{M} states a read for ownership has the same latency as a
\textsf{read} since the line is only present in one cache, requiring no
invalidations.
%
%
The difference in latency impacts the performance of local L1 
accesses where the read latency is $\approx$1-2ns (see
Figure~\ref{fig:Haswell_read_64_E}). It does not significantly influence
accesses to remote caches or memory where latencies are $>$60ns. 
\emph{As predicted by our model and contrary to the common view, \textsf{CAS} has
the same latency as \textsf{FAA}/\textsf{SWP}}, except for the \textsf{E}
and \textsf{M} states on Ivy Bridge, where the latency of \textsf{CAS}
accessing L1 is consistently (by $\approx$2-3ns) \emph{lower} than that of
\textsf{FAA}/\textsf{SWP} (see Figure~\ref{fig:Latency_results_Ivy}). We
attribute this effect to an optimization in the structure of L1 that
detects that no modification will be applied to a cache line, reducing the
latency.

In the \textsf{S}/\textsf{E} states executing an atomic on the data held by
a different core (on the same CPU) is not influenced by the data location
(L1, L2 or L3); see
Fig.~\ref{fig:Haswell_cas_64_E},~\ref{fig:Haswell_cas_64_S},~\ref{fig:Haswell_read_64_E},~\ref{fig:Latency_results_Ivy}.
The data is evicted silently, with neither writebacks nor updating the core
valid bit in L3.  Thus, all the accesses snoop L1/L2, making the latency
identical (as modeled by Eq.~(8)).
%

\textsf{M} cache lines are written back when evicted updating
the core valid bits. Thus, there is no invalidation when reading an
\textsf{M} line in L3 that is not present in any local cache. This explains
why \textsf{M} lines have lower latency in L3 than \textsf{E} lines; cf.
Figures~\ref{fig:Haswell_cas_64_E} and~\ref{fig:Haswell_cas_64_M}.


Remote accesses in the \textsf{M}/\textsf{E} states are by $\approx$50ns
slower than that of another core on the same CPU due to $\mathcal{H}$; see
Fig.~\ref{fig:Latency_results_Ivy}.  For modified cache lines the latency
is different for L3 because the MESIF protocol does not allow for dirty
sharing so the data has to be written to memory incurring $\mathcal{M}$.

In our latency benchmarks \textsf{CAS} does not write to L1. It is not
necessary to invalidate cache lines sharing the data when performing
unsuccessful \textsf{CAS}.  The results indicate that the Intel
architectures do not take advantage of that because a read for ownership
might be issued in any case.


\begin{figure*}[!t]
\centering
 \subfloat[\textsf{CAS}, Modified state]{
  \includegraphics[width=0.3\textwidth]{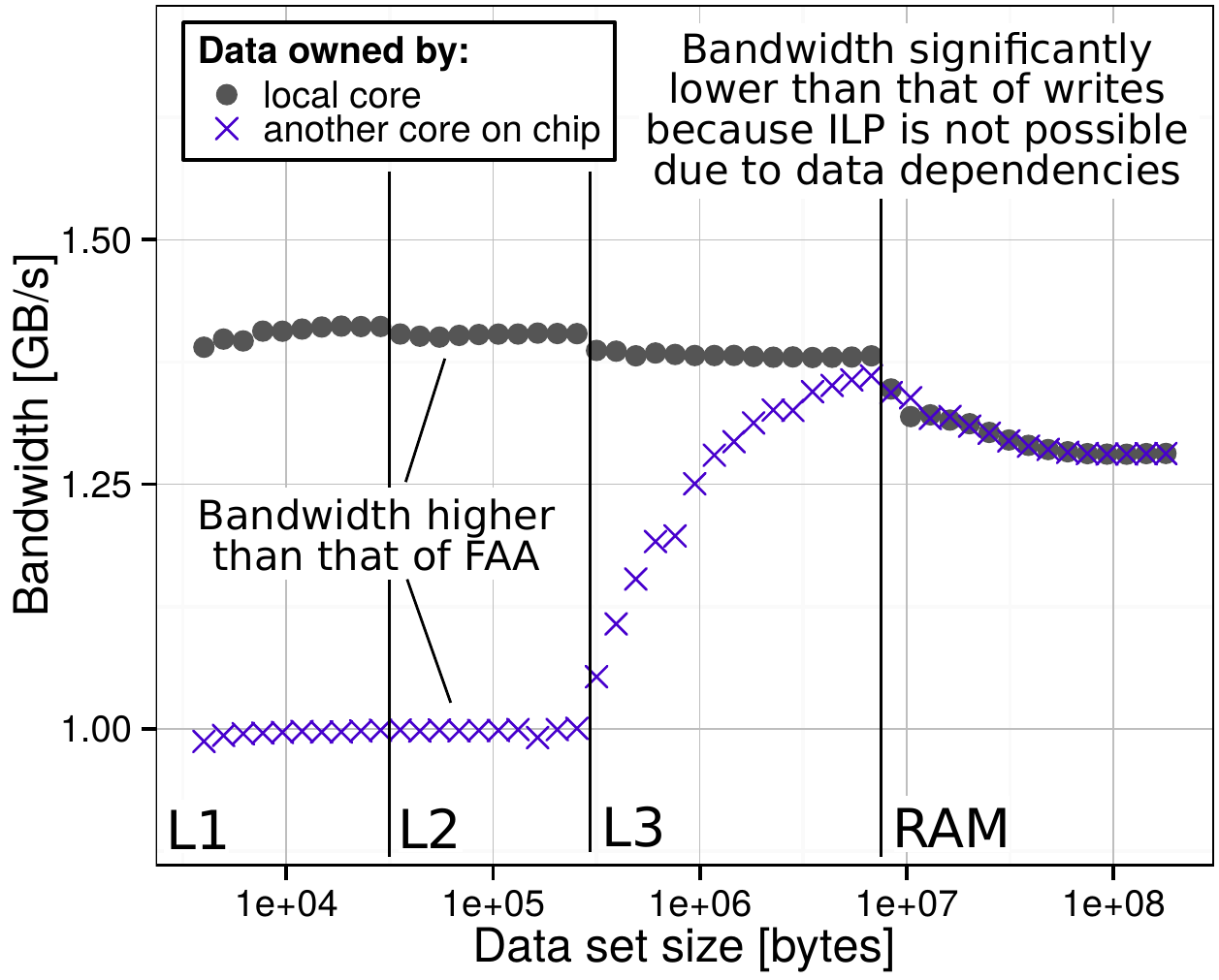}
  \label{fig:Bull_bw_cas_M}
 }
 \subfloat[\textsf{FAA}, Modified state]{
  \includegraphics[width=0.3\textwidth]{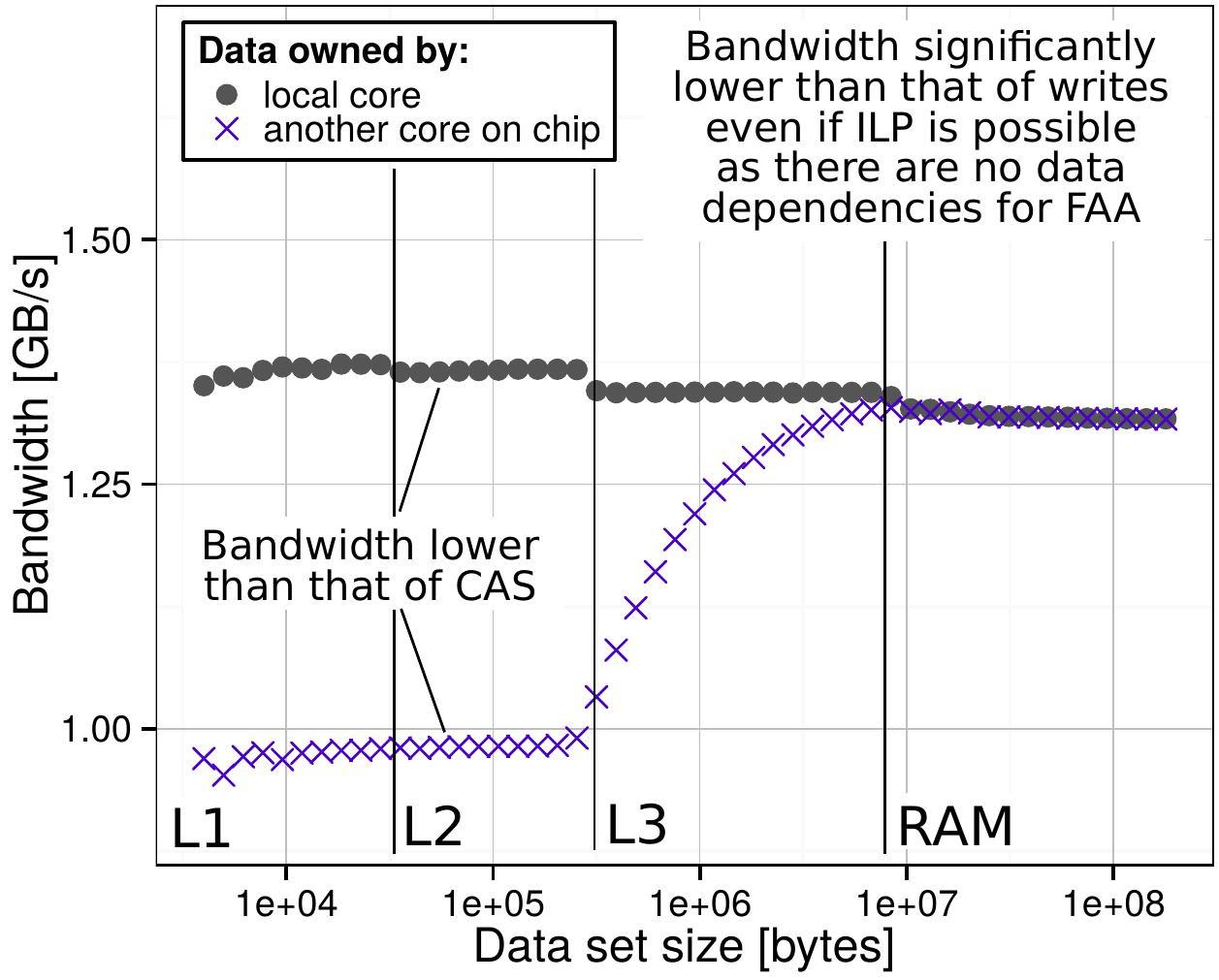}
  \label{fig:Bull_bw_fad_M}
 }
  \subfloat[\textsf{write}, Modified state]{
  \includegraphics[width=0.3\textwidth]{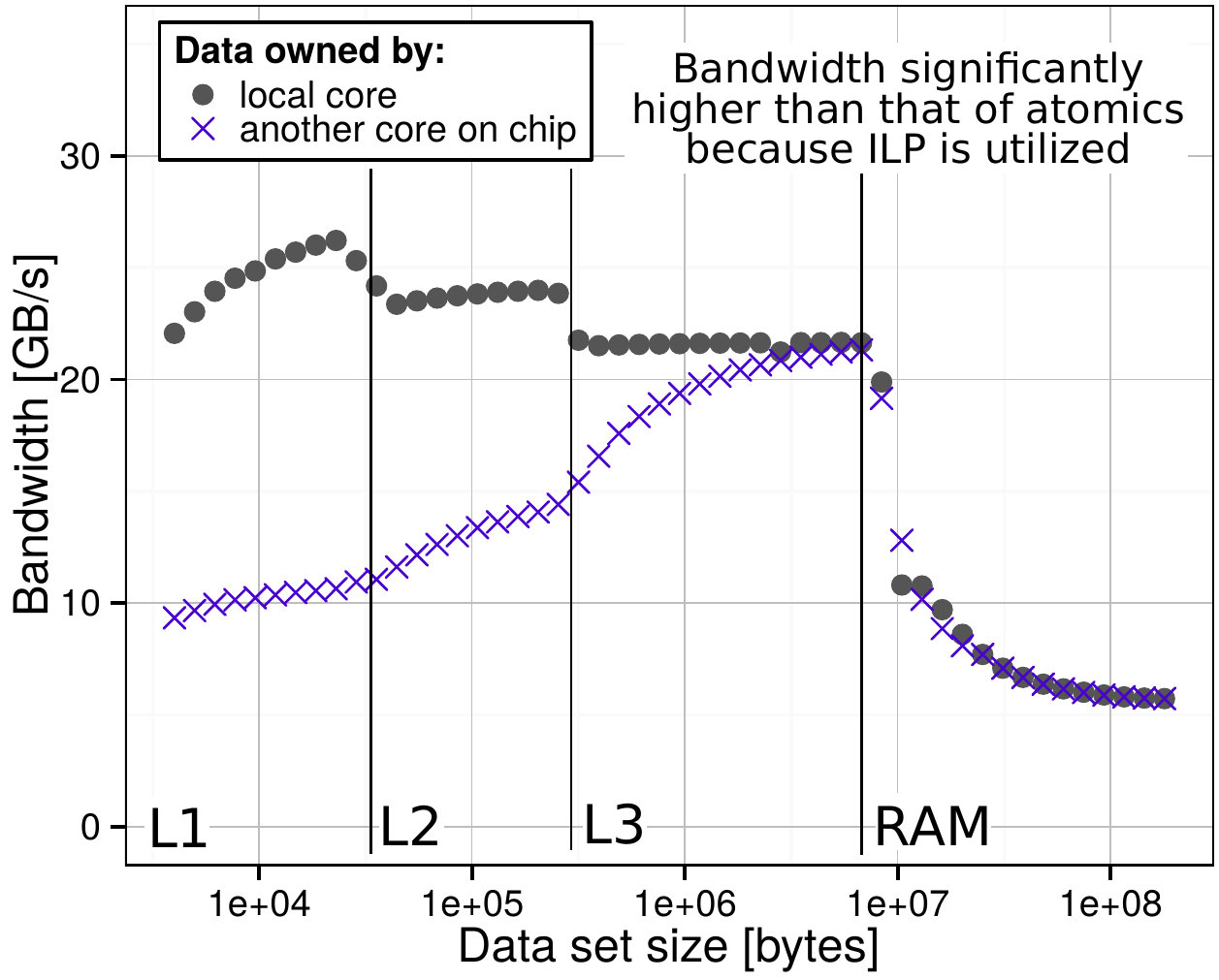}
  \label{fig:Bull_bw_write_M}
 }
\caption{The comparison of the bandwidth of \textsf{CAS},
\textsf{FAA}, and \textsf{writes} on Haswell. The requesting core accesses
its own cache lines (local) and cache lines of a different core from the
same chip (on chip).}
\label{fig:BW_Has}
\end{figure*}

\subsubsection{AMD Bulldozer}

We illustrate the latency results of AMD Bulldozer\footnote{\scriptsize For AMD, we had
to eliminate the effects from AMD hardware prefetchers that always
request multiple consecutive cachelines in case of an L2 miss.
For this, we had to increase the minimal distance between the 
accessed memory addresses. As the size of L1 in Bulldozer is only 16KB, 
this prevented obtaining results for L1. Identical effects were observed
by Molka et al.~\cite{Molka:2014:MMC:2618128.2618129}.}
in Figure~\ref{fig:Latency_results_Bull}.
Atomics are again slower than reads in each case. Yet, the difference
between reads and \textsf{CAS}/\textsf{FAA} is not the same for all the
cache levels. \textsf{CAS} and \textsf{FAA} take $\approx$8ns longer than
reads into the cache of a different core. Yet, for the local cache they
consistently take $\approx$20ns longer than respective reads (cf.
Figures~\ref{fig:Bull_cas_E}/\ref{fig:Bull_fad_E}
and~\ref{fig:Bull_read_E}).  We attribute this surprising result to
variable overheads related to $\mathcal{O}$.

The L3 results indicate that the latency is growing with
the increasing data block size. We conjecture that this effect is caused by
the HT Assist, a unit that uses a part of L3 and works as a
filter for accesses to remote cores~\cite{amd_opt_manual}. The HT Assist
module causes some accesses to L3 to miss and thus to incur higher
latencies. 

Both the \textsf{S} and the \textsf{O} states follow similar performance
patterns (we include the plots in the Appendix).
The latencies of atomics to shared data in the L2 of the requesting
core are similar independent of which cores contain the data; they are
dominated by $\mathcal{H}$ (additional $\approx$62ns). Bulldozer's L3 is
not inclusive and does not have core valid bits. Thus, L3 cannot determine
whether the data is in the L1 or L2 of a different core entailing an
invalidation broadcast. This broadcast has to reach caches on a remote CPU,
generating very high latencies. 



\subsubsection{Intel Xeon Phi}

\begin{figure}[!t]
\centering
 \subfloat[\textsf{CAS}, M/E state]{
  \includegraphics[width=0.23\textwidth]{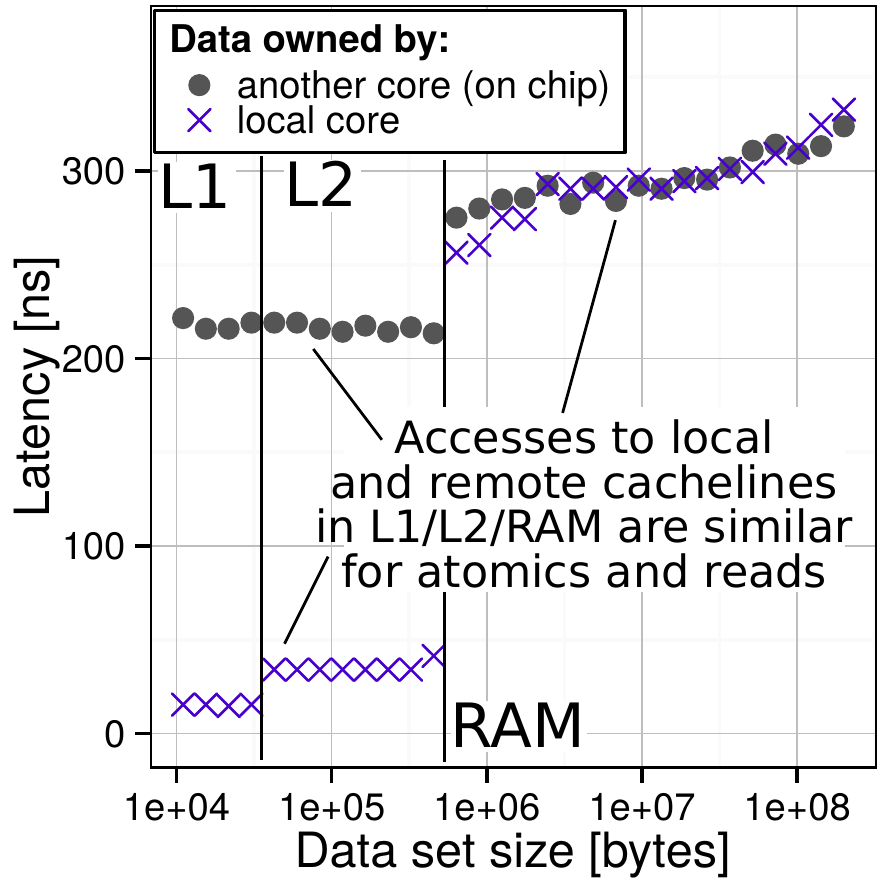}
  \label{fig:mic_cas_M}
 }
   \subfloat[\textsf{CAS}, S state]{
  \includegraphics[width=0.23\textwidth]{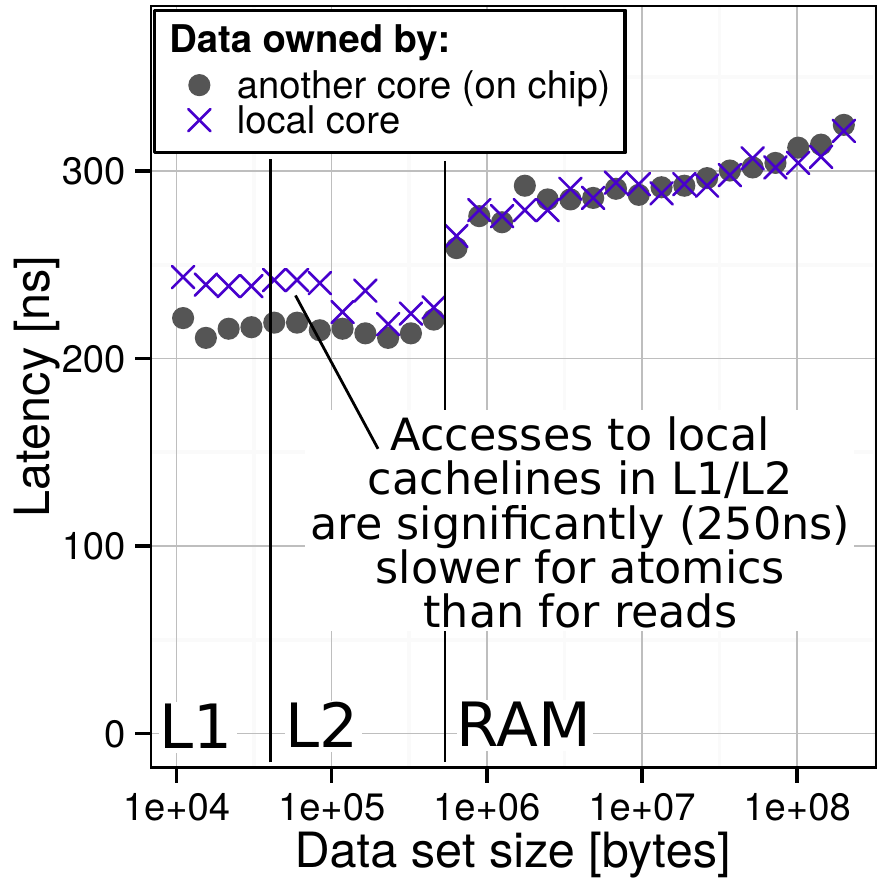}
  \label{fig:mic_cas_SO}
 }
\caption{The comparison of the latency of \textsf{CAS} on Xeon Phi. The
requesting core accesses its own cache lines (local) and cache lines of a
different core from the same chip (on chip).}
\label{fig:Latency_results_Phi}
\end{figure}

Finally, we discuss the latency results for Xeon Phi, see
Figure~\ref{fig:Latency_results_Phi}.
The performance patterns are very similar across all the tested operations
and coherency states as predicted by the model; here we illustrate
\textsf{CAS}.  Similarly to other Intel and AMD systems, atomics introduce
significant overheads over reads for the \textsf{S} state ($\approx$250ns
for L1 accesses).
Yet, contrarily to Haswell, Ivy Bridge, and Bulldozer, \textsf{CAS} is
slower than \textsf{FAA} ($\approx$10ns for local L1 and $\approx$15ns for
remote L1 accesses) while \textsf{FAA} is slower than \textsf{read}
($\approx$2ns for local L1 and $\approx$5ns for remote L1 accesses).

This minor latency differences in accessing cache lines owned by different cores
are due to the design of the Xeon Phi ring-bus: each
of these two accesses requires checking different cache directories.
This is consistent with
previous results for memory accesses as observed by other
studies~\cite{Ramos:2013:MCC:2493123.2462916}.


\subsubsection{Discussion \& Insights}

Our analysis provides novel insights.
It turns out that, contrary to the common
view~\cite{Morrison:2013:FCQ:2442516.2442527}, the latency of
\textsf{CAS}, \textsf{FAA}, and \textsf{SWP} is in most cases identical and
sometimes (L1 on Haswell and L3/memory on AMD) \textsf{CAS} is
\emph{faster} than \textsf{FAA}. Thus, atomics with
different consensus numbers still entail similar overheads. 
As we will show in Section~\ref{sec:number_of_operands_fetched}, additional
overheads in \textsf{CAS} are due to fetching an additional argument from
caches or memory.

The results indicate the correctness of the model assumptions and
predictions. The only significant deviations are caused by factors not
directly related to the cache coherency protocol (TLB misses) and the
overheads from the proprietary HT Assist module on AMD Bulldozer. We also
observe minor ($<$10\%) variations in the latencies to remote caches caused
by system noise.

The analysis suggests several potential improvements for the hardware
implementation of atomics.
For example, we show that unsuccessful \textsf{CAS}es invalidate the copies
of fetched cache lines entailing significant overheads.  Yet, such
operations do not modify the fetched cache line, making the invalidations
unnecessary.
We conjecture that this strategy incorporates the pipelining of atomics,
thus requiring the invalidations.
Another potential strategy would issue invalidations after \textsf{CAS}es
succeed. As unsuccessful \textsf{CAS}es usually constitute a crucial part
of all the issued \textsf{CAS}es in various parallel
designs~\cite{Morrison:2013:FCQ:2442516.2442527}, this might accelerate
some workloads.




\subsection{Bandwidth}

We now analyze a selection of the bandwidth results. Due to space
constraints we illustrate the Haswell results for the \textsf{M} state 
(Figure~\ref{fig:BW_Has}) and only briefly discuss other testbeds.
Here, we compare atomics to writes.
Our goal is to show that atomics do not use ILP even if no
dependencies between successive operations exist.

Similarly to latency, the bandwidth results for Haswell indicate that
\textsf{CAS} is \emph{comparable or more efficient} than \textsf{FAA}
($\approx$0.04 GB/s).
Moreover, the bandwidth is larger in higher level caches (for
the \textsf{E/M} cache lines). Yet, the differences between the levels are
not significant ($\approx$0.05 GB/s) as only the first access to each line
is affected by cache proximity.  Bandwidth (to L3) for the \textsf{E} lines
is lower than for the \textsf{M} lines due to to the silent eviction of the
former.

In each testbed the bandwidth of atomics is 
$\approx$5-30x lower than that of writes because the latter utilize
ILP. Yet, in our design (see Section~\ref{sec:benchmark_structure}) we
specifically enabled the possibility of parallel execution of
\textsf{FAA}/\textsf{SWP}. We conjecture that the hardware implementation
of atomics prevents ILP, limiting performance.





\subsubsection{Discussion \& Insights}


Significantly lower bandwidth of atomics (than that of writes) is due
to the differences in using write buffers.  Cores store to their
write buffers and continue executing further instructions before the previous
writes actually reach the cache (which can take $>$100ns as our
results indicate). The buffer might merge consecutive
writes increasing bandwidth. On the contrary, atomic operations cause the
write buffers to be drained. That means that every atomic is affecting the
cache directly without being merged or buffered.


In addition, our results indicate that atomics do not allow for ILP
whatsoever. Relaxing this restriction in some cases (e.g., for the
independent executions of \textsf{FAA} or \textsf{SWP)} could significantly
improve the bandwidth.

\begin{figure}[!h]
\centering
 \subfloat[CAS (64 bits)]{
  \includegraphics[width=0.23\textwidth]{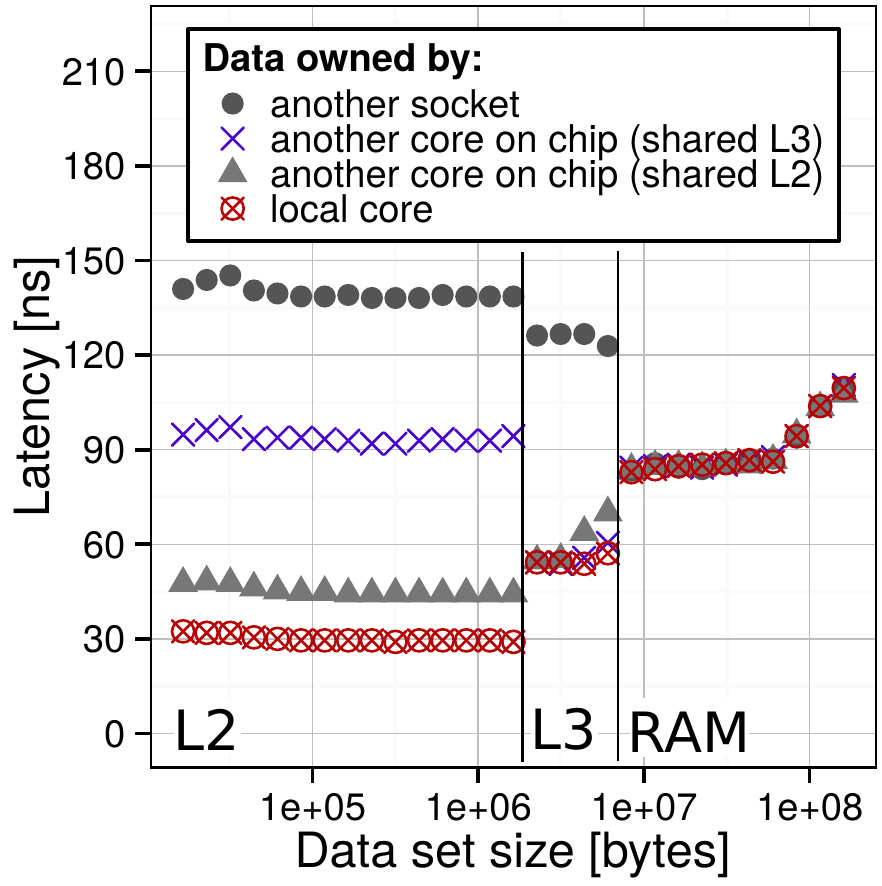}
  \label{fig:Haswell_cas_64_E_64bits}
 }
  \subfloat[CAS (128 bits)]{
  \includegraphics[width=0.23\textwidth]{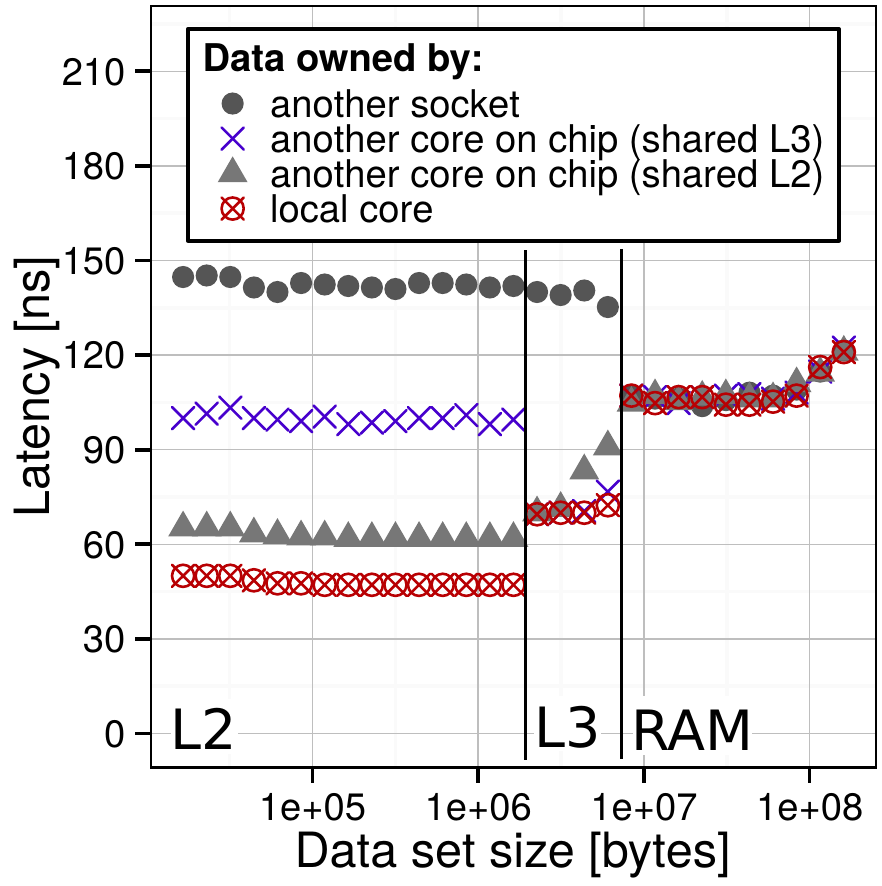}
  \label{fig:Haswell_cas_64_E_128bits}
 }
 \caption{The comparison of the latency of CAS using operands of 64 and 128 bits in size (AMD Bulldozer, the M state).}
 \label{fig:Operand_size_Bulldozer}
\end{figure}

\begin{figure*}
\centering
 \subfloat[Ivy Bridge]{
  \includegraphics[width=0.23\textwidth]{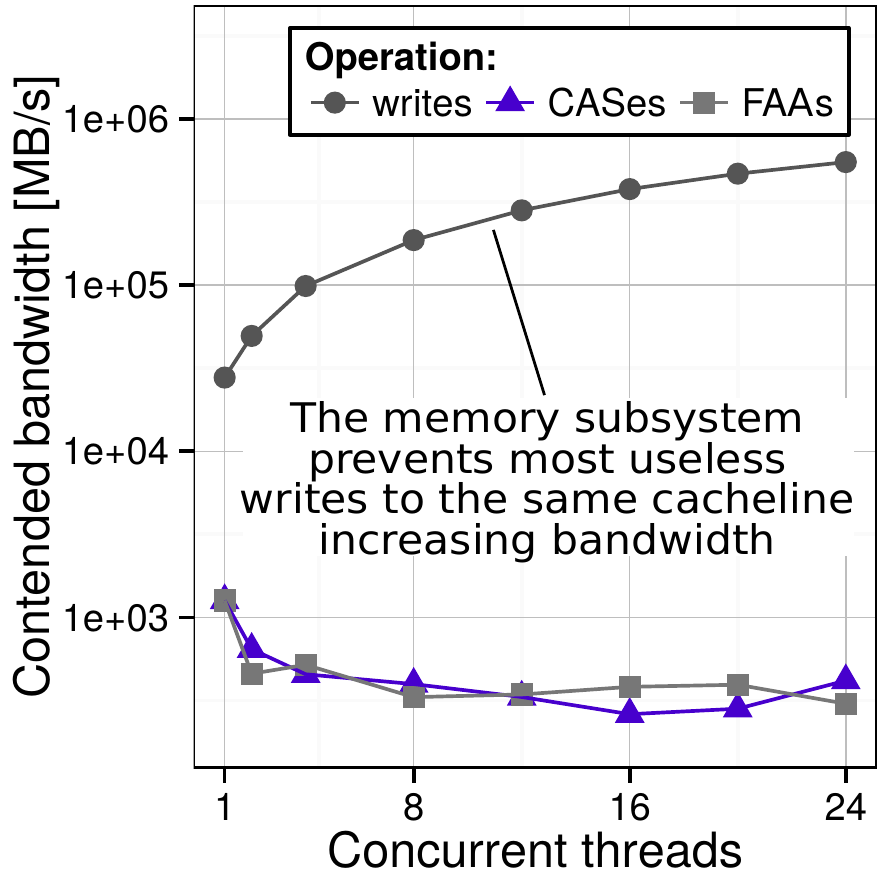}
  \label{fig:cont_bw_ivy}
 }
  \subfloat[Xeon Phi]{
  \includegraphics[width=0.23\textwidth]{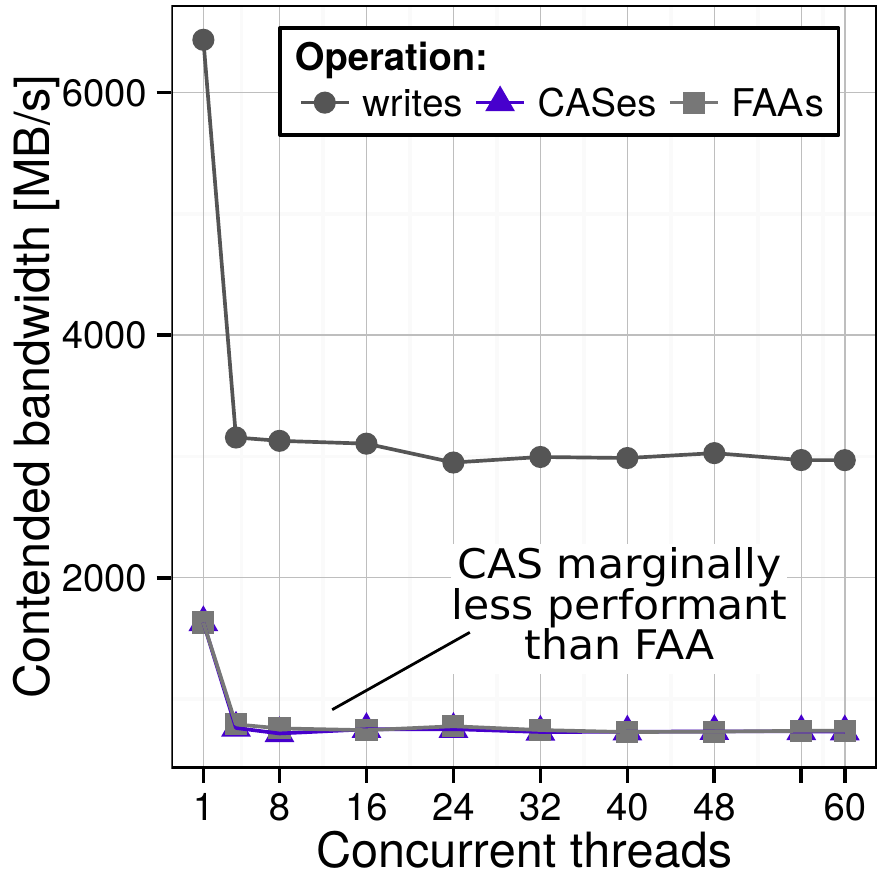}
  \label{fig:cont_bw_mic}
 }
  \subfloat[Bulldozer]{
  \includegraphics[width=0.23\textwidth]{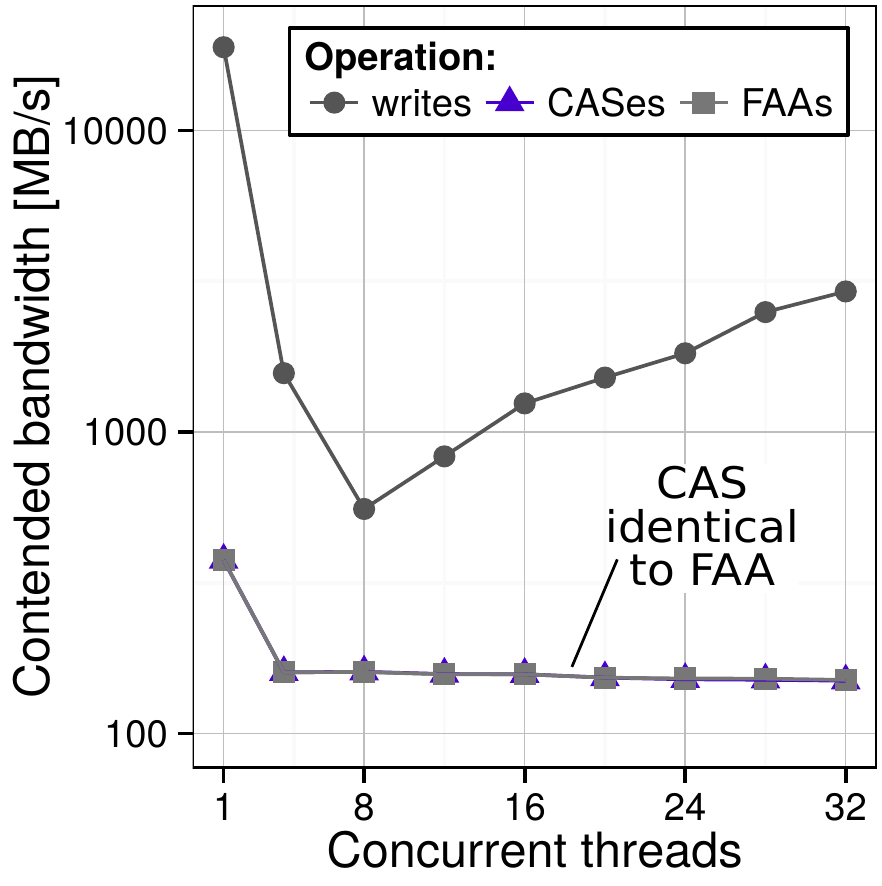}
  \label{fig:cont_bw_amd}
 }
  \subfloat[Bulldozer]{
\includegraphics[width=0.23\textwidth]{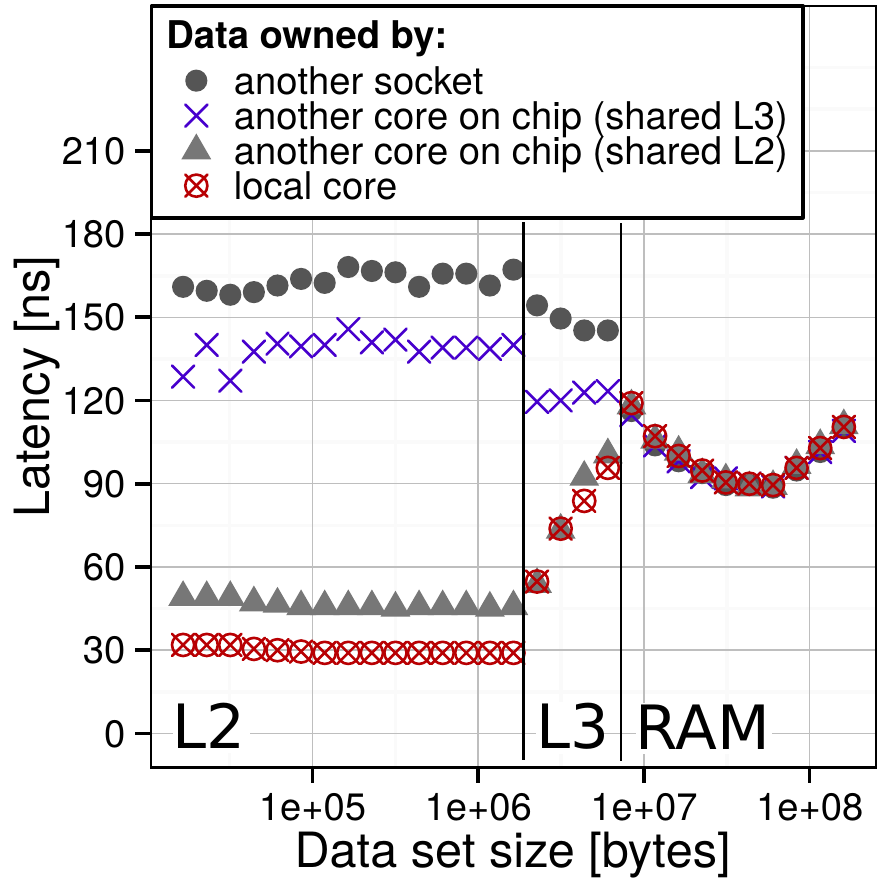}
  \label{fig:cas_operands}
 }

\caption{Figures~\ref{fig:cont_bw_ivy}-\ref{fig:cont_bw_amd} illustrate the
comparison of the contended bandwidth of CAS/FAA/writes on Ivy Bridge,
Bulldozer, and Xeon Phi. Figure~\ref{fig:cas_operands} shows the latency of
CAS that fetches two operands from the memory subsystem on Bulldozer
(Exclusive state).} \label{fig:Bandwidth_Contended}

\end{figure*}

\begin{figure*}
\centering
 \subfloat[Hardware Prefetcher.]{
  \includegraphics[width=0.23\textwidth]{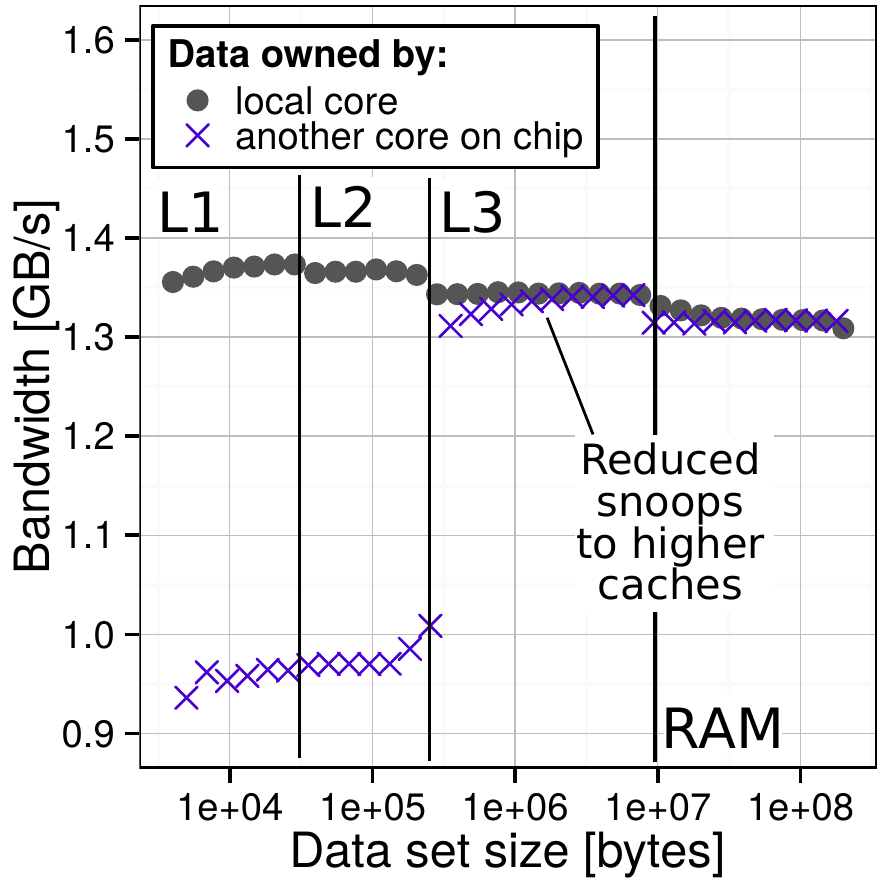}
  \label{fig:hw_prefetcher_bw}
 }
 \subfloat[Adjacent Cache Line Prefetcher.]{
  \includegraphics[width=0.23\textwidth]{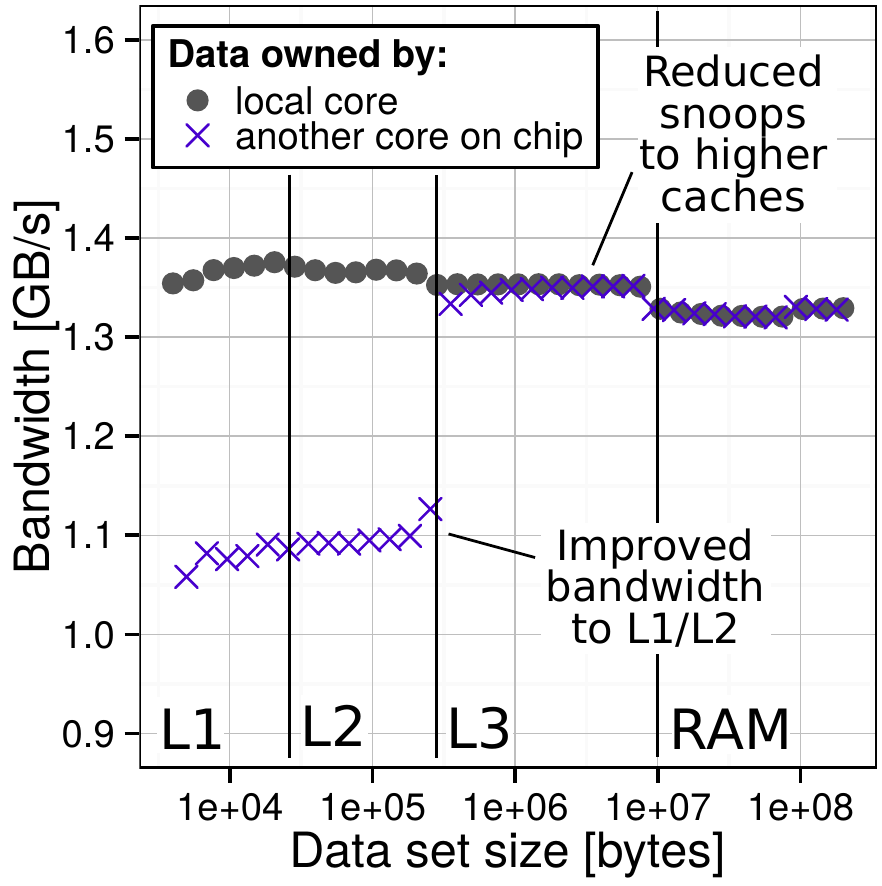}
  \label{fig:line_prefetcher_bw}
 }
  \subfloat[Both prefetchers enabled.]{
  \includegraphics[width=0.23\textwidth]{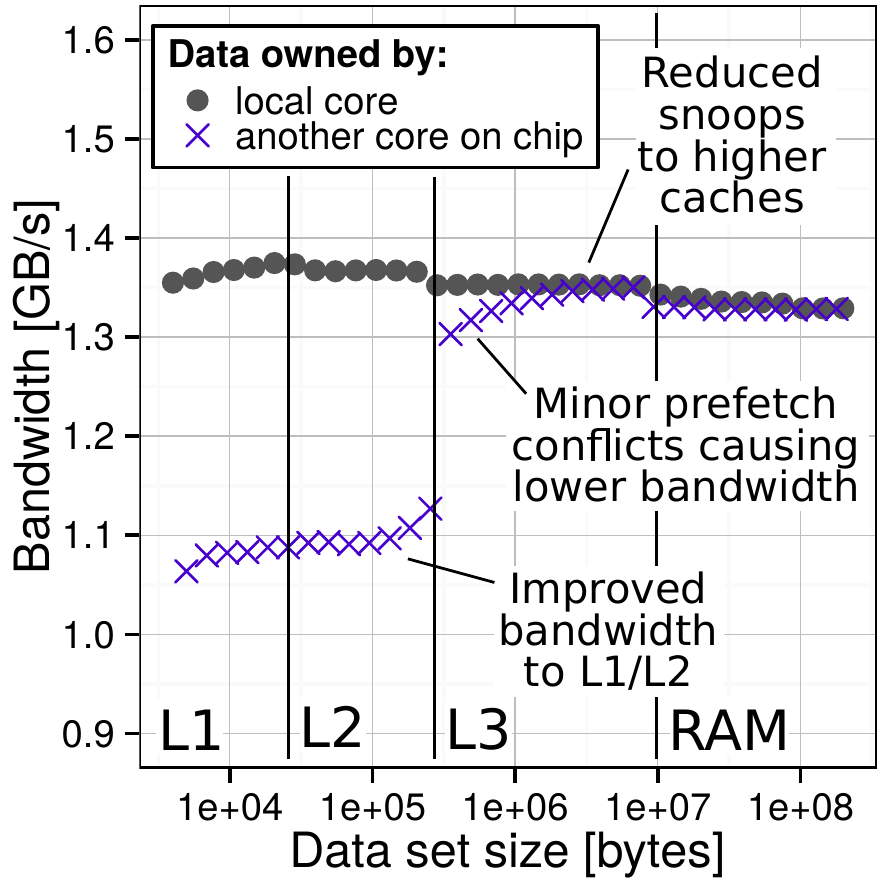}
  \label{fig:hw-line_prefetcher_bw}
 }
   \subfloat[Turbo Boost, EIST, C States.]{
  \includegraphics[width=0.23\textwidth]{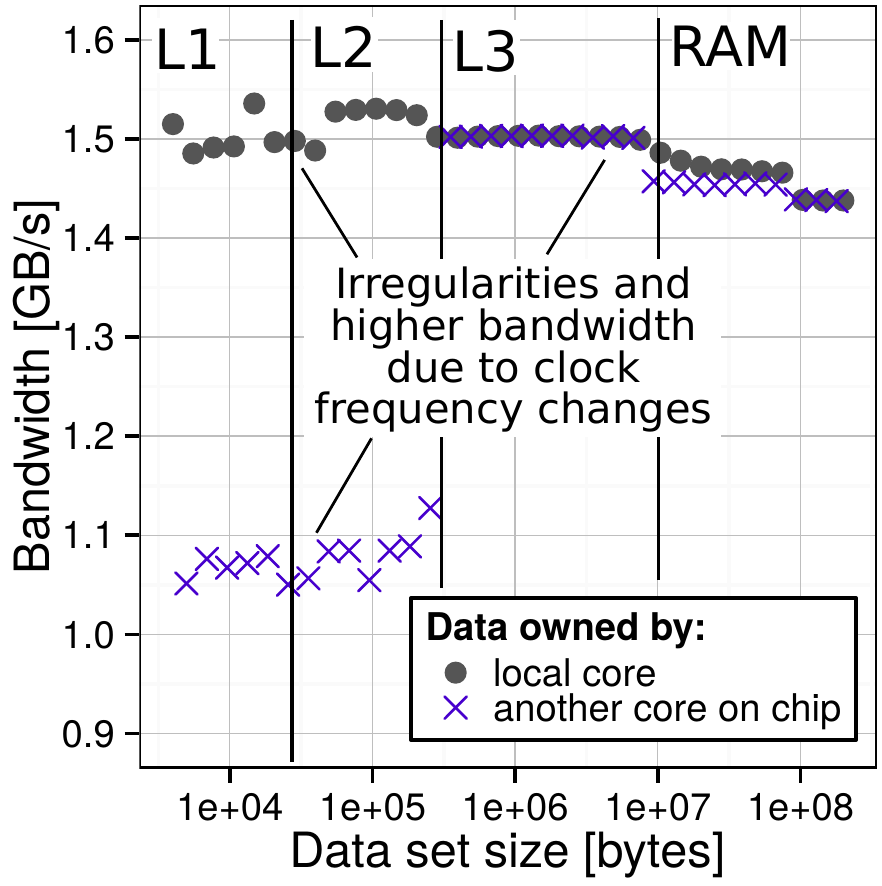}
  \label{fig:mechanisms_bw}
 }
 \caption{The effect on the bandwidth of \textsf{FAA} (accessing cache
lines in the Modified state) coming from prefetchers and various power
efficiency and acceleration mechanisms (Turbo Boost, EIST, C States)
deployed in Intel Haswell.} \label{fig:various_prefetchers_bw}
\end{figure*}

\subsection{Operand Size}

\textsf{CAS} comes with several flavors that differ in the size of the
operands.  We analyze variants that use 64 and 128 bits.
The tested Intel systems provide identical latency in each case. On the
contrary, AMD Bulldozer has lower latency when using 64 bits, see
Figure~\ref{fig:Operand_size_Bulldozer}. The latency difference is
insignificant ($\approx$5ns) when accessing cache of a core that does not
share L2 with the requesting core and close to 20ns for other caches and
memory.
%
%
Using \textsf{CAS} that operates on 64bit operands would thus be desirable
in latency-constrained applications running on AMD Bulldozer.



\subsection{Contention}
\label{sec:contention}

We now evaluate the effect of concurrent threads accessing the same cache
line (using writes and atomics) on the tested manycore systems (Ivy Bridge,
Xeon Phi, Bulldozer); see Figure~\ref{fig:Bandwidth_Contended}. This
benchmark targets the codes with highly contended shared counters and
synchronization variables. 

The bandwidth of writes on Ivy Bridge has almost 100GB/s with eight cores
and is growing steadily with the thread count. These numbers are very close
to the accumulated non-contended bandwidth.
%
%
We observe a similar effect on Haswell and thus we conjecture that both
architectures detect that issued operations access the same cache line in
an arbitrary order, annihilating the need for the actual execution of all
the writes. 
Contrarily to other Intel systems, adding more threads on Xeon Phi quickly
decreases the bandwidth until it converges to $\approx$708 MB/s
(\textsf{CAS}), $\approx$730 MB/s (\textsf{FAA}) and $\approx$2960 MB/s
(writes).

Bulldozer also suffers from the contention. A single thread reaches the
highest bandwidth but additional threads (up to eight) decrease the
performance.  Beyond this point the bandwidth increases steadily, similarly
to Ivy Bridge.

We conclude that all the considered architectures have significantly lower
bandwidth in a contended execution of atomic operations than in a
non-contended case. This may constitute a performance limitation in
state-of-the-art multi- and manycore designs with massive thread-level
parallelism.


\subsection{Number of Operands Fetched}
\label{sec:number_of_operands_fetched}

Here, we show how the performance of \textsf{CAS} changes when two operands
are fetched from the memory subsystem. We analyze the latency results for
Bulldozer, see Figure~\ref{fig:cas_operands} for the \textsf{E} state.
It turns out that additional reads and invalidations impact the latency
only marginally because of the pipelined execution of
\textsf{CAS}es/\textsf{read}s, requiring additional $\approx$2-4ns and
$\approx$15-30ns for local and remote accesses, respectively.

Surprisingly, the latency of accessing \textsf{M} cache lines is not affected
and is similar to the one from Figure~\ref{fig:Bull_cas_E}.
This effect is caused by the proprietary AMD optimization called the
\emph{MuW} state~\cite{lepak2014method}.  It immediately invalidates the
accessed \textsf{M} cache line and allows the requesting core to modify it
without further actions, limiting remote invalidations and improving the
performance even further.

\subsection{Prefetchers and Other Mechanisms}

State-of-the-art architectures host different mechanisms that impact the
CPU performance.  For example, Haswell deploys Hardware Prefetcher
(prefetching data/instructions after successive L3 misses or after
detecting cache hit patterns), Adjacent Cache Line Prefetcher
(unconditional prefetching of two additional cache lines), and several
mechanisms that may affect the clock frequency and power efficiency (Turbo
Boost, EIST, and C States).
We now illustrate how these mechanisms impact the latency and bandwidth of
atomics. We select Haswell as the testbed and we skip the
latency results because they are only marginally ($\approx$1\%
difference) affected. The bandwidth results are illustrated in
Figure~\ref{fig:various_prefetchers_bw}.
Any of the prefetchers improves bandwidth for L3 cache accesses by reducing
the effect of snooping (improvement up to $\approx$0.3 GB/s).
Interestingly, if both are enabled, they negligibly conflict with each
other reducing bandwidth to L3.  Adjacent Cache Line Prefetcher
additionally accelerates atomics to L1/L2 (up to $\approx$0.135 GB/s).
Turbo Boost, EIST, and C States impact the clock frequency and thus both
introduce irregularities in the results and improve the bandwidth of L3,
RAM, and remote L1/L2 accesses by $\approx$0.15 GB/s.

\subsection{Unaligned Operations}

Finally, we analyze the performance of reads and atomics when accessing unaligned words
that span two consecutive cachelines. We present the results for \textsf{CAS} (M state) in Figure~\ref{fig:unaligned};
\textsf{read} and \textsf{FAD} are included in the Appendix.
Unaligned reads suffer at most 20\% of performance loss compared to aligned operations.
Contrarily, atomics suffer from significant latency increases; \textsf{CAS} reaches
up to 750ns. We conjecture this change of performance is due to the fact that the CPU
locks the whole bus as soon as a operation starts to access more than a single cache line.

\begin{figure}[!h]
\centering
\vspace{-0.5em}
\subfloat[Unaligned CAS (E state).]{
    \includegraphics[width=0.23\textwidth]{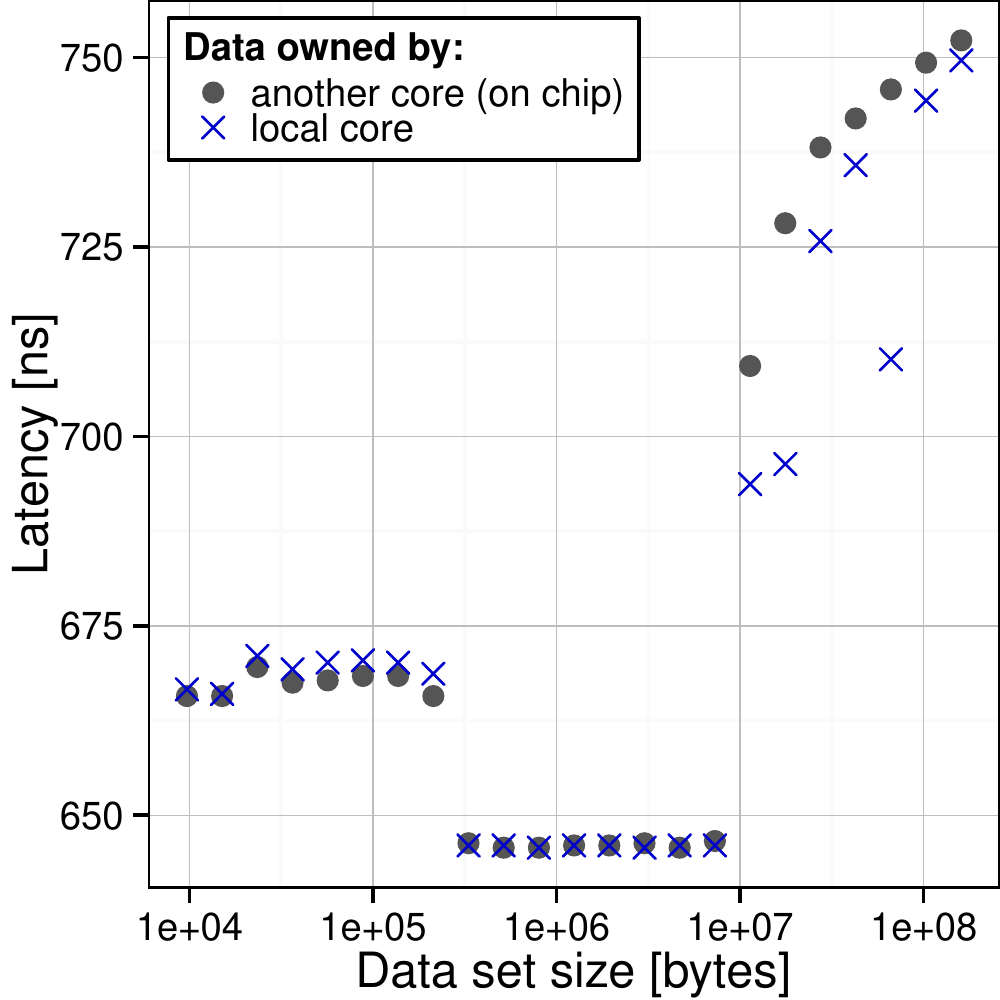}
      \label{fig:unaligned}
}
\subfloat[BFS designed with SWP/CAS.]{
    \includegraphics[width=0.23\textwidth]{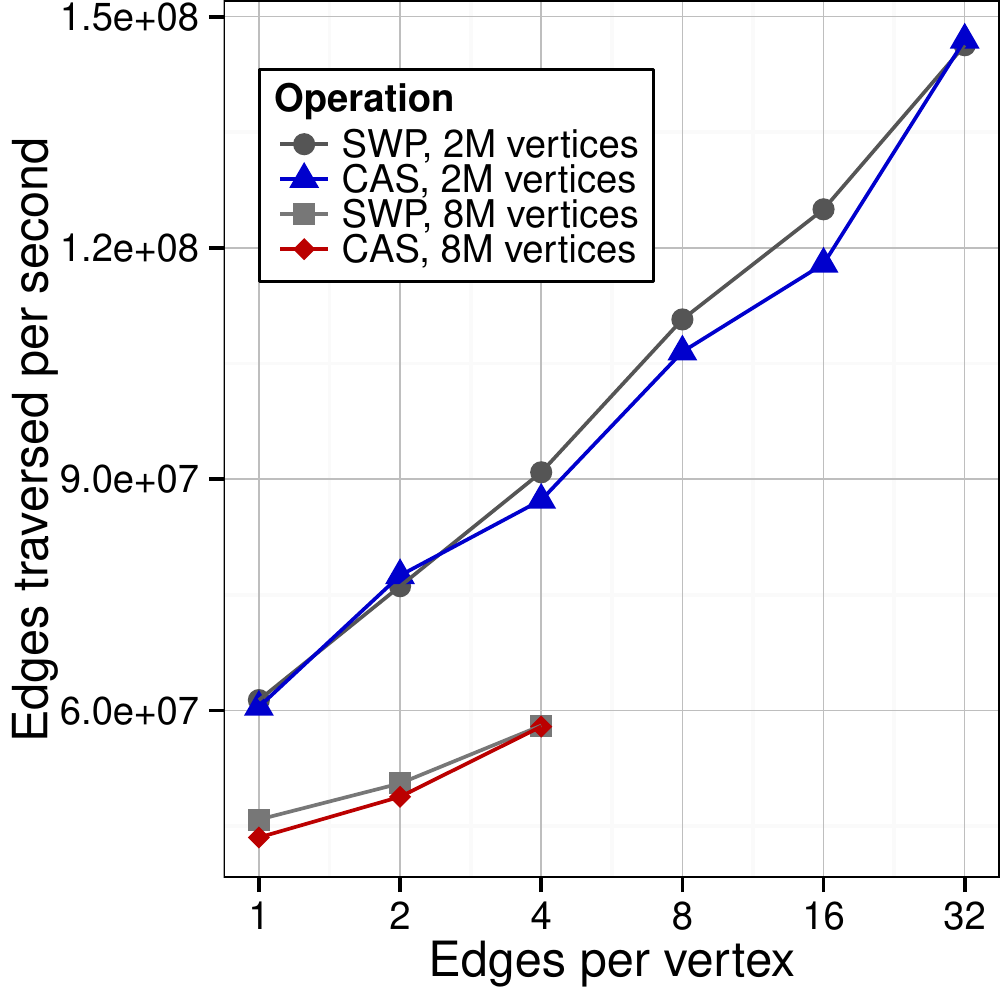}
      \label{fig:bfs}
}
\caption{The analysis of unaligned atomics and BFS.}
 \vspace{-1.0em}
 \end{figure}


%
%

%


\section{Discussion}
\label{sec:disc_and_app}

Our model and benchmarks provide insights into the latency and bandwidth of
atomics; they also identify various performance issues. We now proceed to
discuss how the insights can simplify parallel programming. Then, we
propose several solutions to the identified performance issues.

\subsection{Simplifying Parallel Programming}

Our analysis illustrates that all atomics have comparable
latency and bandwidth. We argue that the only significant difference
between atomics that matters for performance is the semantics (e.g.,
\textsf{CAS} introduces the notion of wasted work). This conclusion
reduces the complexity of designing parallel lock-free algorithms.

As an example, consider a parallel synchronous Breadth First
Search (BFS) graph traversal included in the well-known
Graph500 benchmark~\cite{murphy2010introducing}.
A key part of this algorithm is a concurrent array called
\texttt{bfs\_tree} that is constructed during the traversal. Initially,
each array cell equals $-1$. After the traversal, given a graph with $n$
vertices labeled from $0$ to $n-1$, \texttt{bfs\_tree[v]} contains the
label of the parent of vertex $v$ as determined by the traversal.

Now, concurrent accesses to the array cells can be performed with
\textsf{CAS}, \textsf{SWP}, or \textsf{FAA}. Our model and analysis clearly
indicate that the latency and bandwidth of these actions are almost
identical. Thus, the only significant factor is the semantics.  As
\textsf{FAA} always succeeds, it might happen that some threads updating
the same cell would both increase the value of the cell.  This would
require a complex scheme in which the effects of some of the issued
operations would be reverted.
On the contrary, \textsf{CAS} and \textsf{SWP} can be used to design
simpler protocols. However, as \textsf{CAS} introduces wasted work, we
expect it to generate some additional overheads. We illustrate the results
of a BFS traversal (4 concurrent threads) using both of these actions in Figure~\ref{fig:bfs}.  A
traversal is performed on Kronecker
graphs~\cite{Leskovec:2010:KGA:1756006.1756039} that model heavy-tailed
real-world graphs. The largest graphs fill the whole available memory.  The
results illustrate that \textsf{SWP} results in more edges traversed per
second.

\subsection{Addressing the Shortages of the Tested Architectures}

Designing feasible solutions to the identified issues would require access
to the details of the benchmarked architectures. Unfortunately, such
details are Intellectual Property (IP) of AMD/Intel that are beyond our
access.
Thus, we exclude microarchitectural details and we focus on general
schemes that we believe could be easily incorporated into current
architectures.
The issue of insufficient information on microarchitectural features from
hardware vendors has already been indicated by Mytkowicz et al.~\cite{Mytkowicz:2009:PWD:1508244.1508275}.

On Bulldozer, atomics accessing O/S cachelines always trigger invalidations
to remote dies because the architecture lacks a method to track which
caches share a cacheline. 
%
%
We now discuss two alterations to the Bulldozer design; each would prevent
unnecessary invalidations to remote dies. 

\subsubsection{Extending MOESI}

We extend MOESI with two new cache coherency states named
\textbf{O}wned-\textbf{L}ocal (OL) and \textbf{S}hared-\textbf{L}ocal (SL).
When a cacheline in the E state is read from a core on the same die it
enters the SL state in both caches, rather than the S state.
Similarly, when a cacheline in the M state is read by a core on the same
die, it enters the OL/SL state instead of O/S.  The SL and OL states
indicate that the cacheline is only cached by the local die. Writing to an
OL/SL cacheline requires no invalidations to a remote die.
However, when an OL/SL cacheline is read by a core on a remote die, all the
copies of that cacheline will transition from SL to S or from OL to O
indicating that remote invalidations will be necessary when modifying the
cache line.

The introduced states prevent unnecessary invalidations when writing to
cachelines shared by cores on the same die. A potential disadvantage is
that reading an SL or OL cacheline from a remote die might be slightly
slower due to multiple transitions. However, this could be addressed with a
careful overlap of the cacheline transfer and the transitions on a die.

\subsubsection{Extending HT Assist}

One could also eliminate unnecessary remote invalidations on AMD by using
HT Assist to track S and O cachelines that are only present on one
die. For this, a portion of L3 on each die would be dedicated to HT
Assist to track the cachelines that most recently changed to the S or O
state. Upon a read of an S/O cacheline (issued by a remote core), the respective entry in HT Assist
would be removed. Upon a write (by a remote core), the HT Assist
would be probed to determine whether to issue remote invalidations.
Probing HT Assist does not increase the latency of local writes as L3
is consulted in any case.

\subsubsection{Enabling ILP for Atomics}

Atomics act as barriers and they require write buffers to be drained before they are executed~\cite{intel_arch_opt_manual,intel_soft_opt_manual}.  
To enable ILP, an additional set of \emph{relaxed} atomics could be introduced. For this, an instruction prefix called \textsf{FastLock} could be added to the instruction set. The \textsf{FastLock} prefix would enable reordering issued atomic operations as long as non-overlapping memory regions are affected. The non-relaxed semantics would be available with the original lock prefix.


\section{Related Work}
\label{sec:rw}

To the best of our knowledge, there exists no detailed performance analysis
of atomic operations. A brief discussion that compares the contention of
Compare-And-Swap and Fetch-And-Add can be found in the first part of the
work by Morrison et al.~\cite{Morrison:2013:FCQ:2442516.2442527}.
This work uses the comparison to motivate the proposed parallel queue
that extensively utilizes Fetch-And-Add.
It differs from the study in this paper as it only illustrates
the lower performance of \textsf{CAS} caused by the semantics that
introduce wasted work.
Another work by David et al.~\cite{David:2013:EYA:2517349.2522714}
analyzes the performance of atomics. Yet, this study
targets a broad range of synchronization mechanisms and does not provide
an in-depth analysis of both the latency and bandwidth of atomics in the
context of complex multilevel memory hierarchies.

A methodology for benchmarking the latency and bandwidth
of reads and writes accessing different levels of the caching
hierarchy in the NUMA systems was conducted by Molka et al.~\cite{Molka:2009:MPC:1636712.1637764}.
A comparison of the performance of memory accesses 
on Intel Nehalem and AMD Shanghai was performed by Molka et al.~\cite{Hackenberg:2009:CCA:1669112.1669165};
a similar study targets the AMD Bulldozer and Intel Sandy Bridge microarchitectures~\cite{Molka:2014:MMC:2618128.2618129}.
Other analyses on the performance of the memory subsystems include the work by Babka et al.~\cite{Babka:2009:ICP:1506661.1506668},
Pend et al.~\cite{Peng:2008:MHP:1399642.1399665},
and Hristea et al.~\cite{Hristea:1997:MMH:509593.509638}.
Our works differs from these studies as it specifically targets atomic operations,
providing several insights into the performance relationships between atomics and the utilized caching hierarchy.

There exist numerous works proposing concurrent codes and data structures that use atomics for synchronization.
Examples include a queue by Morrison at el.~\cite{Morrison:2013:FCQ:2442516.2442527}, a hierarchical lock by Luchangco et al.~\cite{Luchangco:2006:HCQ:2135763.2135859},
and a queue by Michael and Scott~\cite{Michael:1996:SFP:248052.248106}. Many fundamental
structures and designs can be found in a book by Herlihy and Shavit~\cite{herlihy2012art}.

Finally, the considered architectures and cache coherency protocols are extensively
described in various manuals and papers~\cite{jain2013haswell,intel_arch_opt_manual,intel_soft_opt_manual,amd_opt_manual,goodman2005forward,suh2004supporting}.
Several performance models targeting on-chip communication have been introduced, for example
a model by Ramos and Hoefler~\cite{Ramos:2013:MCC:2493123.2462916}. The model proposed in this paper
differs from that work because it specifically targets latency and bandwidth of atomic
operations in the onnode environment.

\section{Conclusion}
\label{sec:conclusion}

Atomic operations are used in numerous parallel data structures,
applications, and libraries. Yet, there exists no evaluation that would
illustrate tradeoffs and relationships between the performance of atomics
and various characteristics of multi- and manycore environments.

In this work we propose a performance model and provide a detailed
evaluation of the latency and bandwidth of several atomic operations
(\textsf{Compare-And-Swap}, \textsf{Fetch-And-Add}, \textsf{Swap}) that
validates the model.  The selected atomics are widely utilized in various
parallel codes such as graph traversals, shared counters, spinlocks, and
numerous data structures.
Our performance insights include the observation that \textsf{CAS} and
\textsf{FAA} have in principle identical latency and the only difference is
related to the number of operands to be fetched and the semantics of
\textsf{CAS} that introduce the notion of the ``wasted work''.
Another insight is that the atomics prevent 
any instruction level parallelism, significantly limiting the bandwidth (up
to 30x in comparison with simple writes), even if there are no dependencies
between the successive operations.  Our analysis can thus be used for
designing more efficient parallel systems. 

The results also indicate several potential improvements in the design of
the caching hierarchy. For example, the AMD Bulldozer architecture limits
performance with invalidations issued to remote CPUs even if the respective
cache line is stored only in local caches.  Eliminating such invalidations
would significantly accelerate atomic operations accessing cache lines in
the shared state.  Here, we discuss two general solutions to this problem. 

Finally, we illustrate that our analysis simplifies the design of parallel
algorithms.
Our study and data can be used by architects and engineers to develop more
performant memory subsystems that would offer even higher speedups for
parallel workloads.

{\small
\vspace{0em}\section*{Acknowledgements}
We thank the CSCS and Euler teams
granting access to the Monte Rosa and Euler machines,
and for their excellent technical support.
We thank Gregorz Kwaśniewski for his help with Xeon Phi.}
\appendix

\section{Appendix: Full Results}

\begin{figure*}[t]
\subfloat[\textsf{CAS}, the Exclusive state]{
  \includegraphics[width=0.35\textwidth]{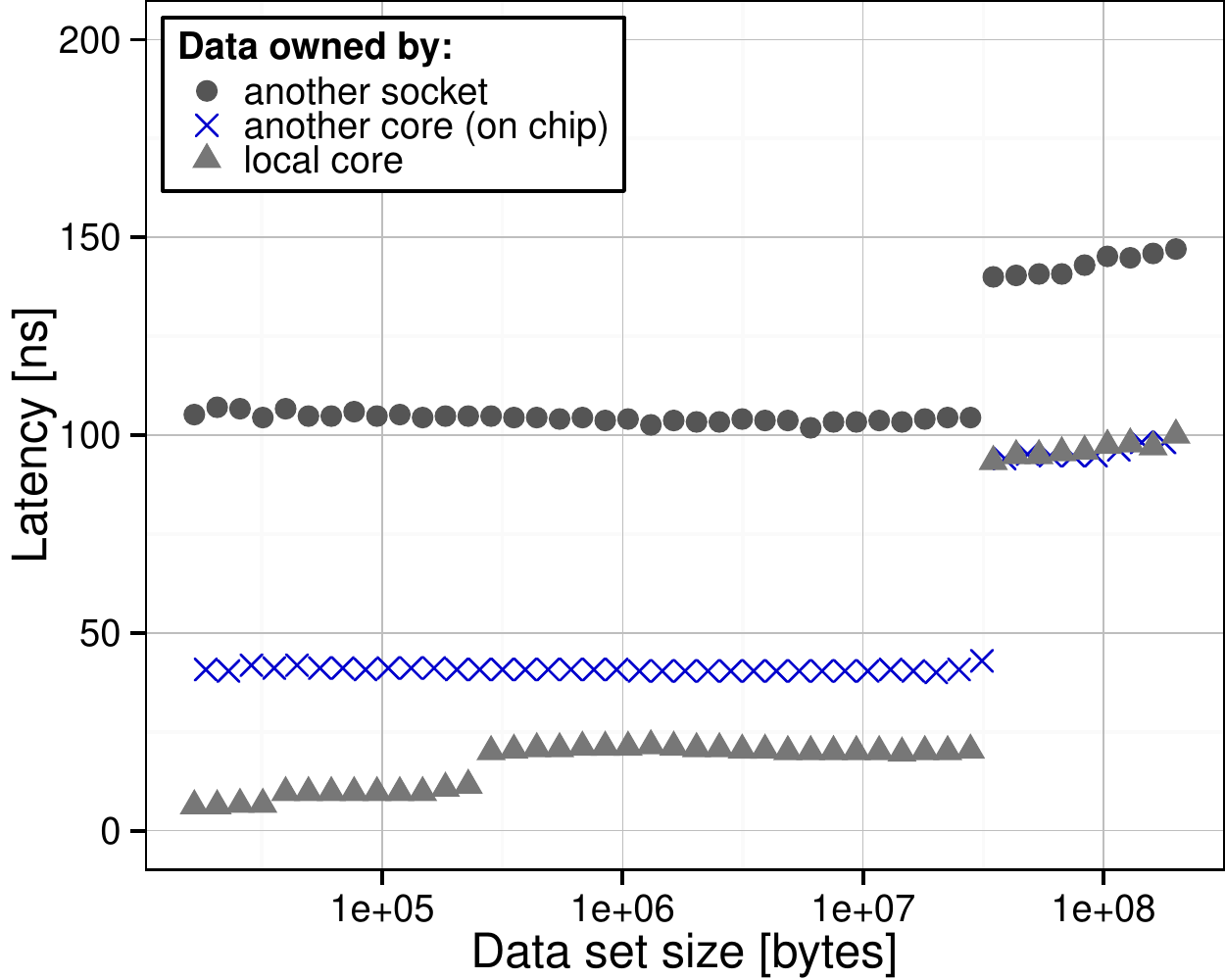}
 }
 \subfloat[\textsf{CAS}, the Modified state]{
  \includegraphics[width=0.3\textwidth]{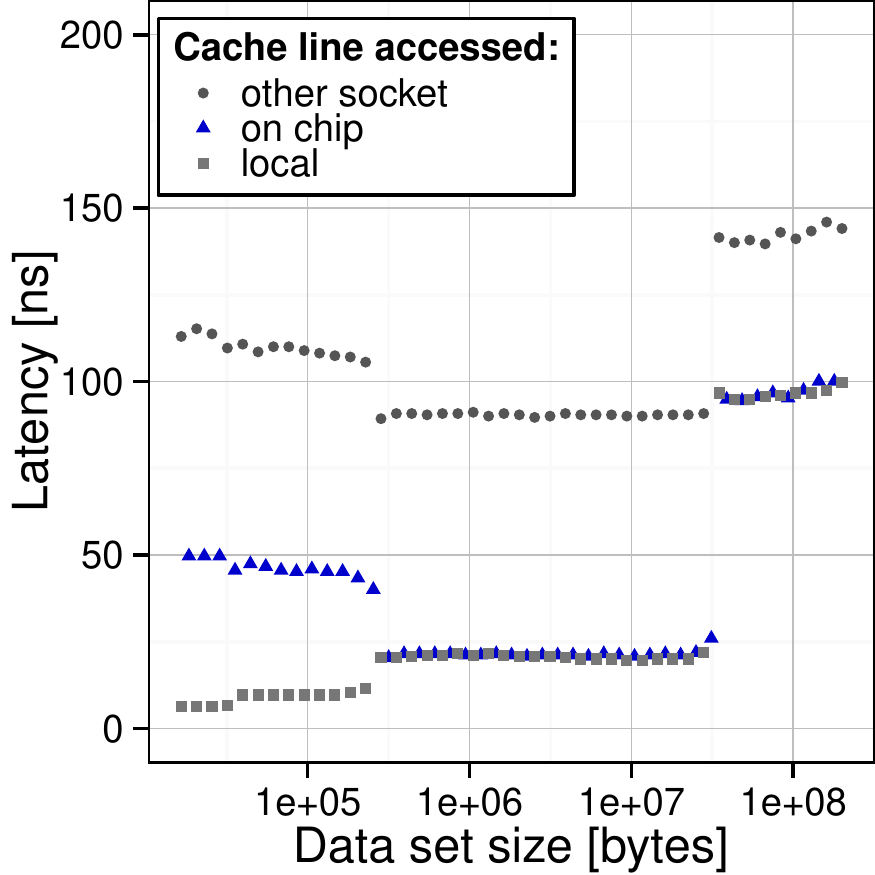}
 }
 \subfloat[\textsf{CAS}, the Shared state]{
  \includegraphics[width=0.3\textwidth]{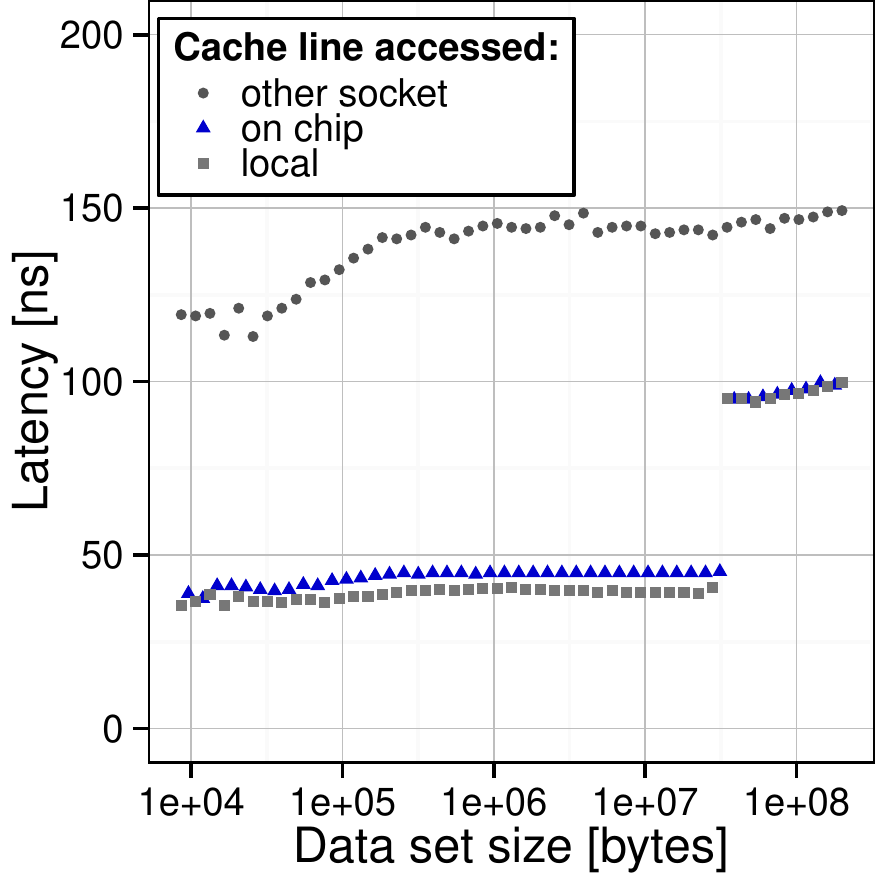}
 }\\
 \subfloat[\textsf{FAA}, the Exclusive state]{
  \includegraphics[width=0.4\textwidth]{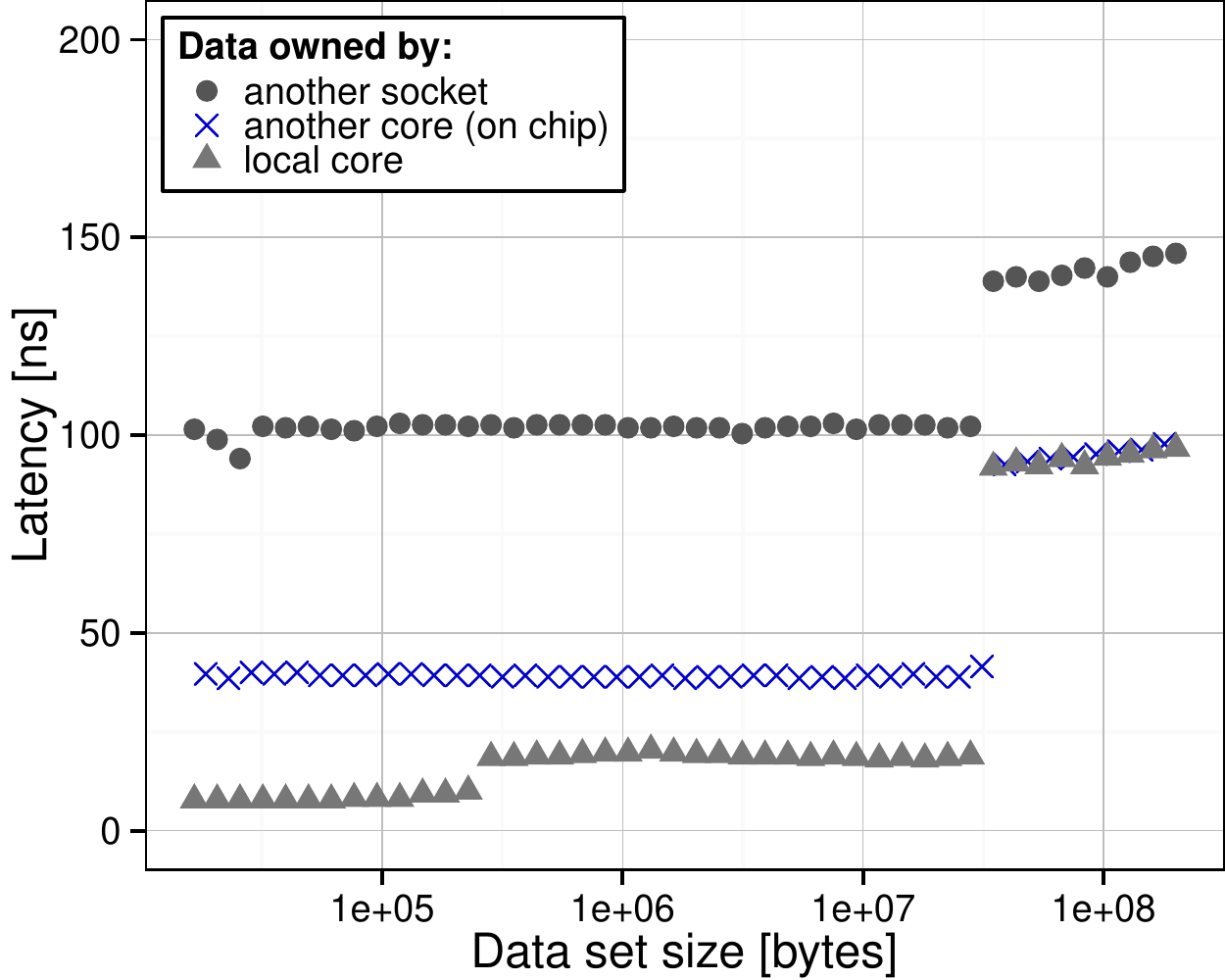}
 }
 \subfloat[\textsf{FAA}, the Modified state]{
  \includegraphics[width=0.33\textwidth]{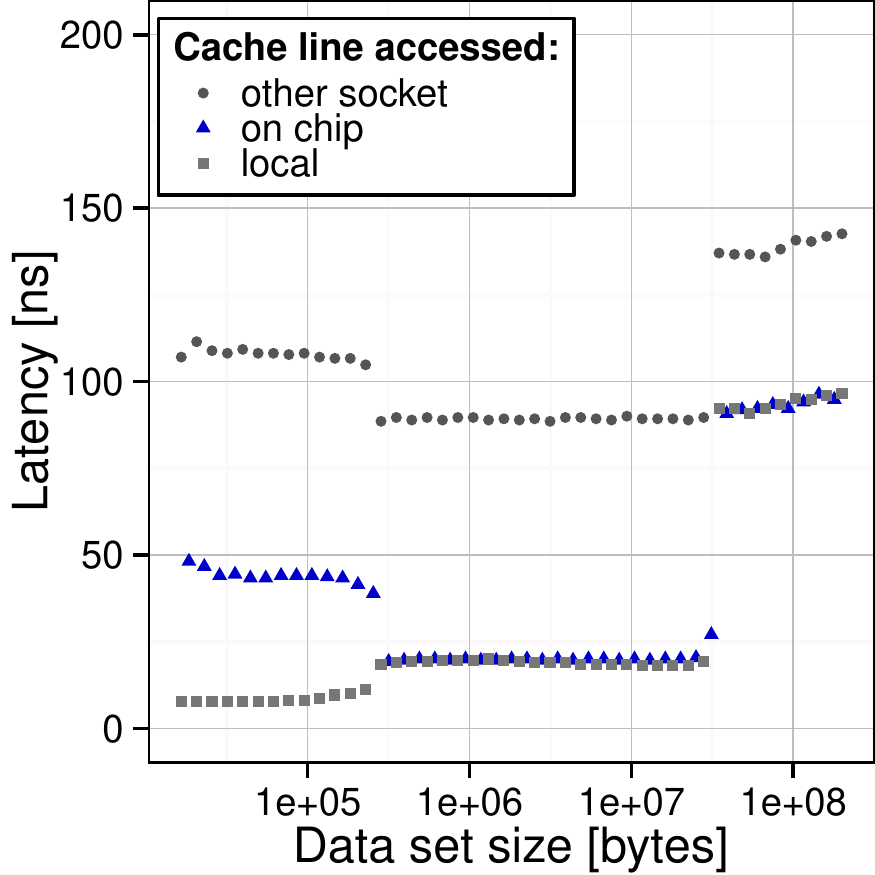}
 }\\
  \subfloat[\textsf{read}, the Exclusive state]{
  \includegraphics[width=0.35\textwidth]{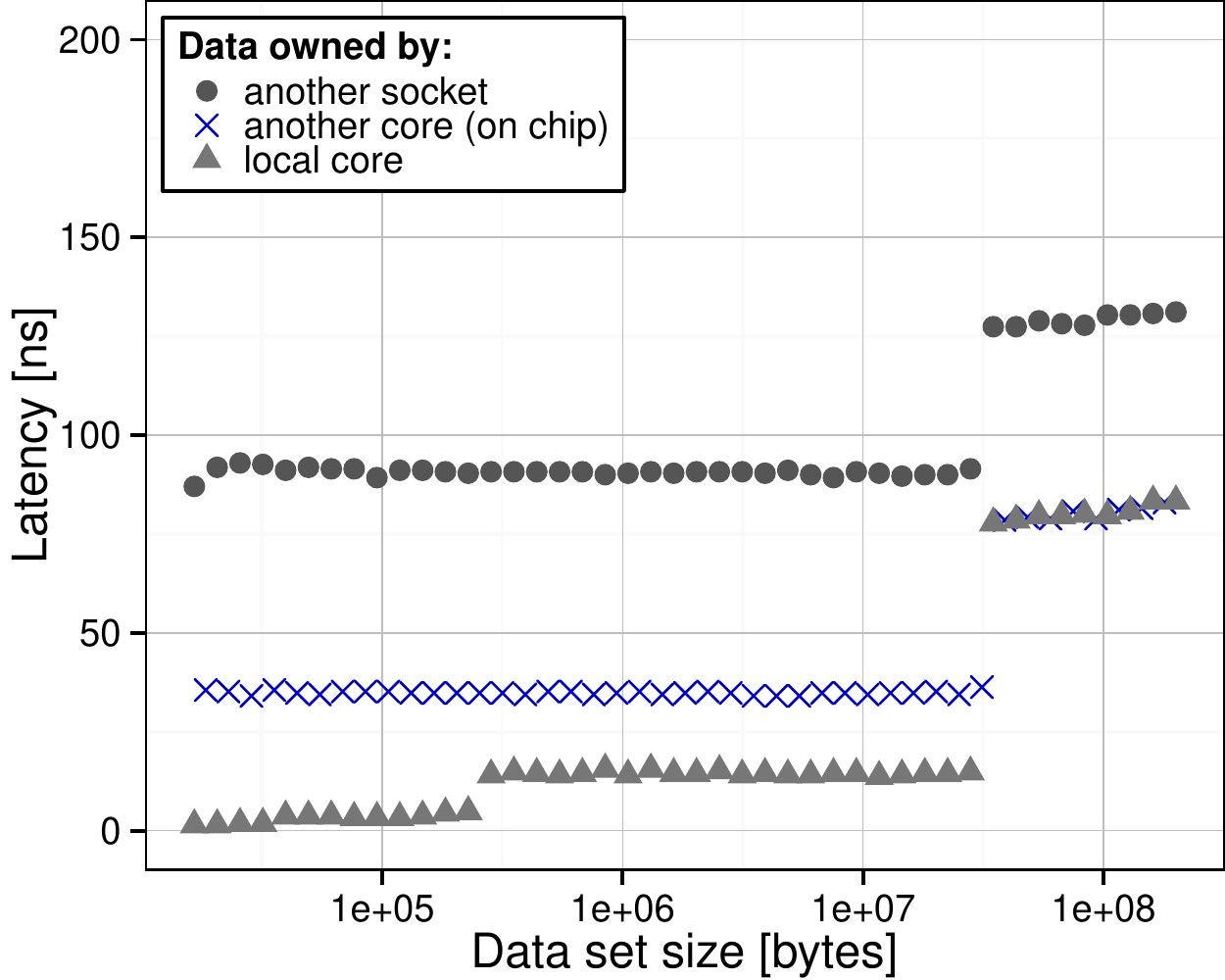}
 }
  \subfloat[\textsf{read}, the Modified state]{
  \includegraphics[width=0.3\textwidth]{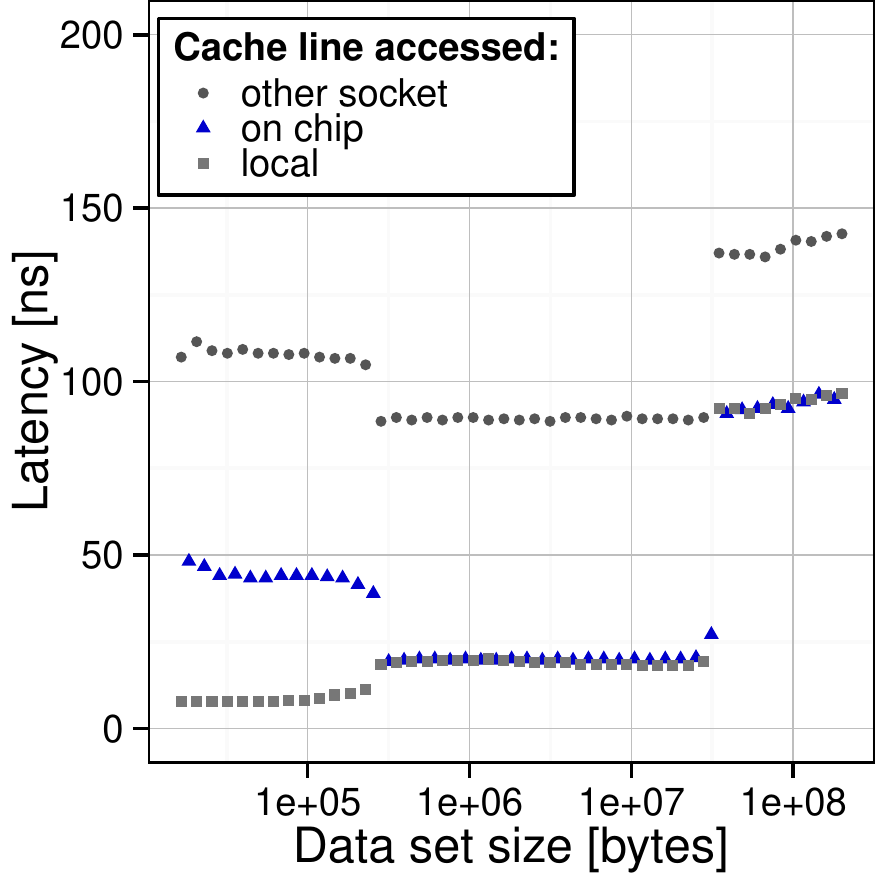}
 }
  \subfloat[\textsf{read}, the Shared state]{
  \includegraphics[width=0.3\textwidth]{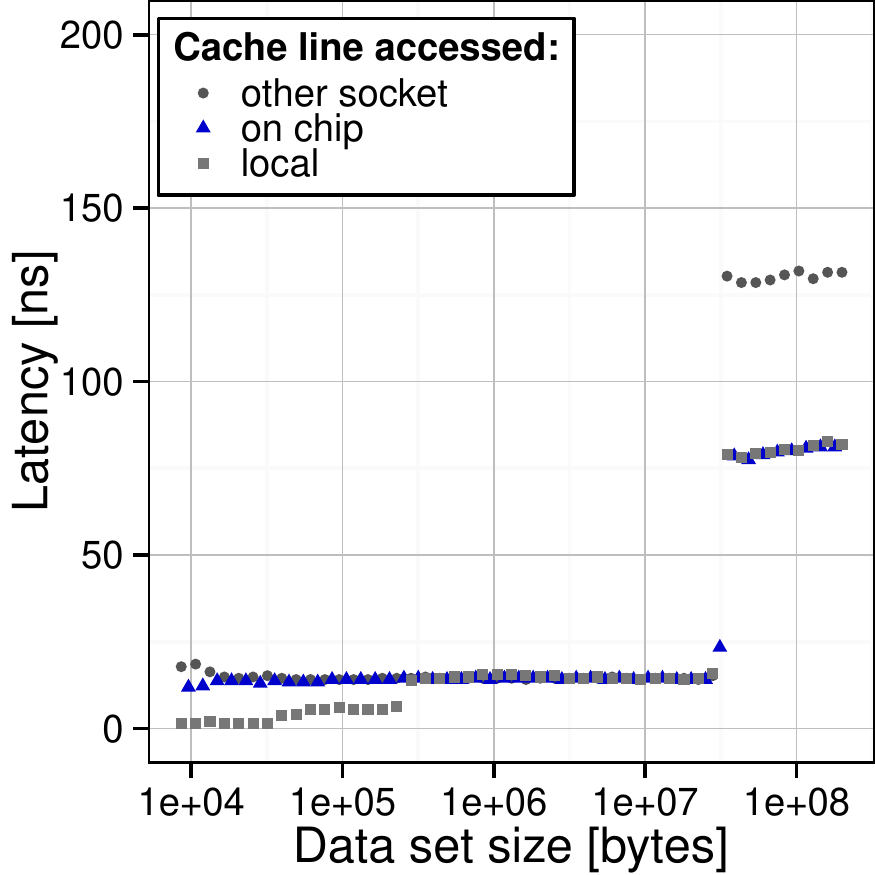}
 } 
\caption{The comparison of the latency of \textsf{CAS}, \textsf{FAA}, and \textsf{read} on Ivy Bridge. The requesting core accesses its own cache lines (local),  cache lines of a different core from the same chip (on chip), and cache lines of a different core from a different socket (other socket).}
\label{fig:Latency_results_Ivy___APP}
\end{figure*}

\begin{figure*}[!t]
 \subfloat[\textsf{CAS}, the Exclusive state]{
  \includegraphics[width=0.23\textwidth]{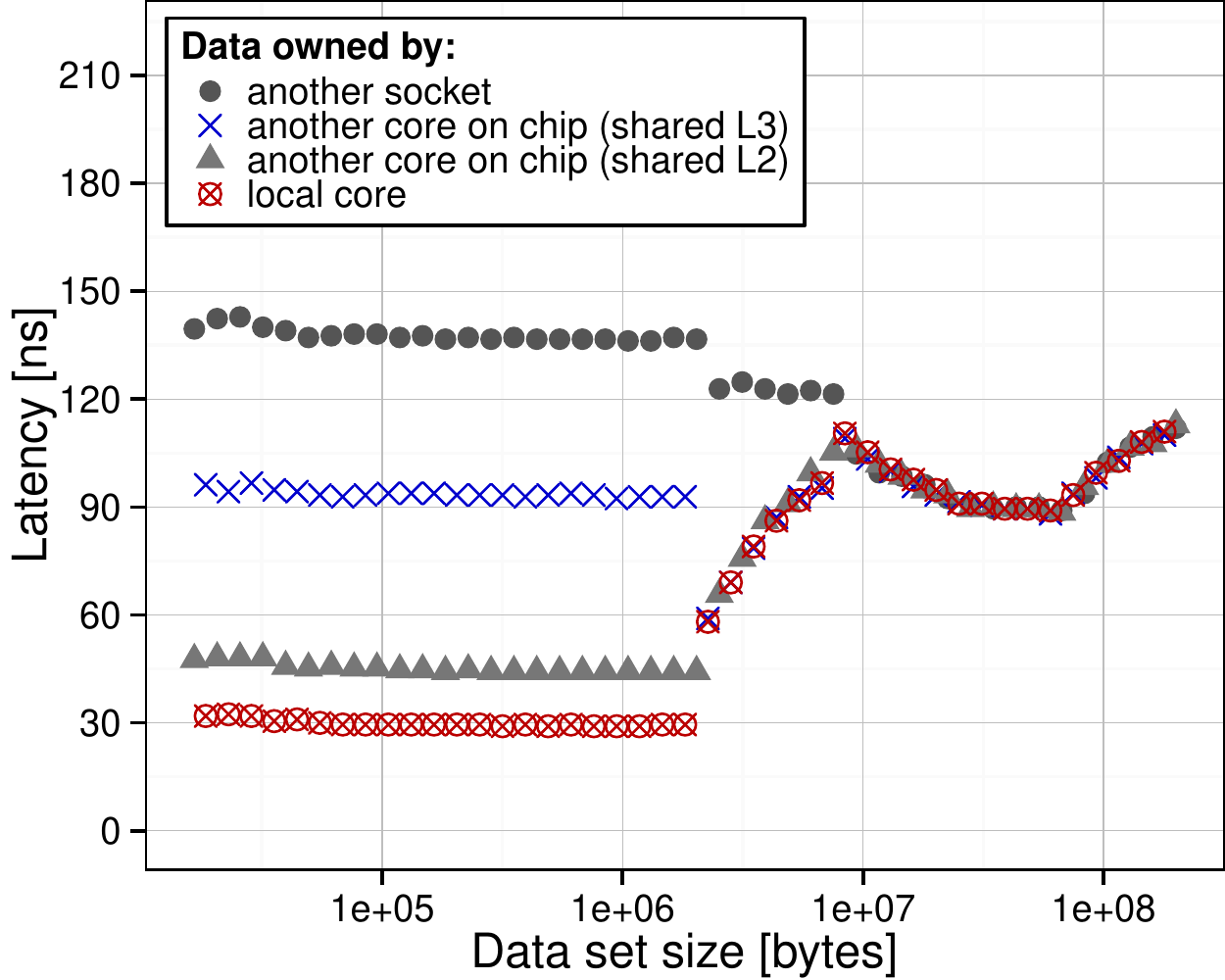}
 }
 \subfloat[\textsf{CAS}, the Modified state]{
  \includegraphics[width=0.23\textwidth]{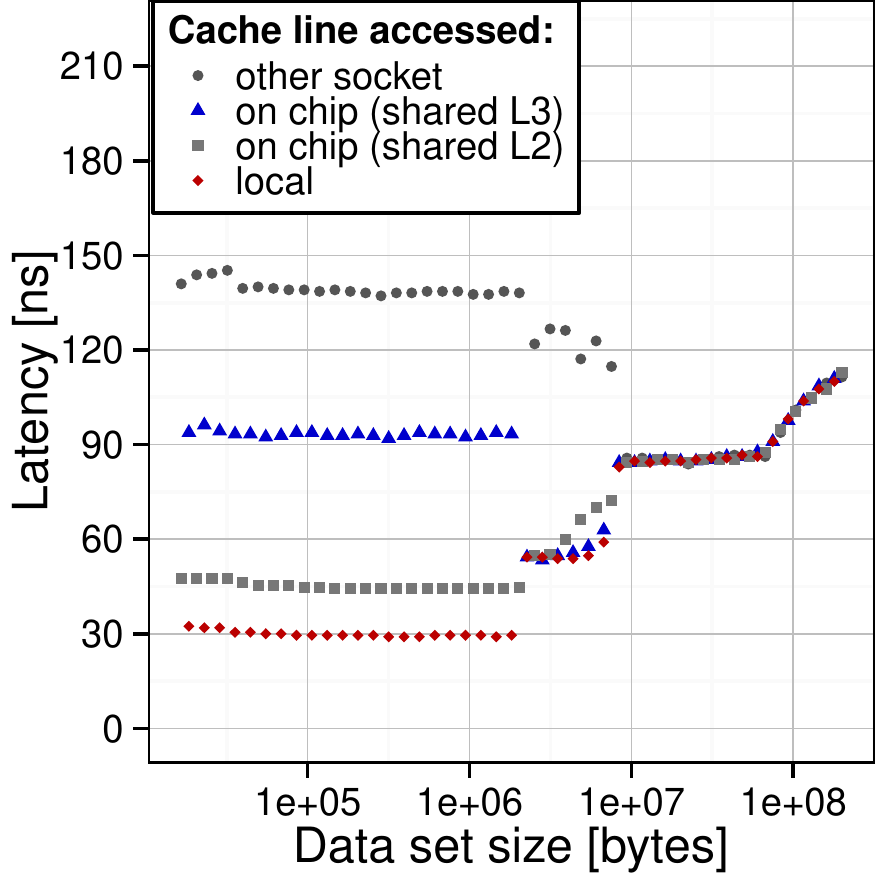}
 }
 \subfloat[\textsf{CAS}, the Shared state]{
  \includegraphics[width=0.23\textwidth]{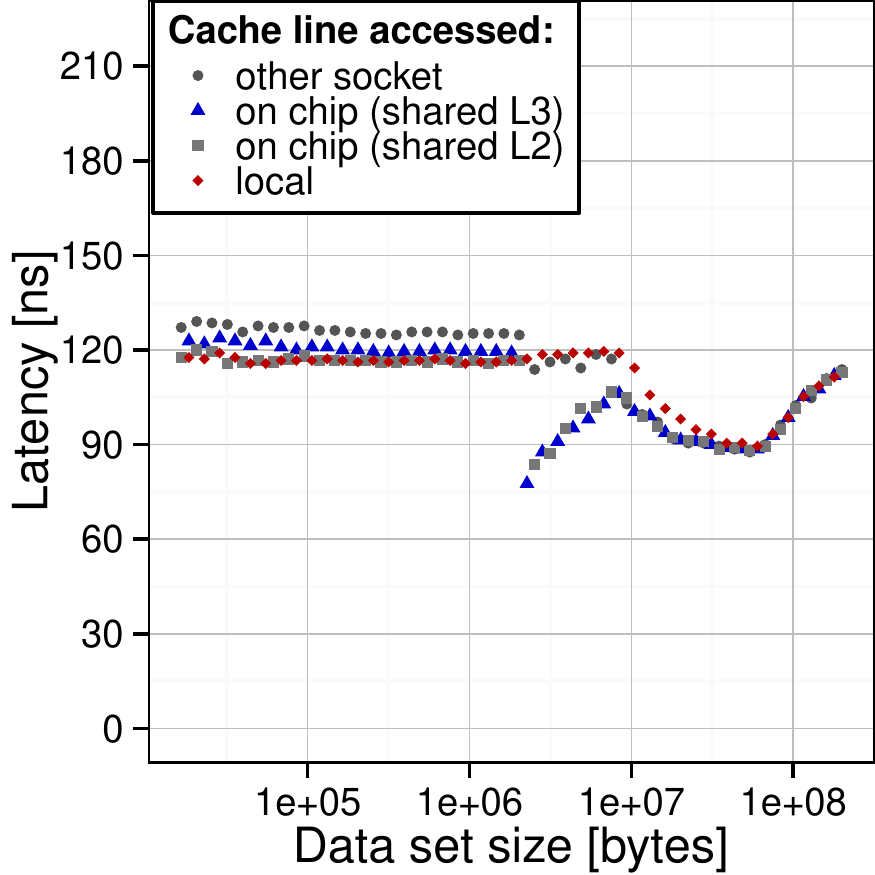}
 }
 \subfloat[\textsf{CAS}, the Owned state]{
  \includegraphics[width=0.23\textwidth]{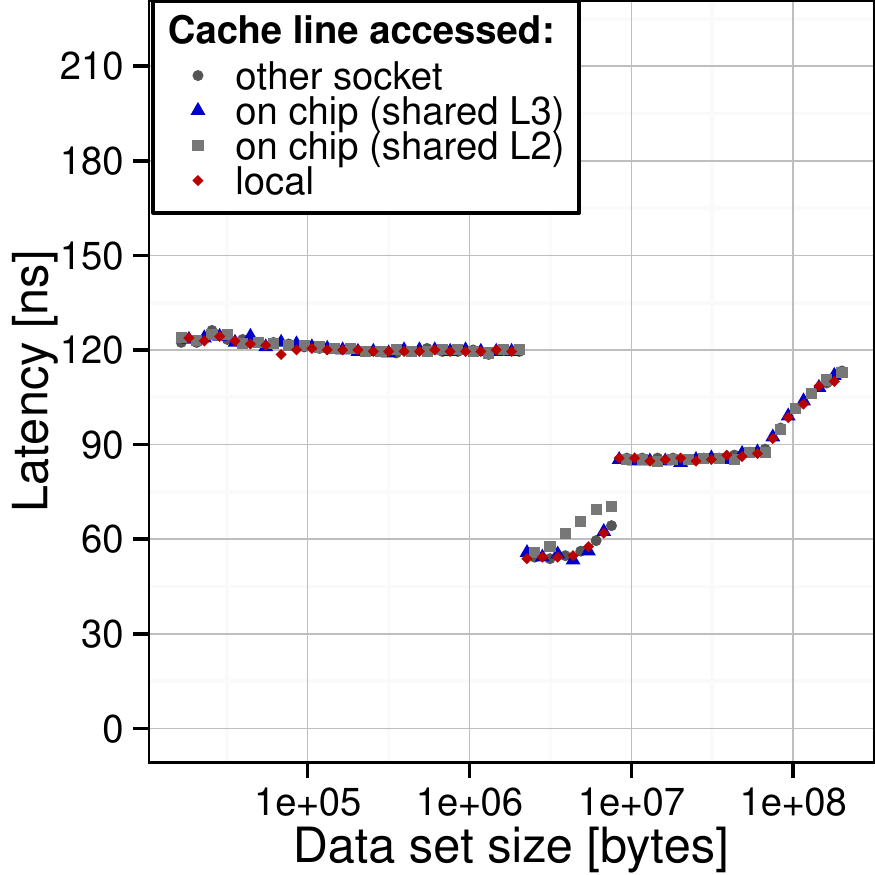}
 }\\
 \subfloat[\textsf{FAA}, the Exclusive state]{
  \includegraphics[width=0.23\textwidth]{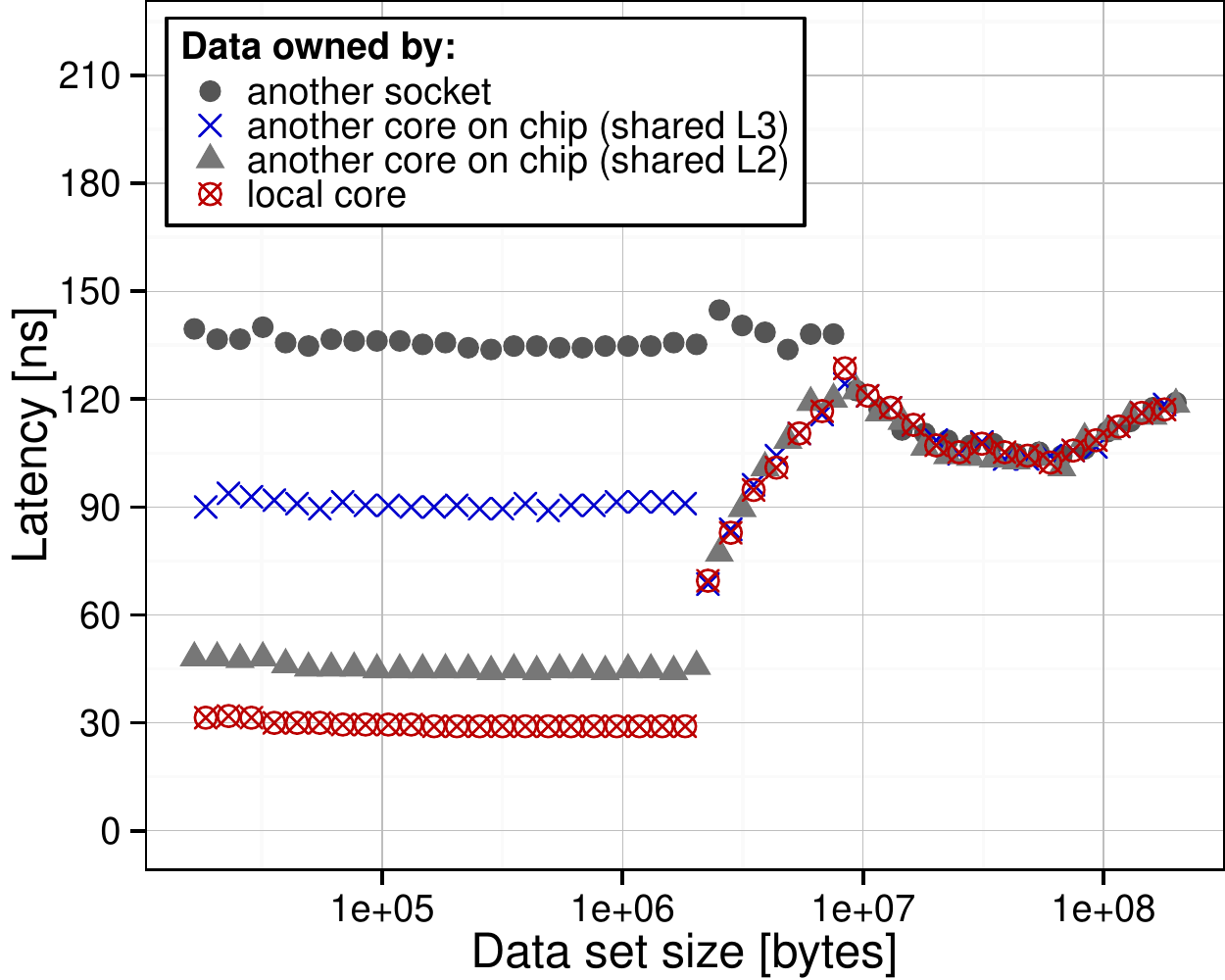}
 }
 \subfloat[\textsf{FAA}, the Modified state]{
  \includegraphics[width=0.23\textwidth]{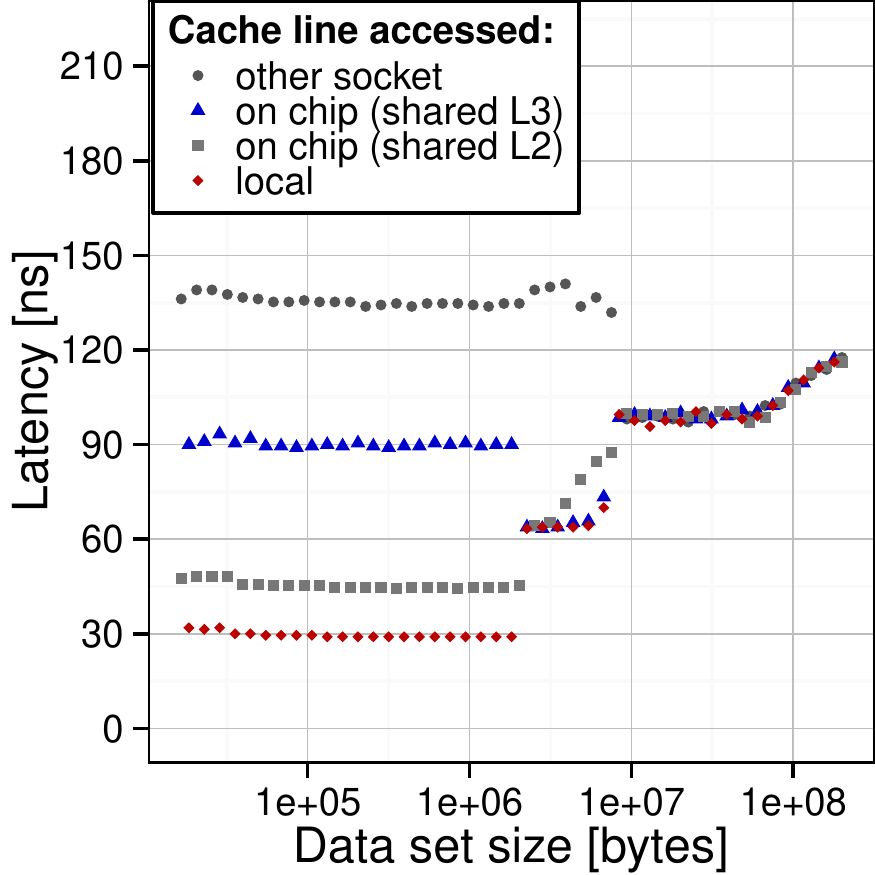}
 }
 \subfloat[\textsf{FAA}, the Shared state]{
  \includegraphics[width=0.23\textwidth]{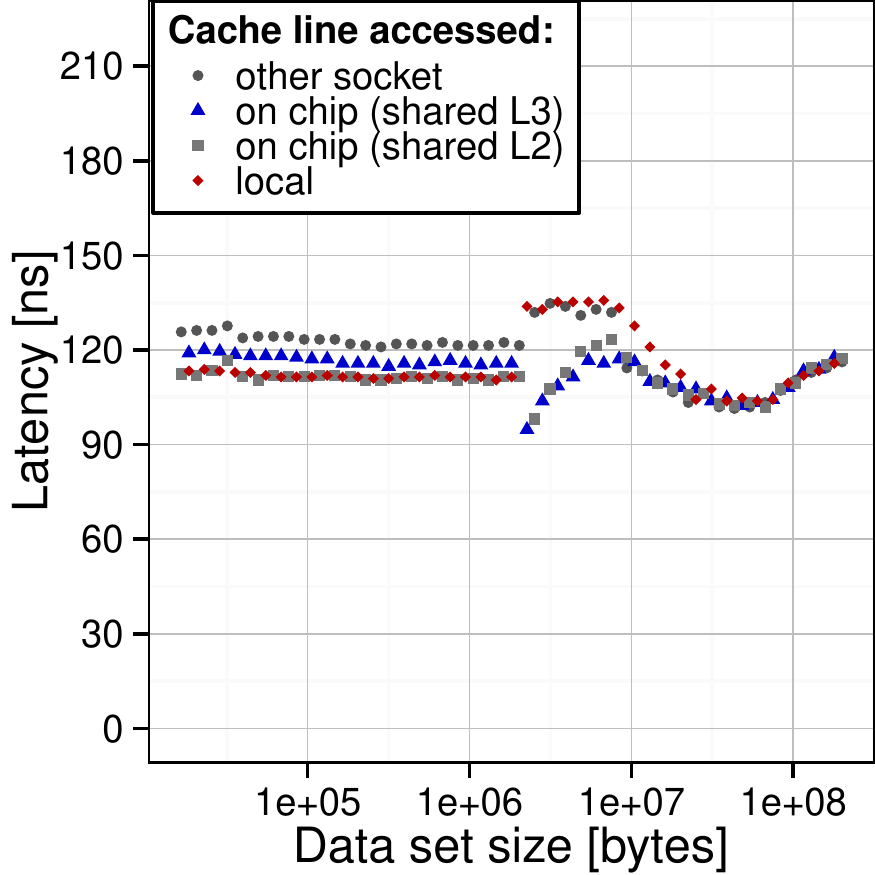}
 }
 \subfloat[\textsf{FAA}, the Owned state]{
  \includegraphics[width=0.23\textwidth]{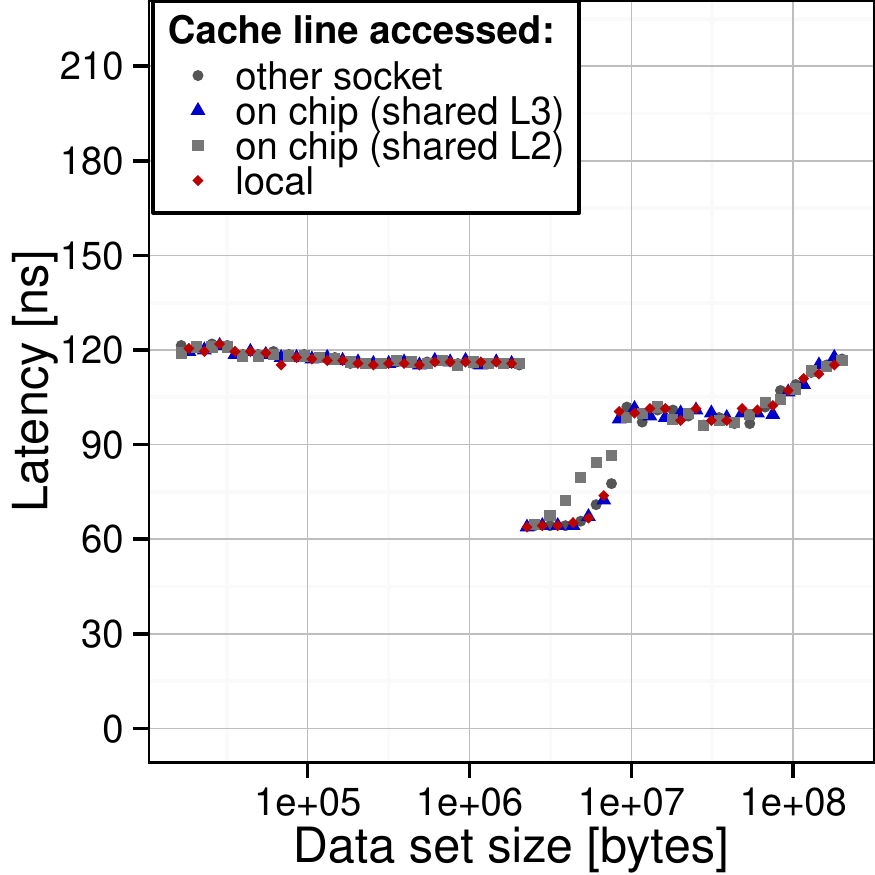}
 }\\
  \subfloat[\textsf{SWP}, the Exclusive state]{
  \includegraphics[width=0.23\textwidth]{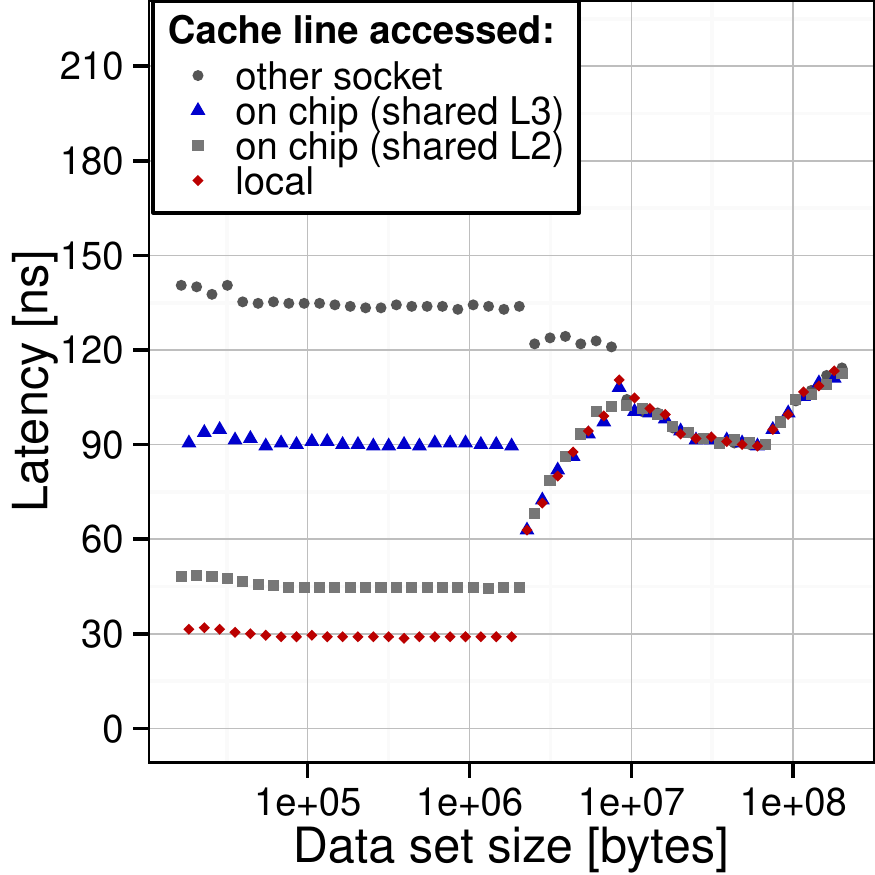}
 }
  \subfloat[\textsf{SWP}, the Modified state]{
  \includegraphics[width=0.23\textwidth]{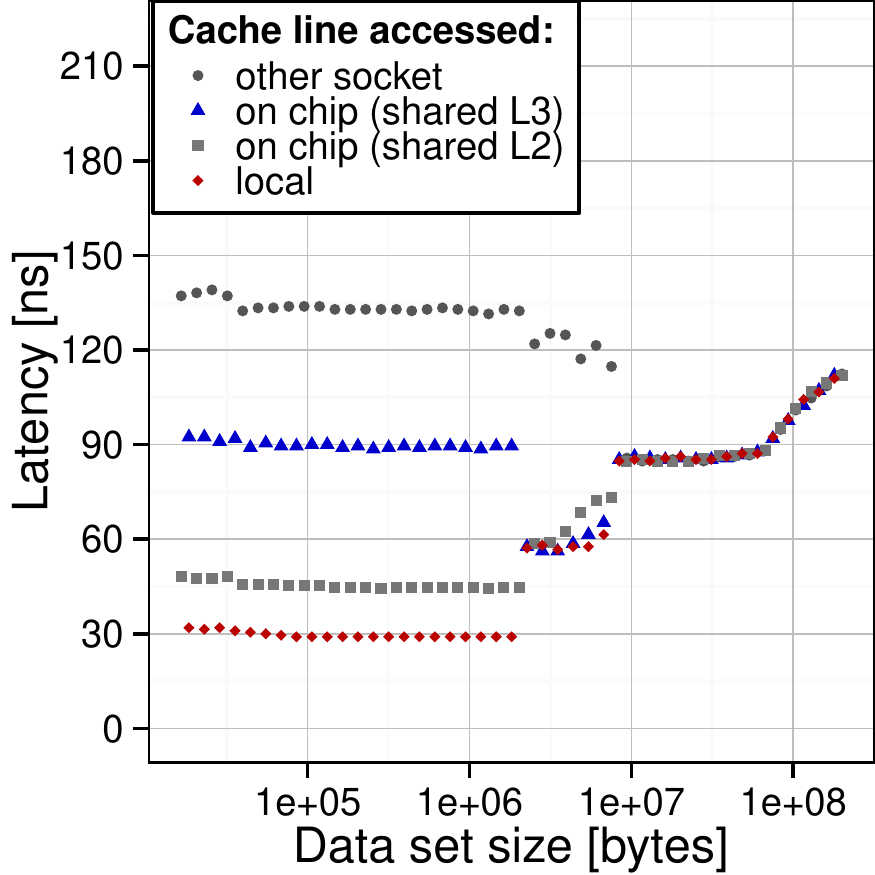}
 }
  \subfloat[\textsf{SWP}, the Shared state]{
  \includegraphics[width=0.23\textwidth]{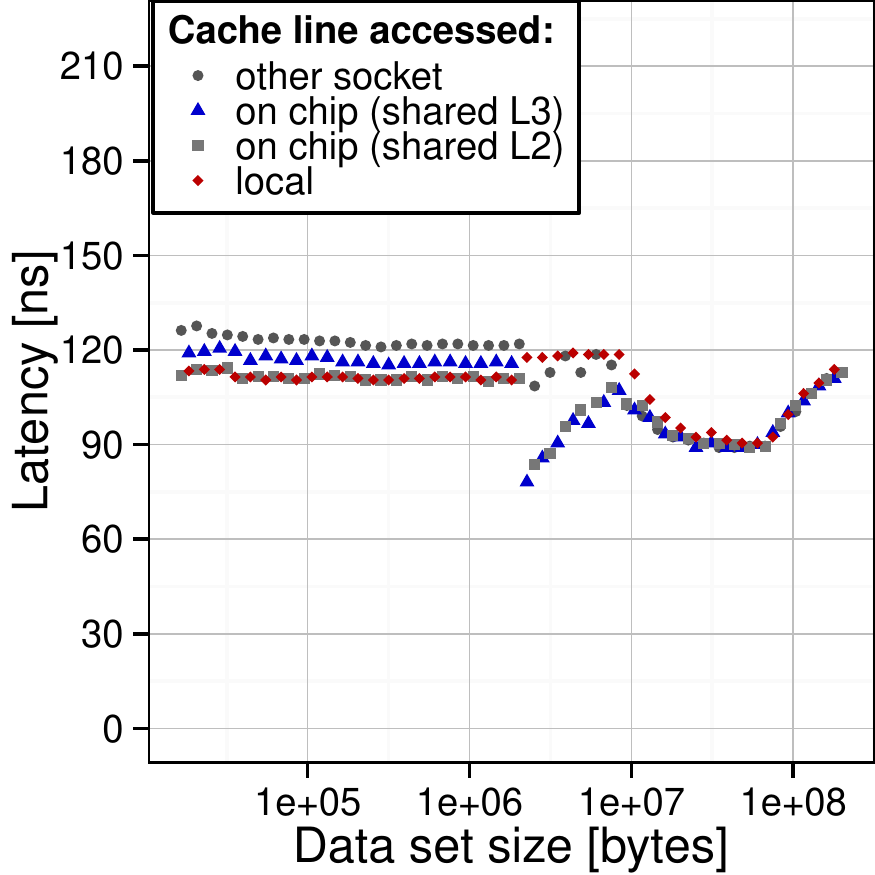}
 }
  \subfloat[\textsf{SWP}, the Owned state]{
  \includegraphics[width=0.23\textwidth]{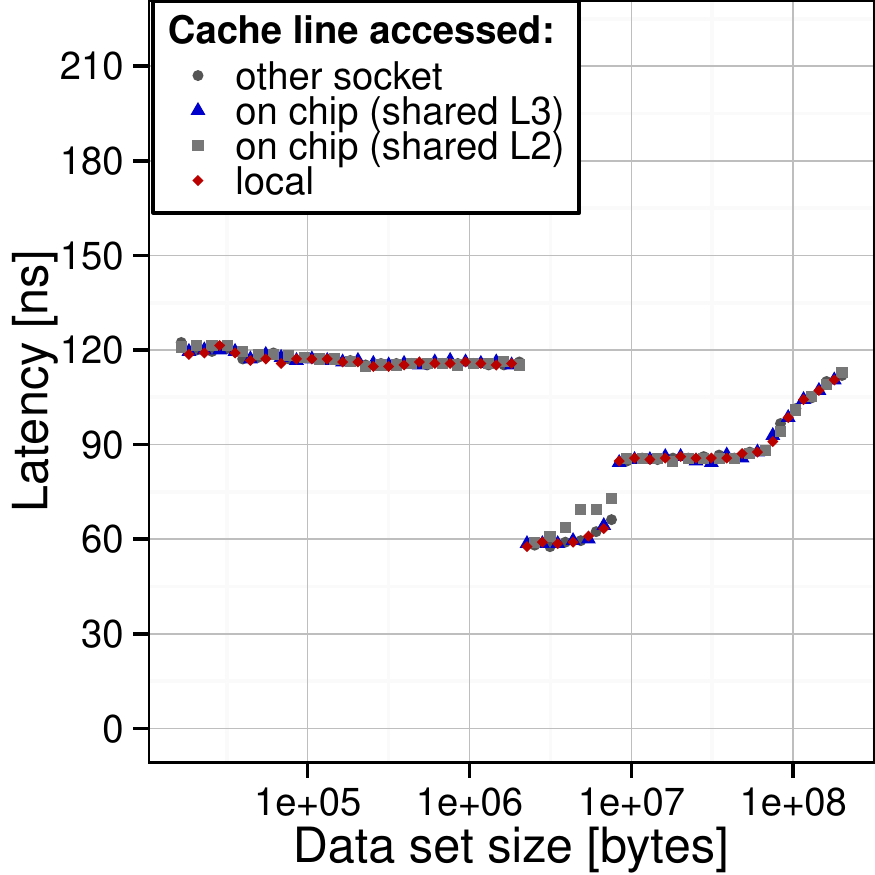}
 }\\
  \subfloat[\textsf{read}, the Exclusive state]{
  \includegraphics[width=0.23\textwidth]{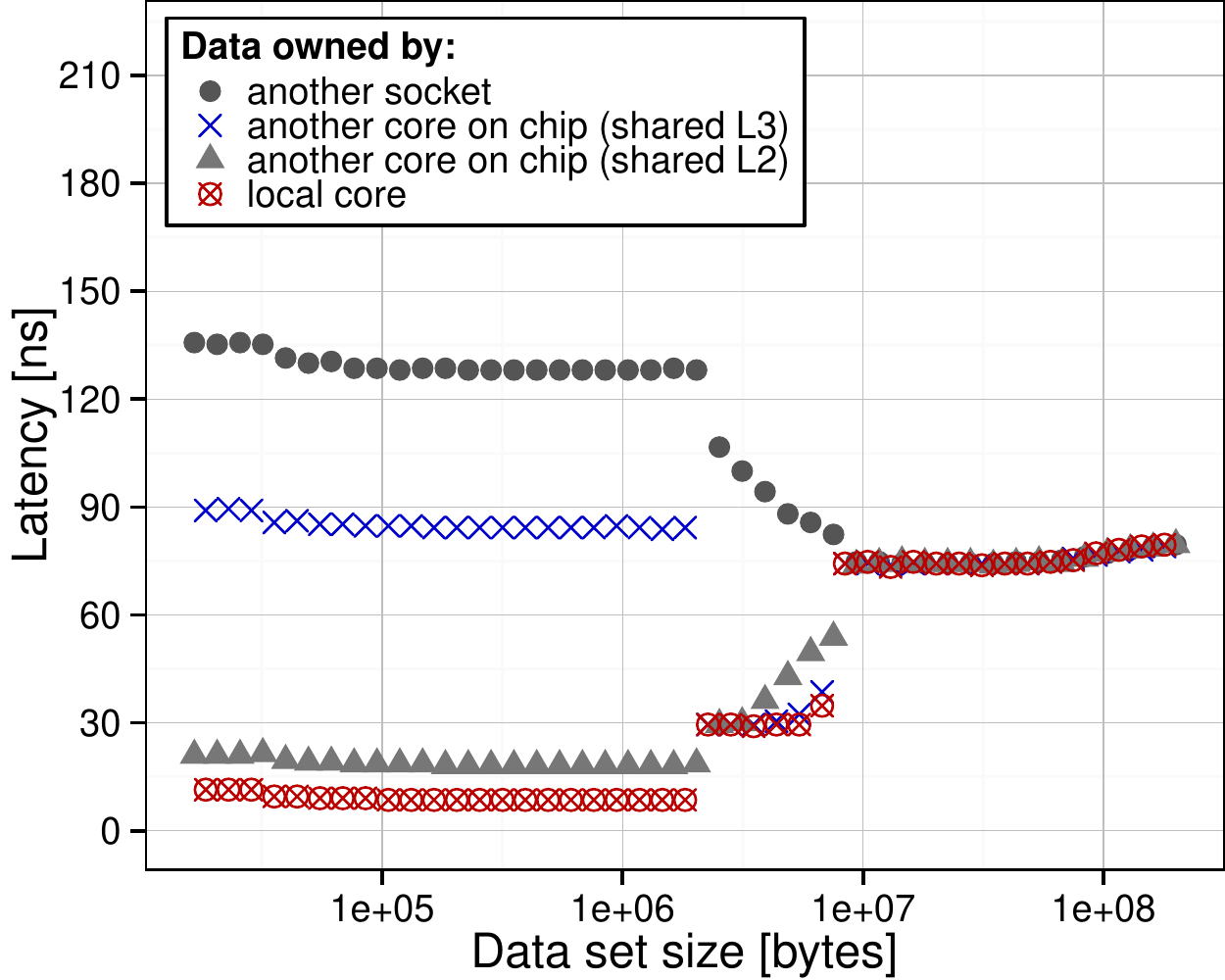}
 }
  \subfloat[\textsf{read}, the Modified state]{
  \includegraphics[width=0.23\textwidth]{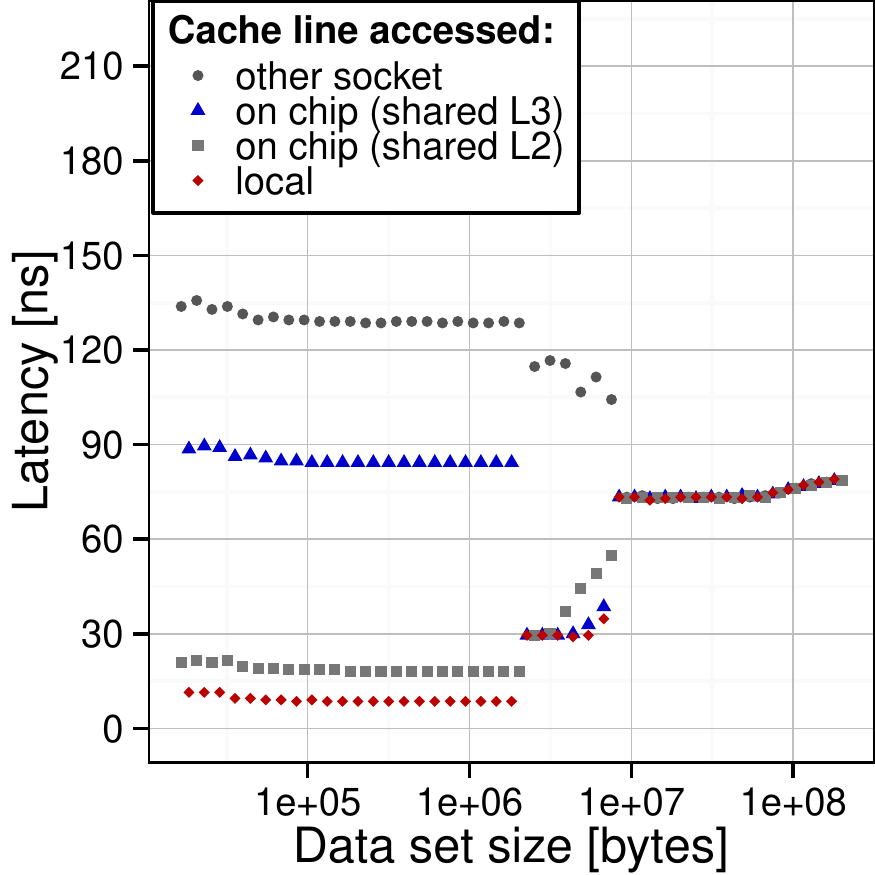}
 }
  \subfloat[\textsf{read}, the Shared state]{
  \includegraphics[width=0.23\textwidth]{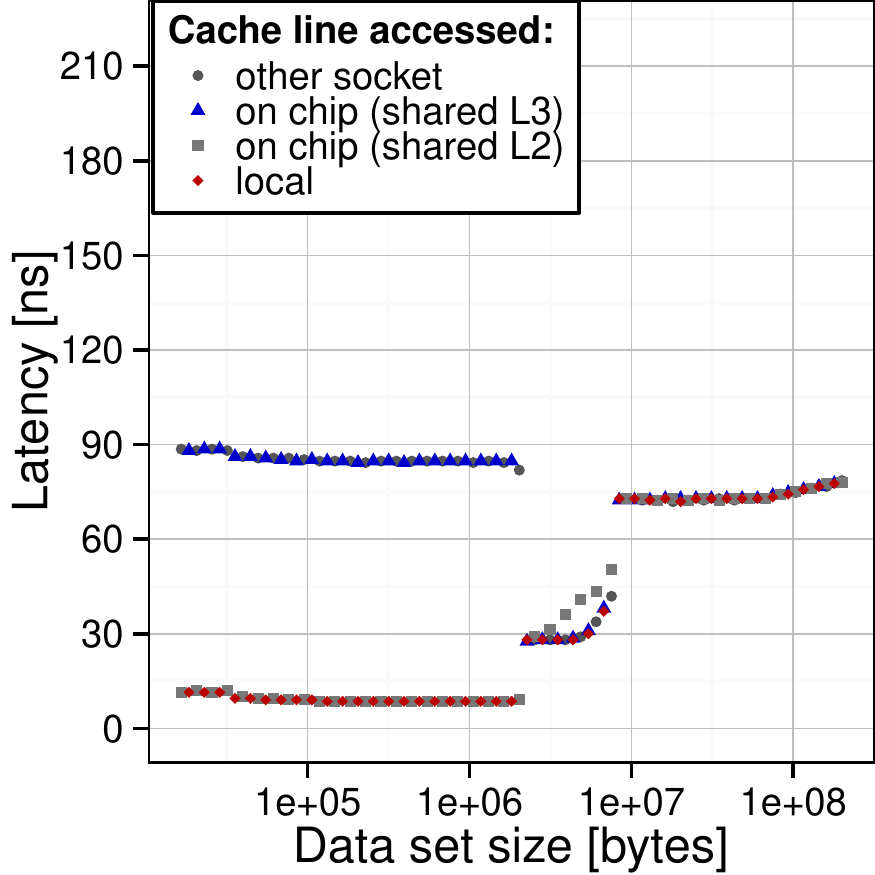}
 }
  \subfloat[\textsf{read}, the Owned state]{
  \includegraphics[width=0.23\textwidth]{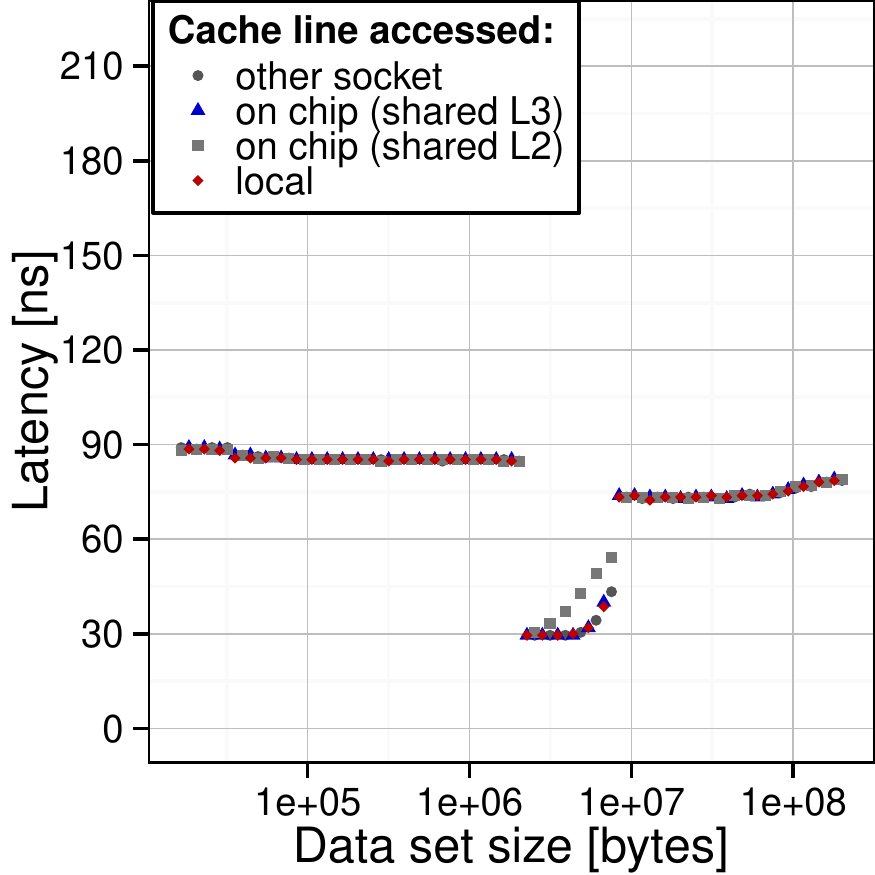}
 }
\caption{The comparison of the latency of \textsf{CAS}, \textsf{FAA}, \textsf{SWP}, and \textsf{read} on Bulldozer. The requesting core accesses its own cache lines (local),  cache lines of different cores that share L2 and L3 with the requesting core (on chip, shared L2 and L3, respectively), and cache lines of a different core from a different socket (other socket).}
\label{fig:Latency_results_Bull___APP}
\end{figure*}

\begin{figure*}[t]
 \subfloat[\textsf{CAS}, the Exclusive state]{
  \includegraphics[width=0.33\textwidth]{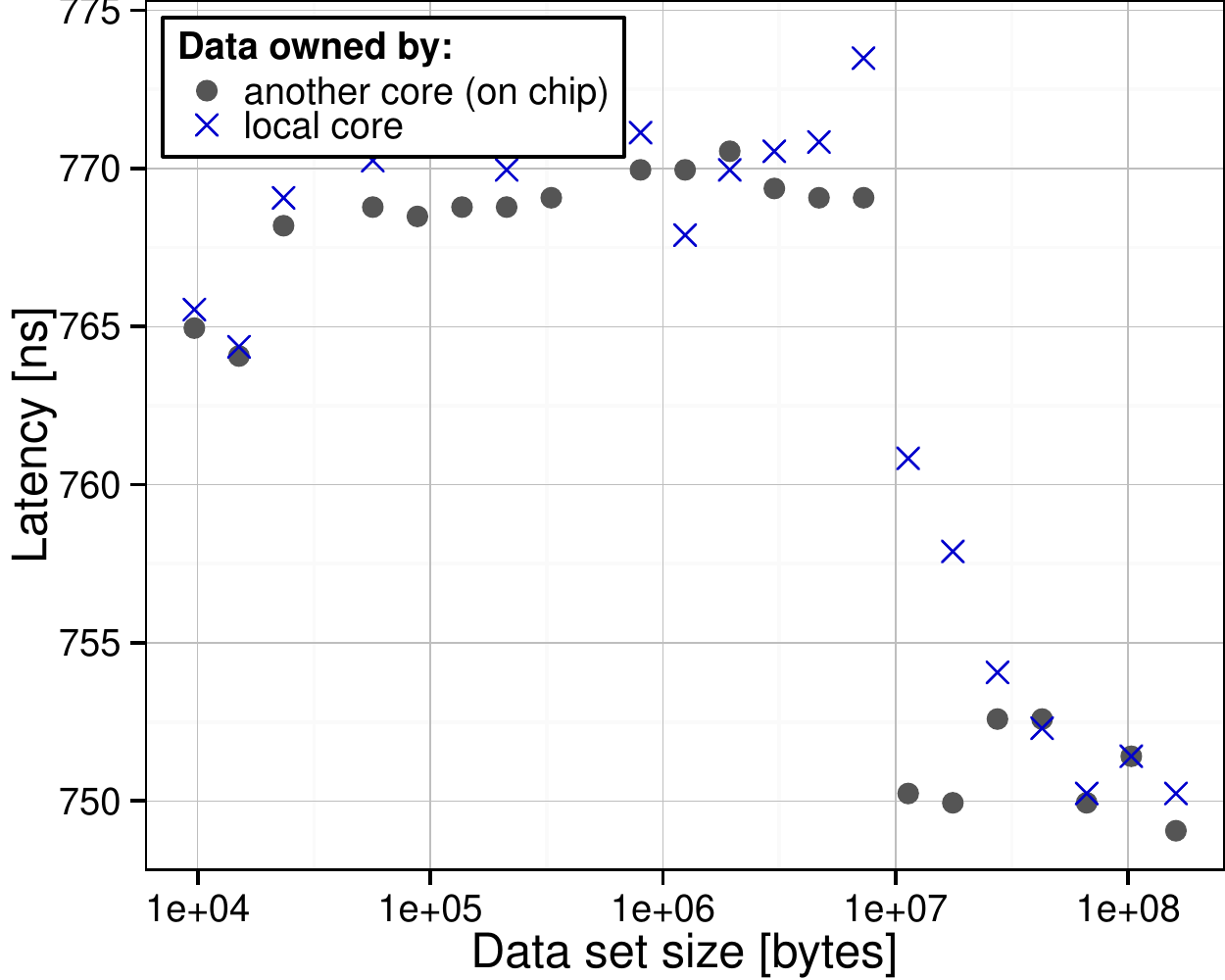}
 }
 \subfloat[\textsf{CAS}, the Modified state]{
  \includegraphics[width=0.33\textwidth]{lat_haswell_cas_M____UNALIGNED-eps-converted-to.pdf}
 }\\
 \subfloat[\textsf{FAA}, the Exclusive state]{
  \includegraphics[width=0.33\textwidth]{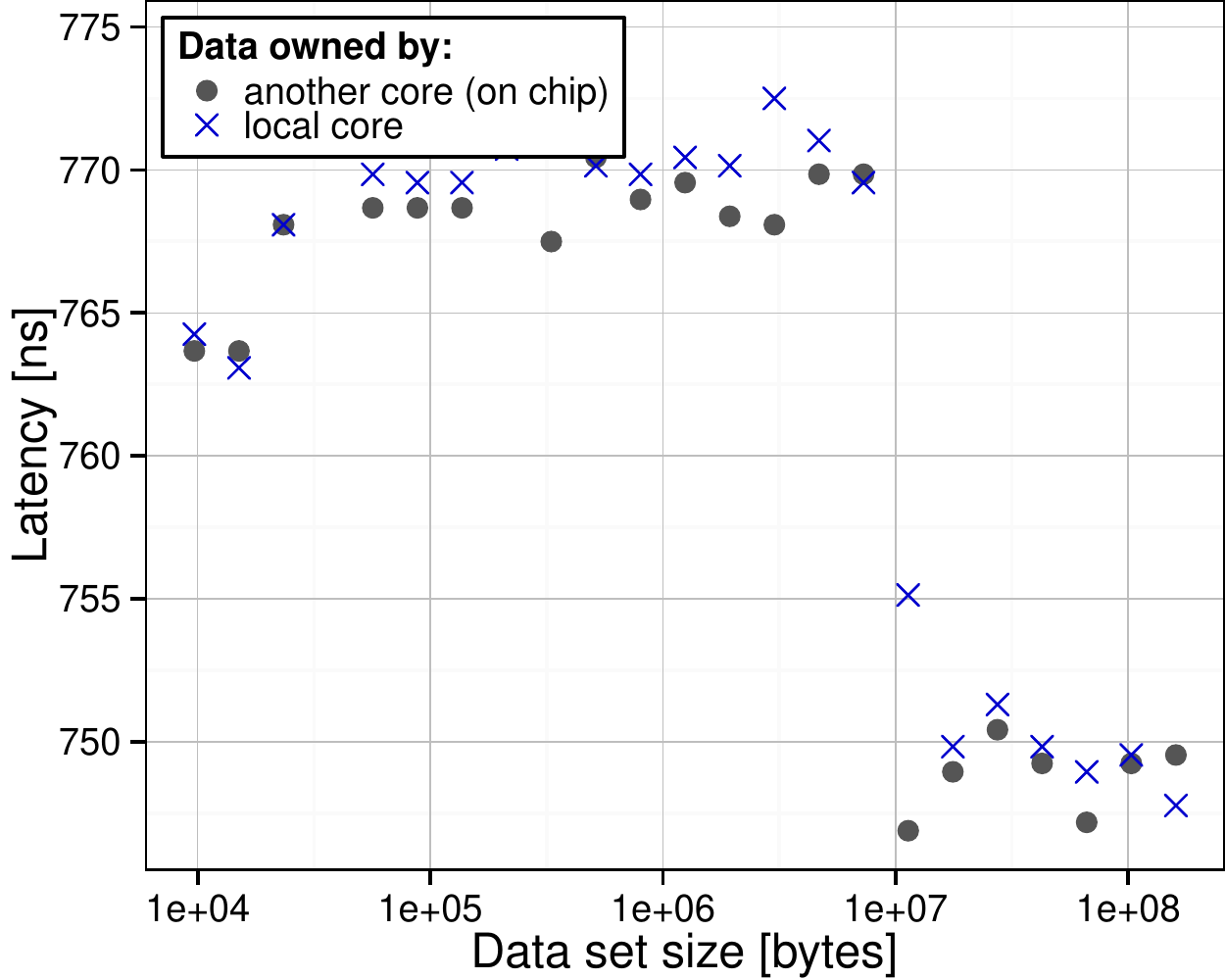}
 }
 \subfloat[\textsf{FAA}, the Modified state]{
  \includegraphics[width=0.33\textwidth]{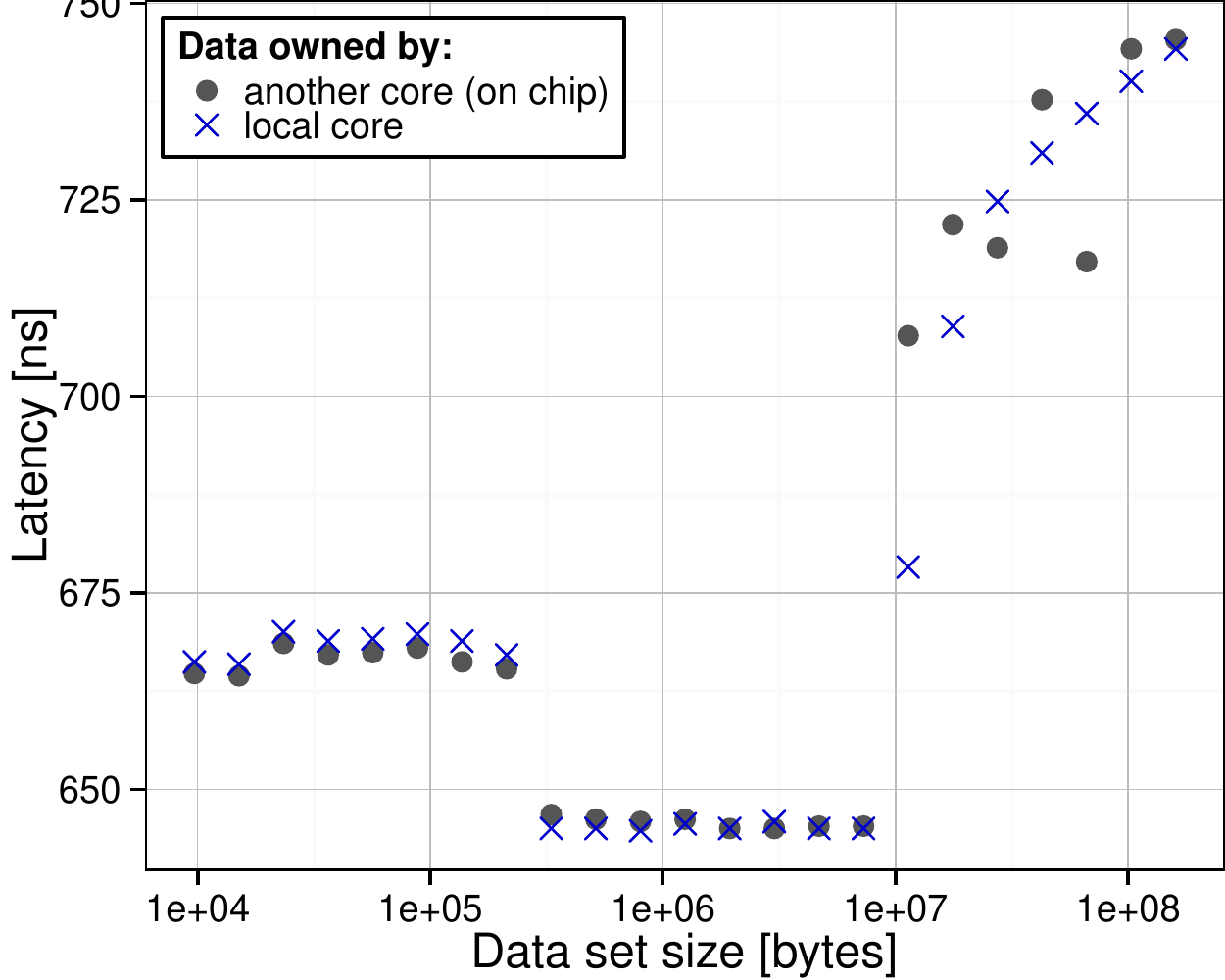}
 }\\
  \subfloat[\textsf{read}, the Exclusive state]{
  \includegraphics[width=0.33\textwidth]{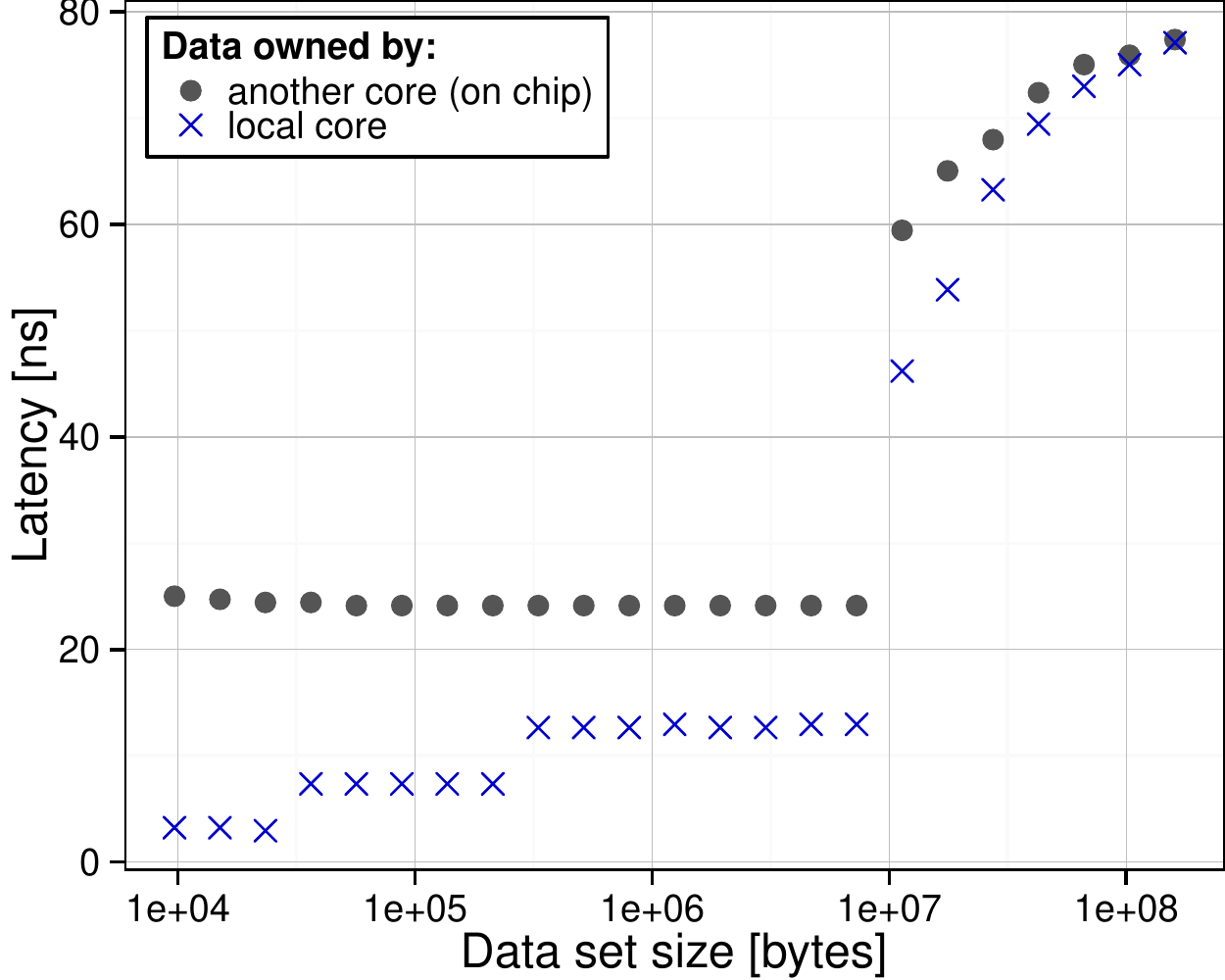}
 }
  \subfloat[\textsf{read}, the Modified state]{
  \includegraphics[width=0.33\textwidth]{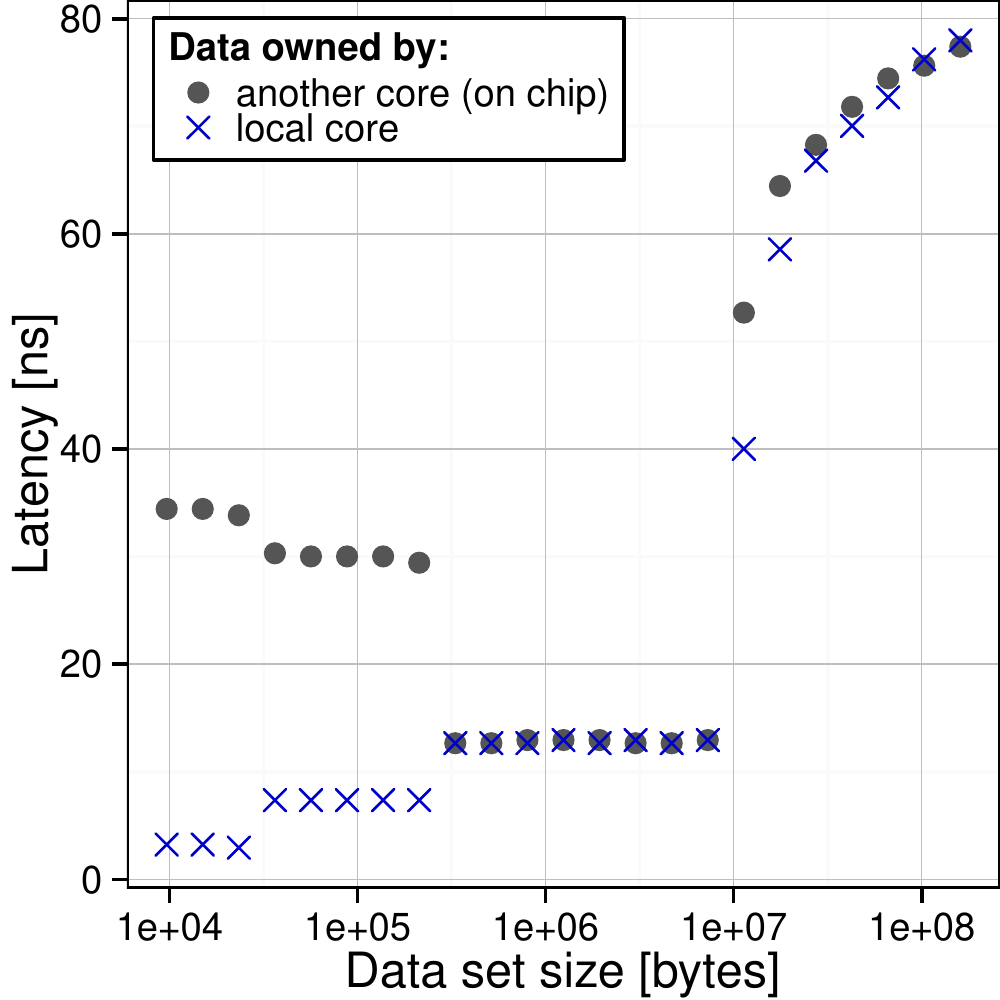}
 }
\caption{The comparison of the latency of \ul{unaligned} \textsf{CAS}, \textsf{FAA}, and \textsf{read} on Haswell. The requesting core accesses its own cache lines (local) and cache lines of a different core from the same chip (on chip)}
\label{fig:Latency_results_Haswell_un___APP}
\end{figure*}

\begin{figure*}[t]
 \subfloat[\textsf{CAS}, the Exclusive state]{
  \includegraphics[width=0.22\textwidth]{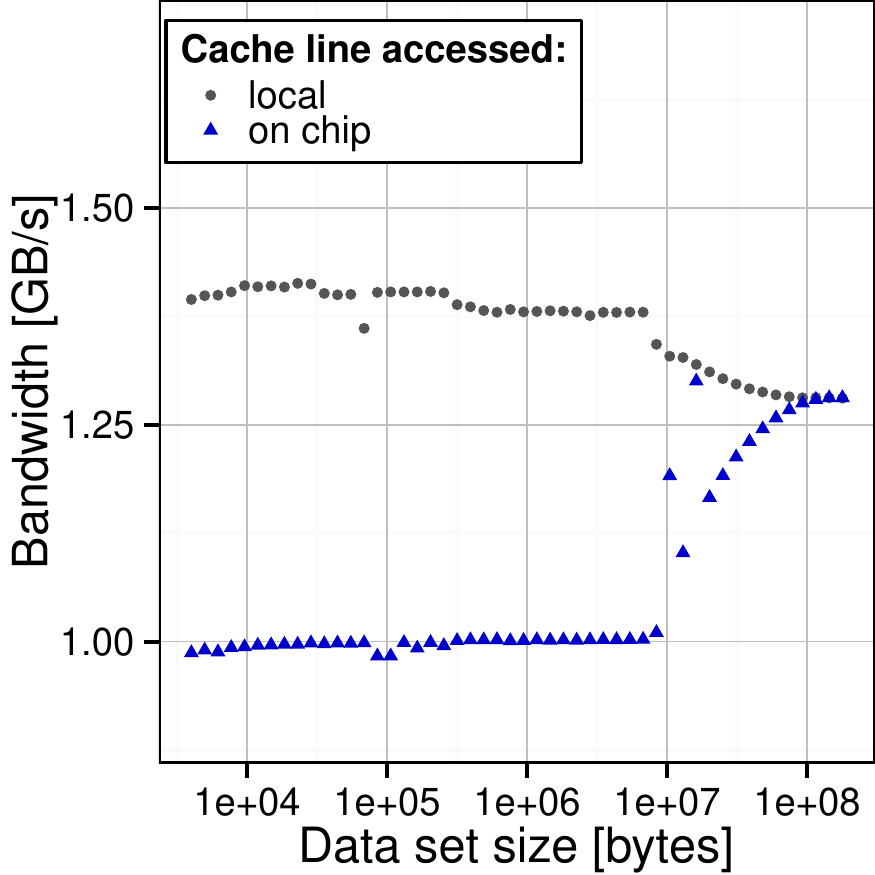}
 }
 \subfloat[\textsf{FAA}, the Exclusive state]{
  \includegraphics[width=0.22\textwidth]{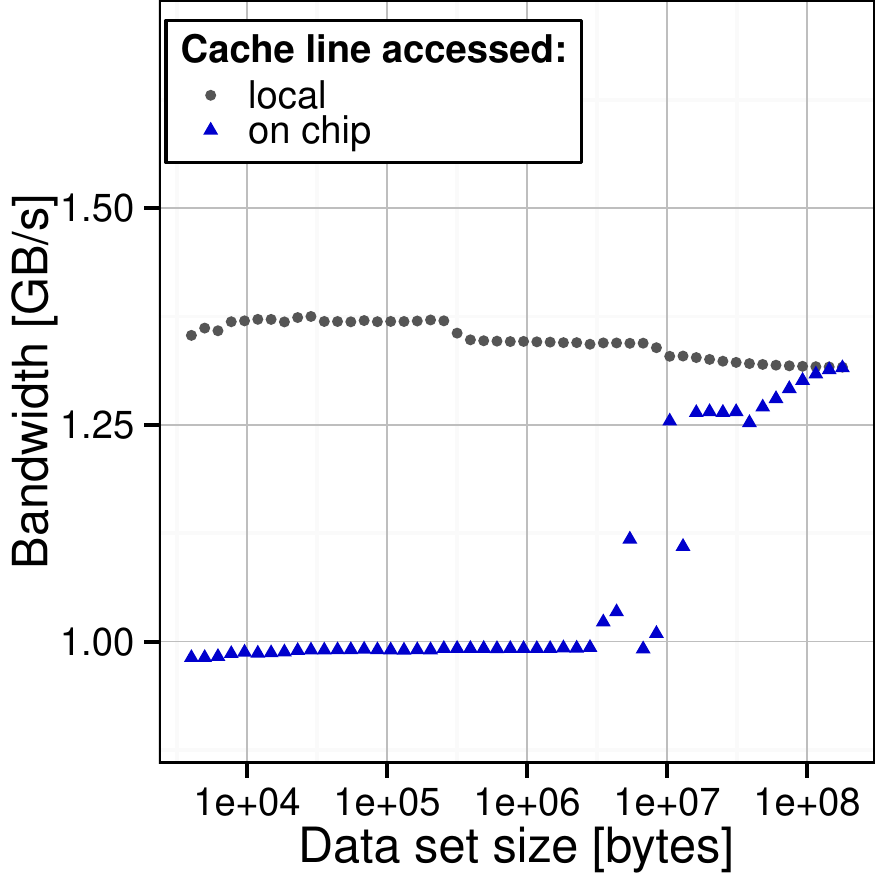}
 }
  \subfloat[\textsf{SWP}, the Exclusive state]{
  \includegraphics[width=0.22\textwidth]{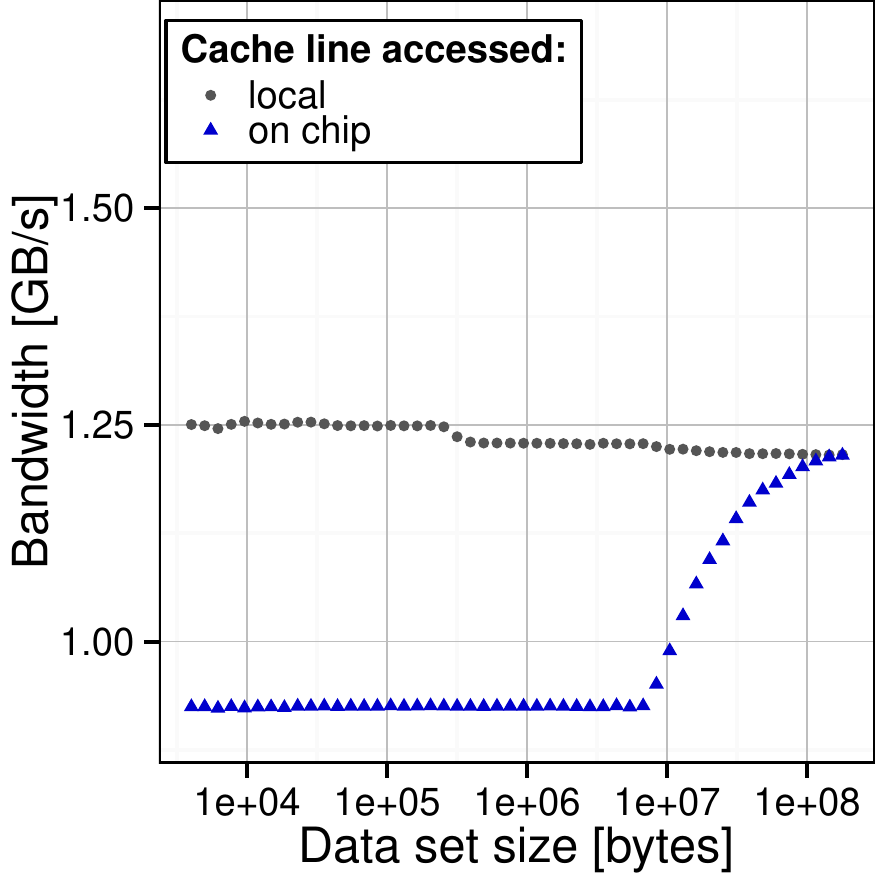}
 }
  \subfloat[\textsf{read}, the Exclusive state]{
  \includegraphics[width=0.22\textwidth]{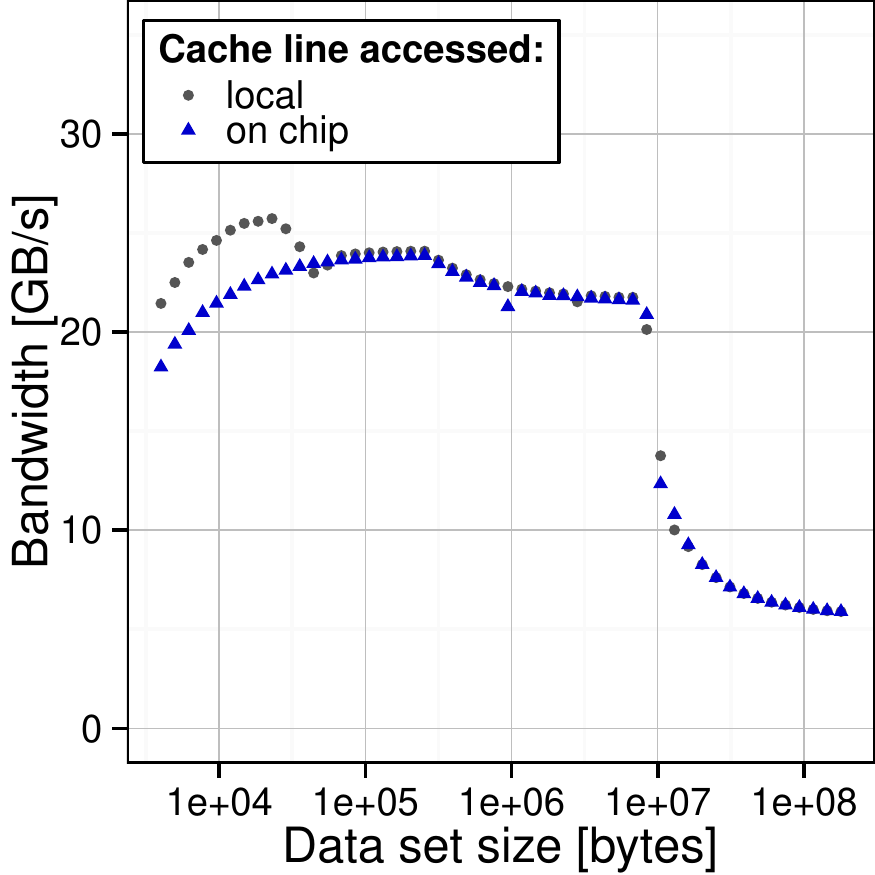}
 }\\
 \subfloat[\textsf{CAS}, the Modified state]{
  \includegraphics[width=0.22\textwidth]{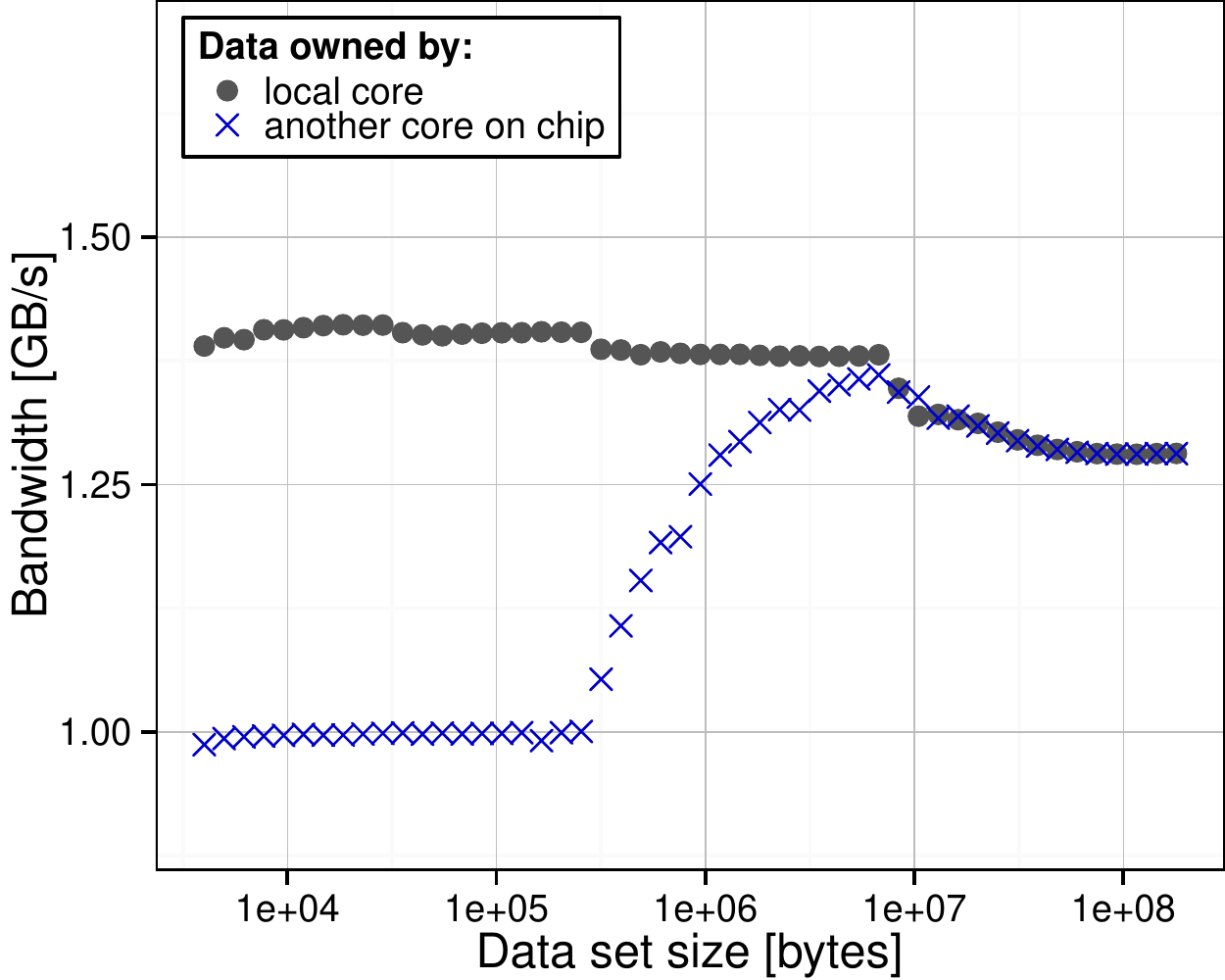}
 }
 \subfloat[\textsf{FAA}, the Modified state]{
  \includegraphics[width=0.22\textwidth]{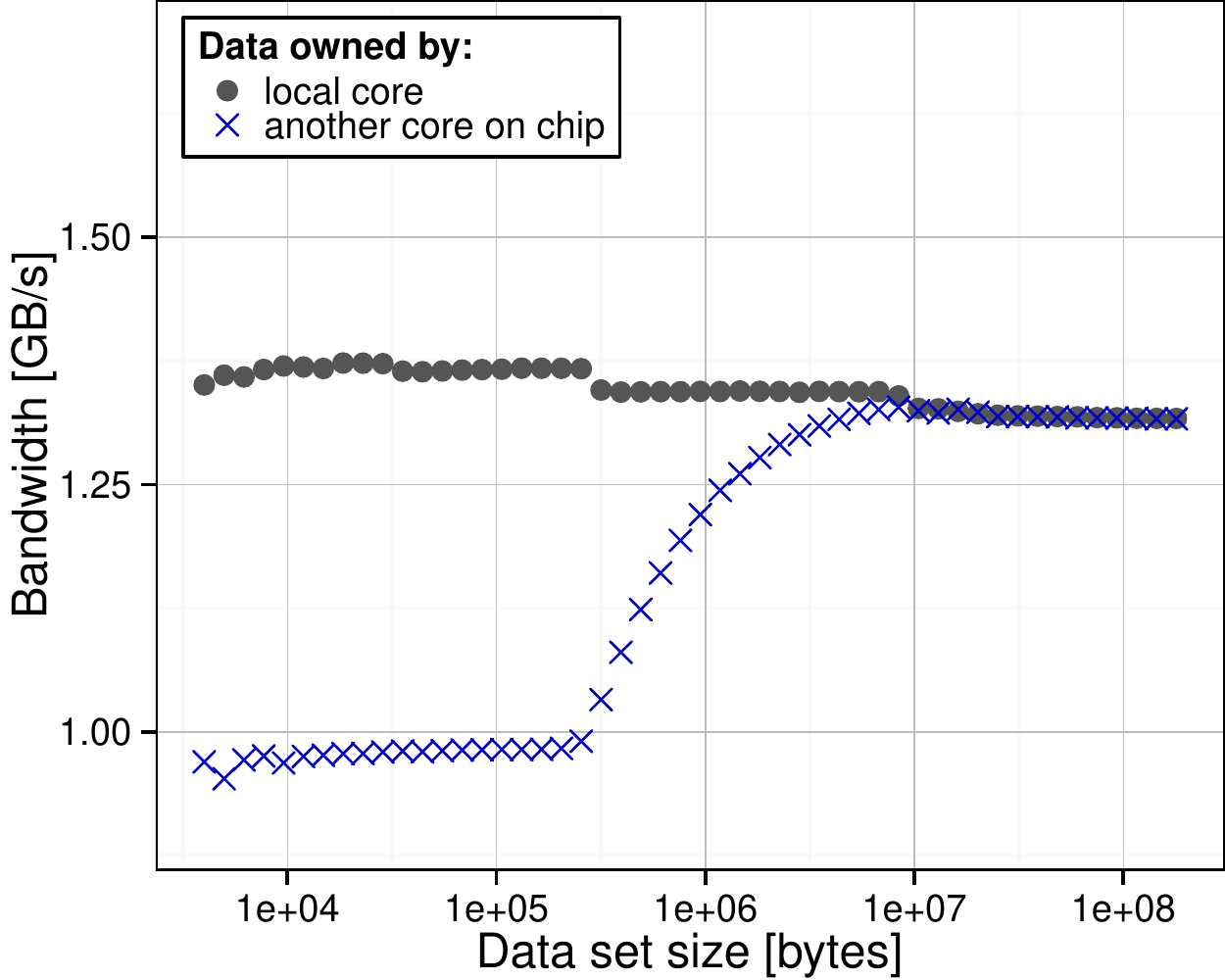}
 }
  \subfloat[\textsf{SWP}, the Modified state]{
  \includegraphics[width=0.22\textwidth]{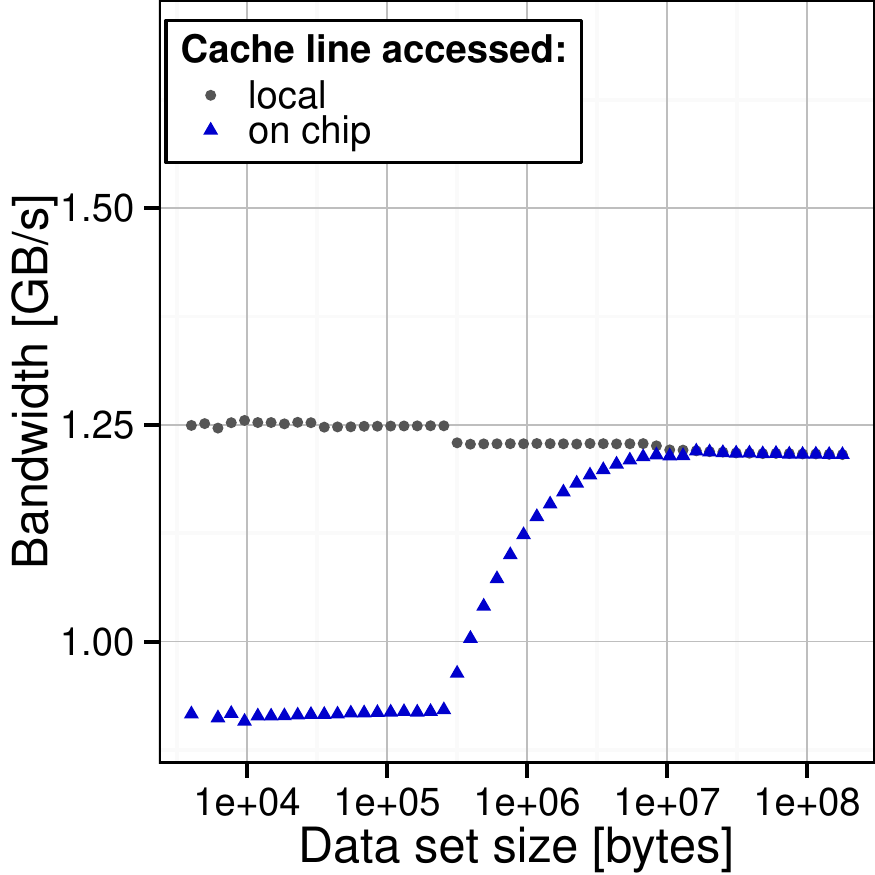}
 }
  \subfloat[\textsf{read}, the Modified state]{
  \includegraphics[width=0.22\textwidth]{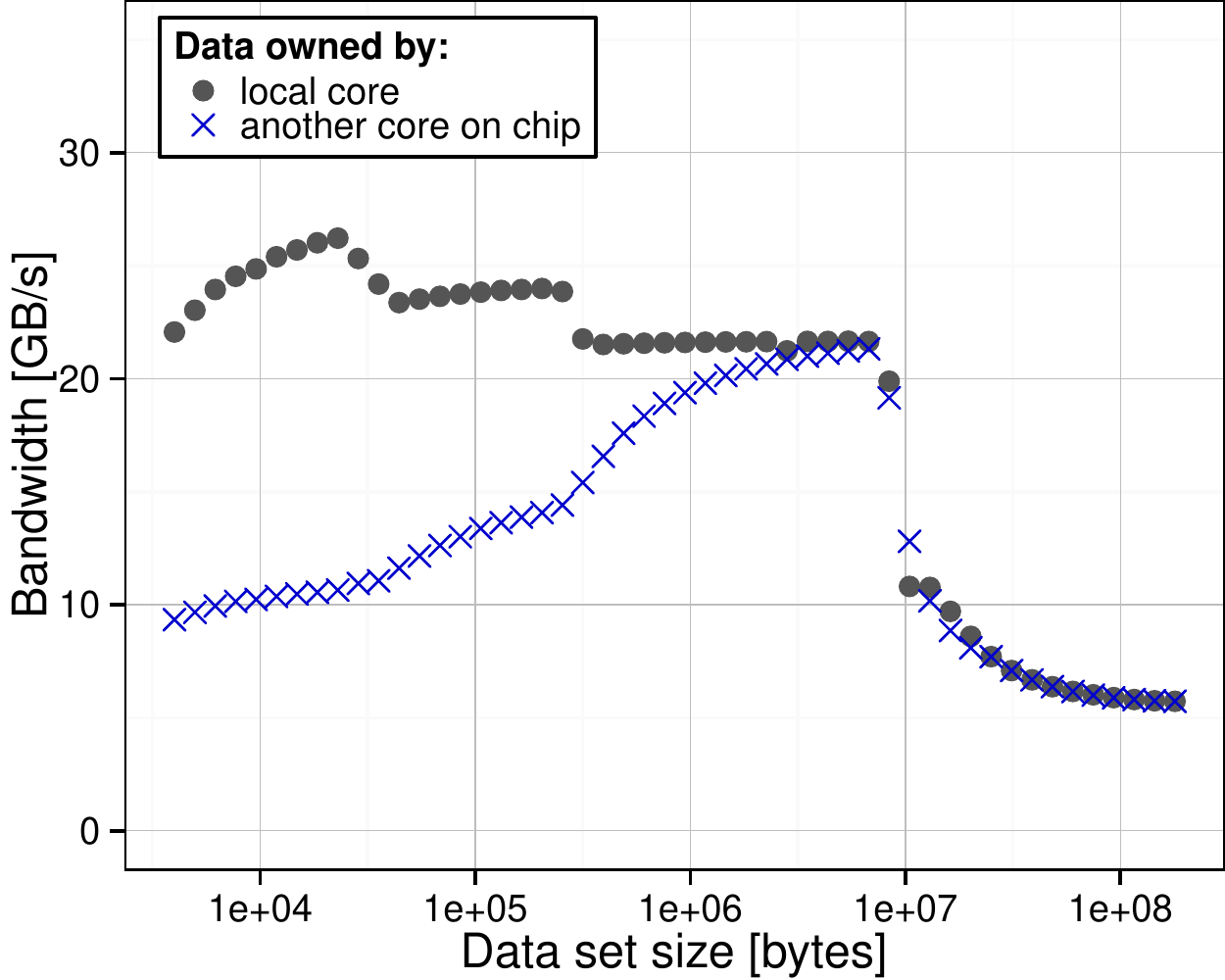}
 }\\
 \subfloat[\textsf{CAS}, the Shared state]{
  \includegraphics[width=0.22\textwidth]{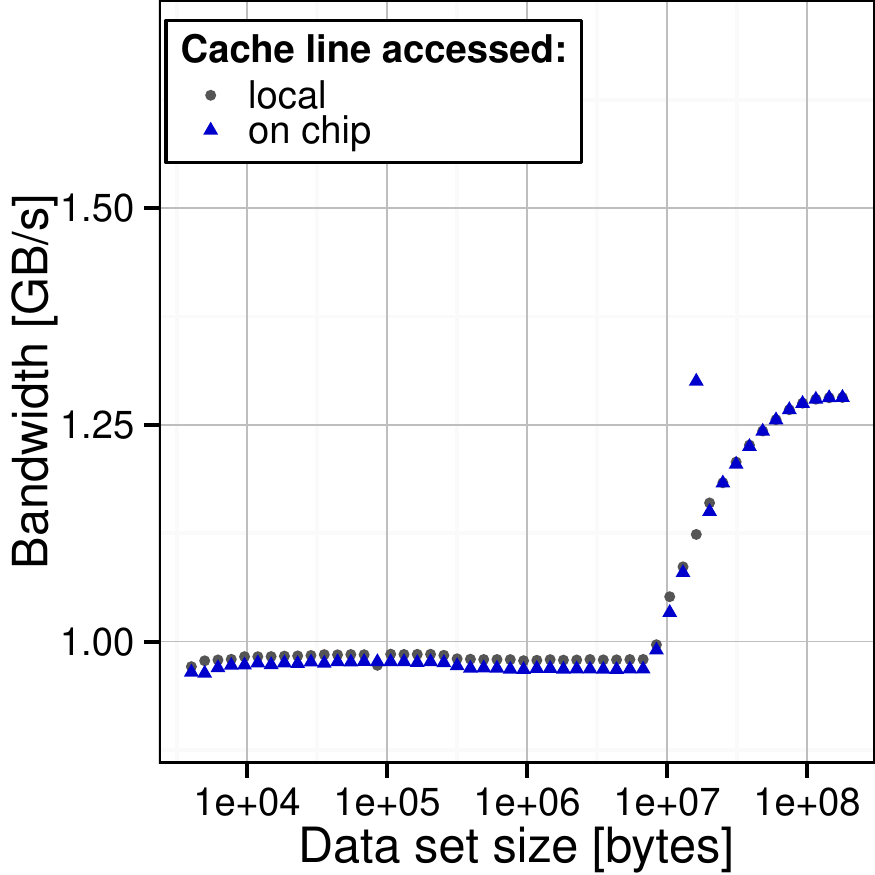}
 }
 \subfloat[\textsf{FAA}, the Shared state]{
  \includegraphics[width=0.22\textwidth]{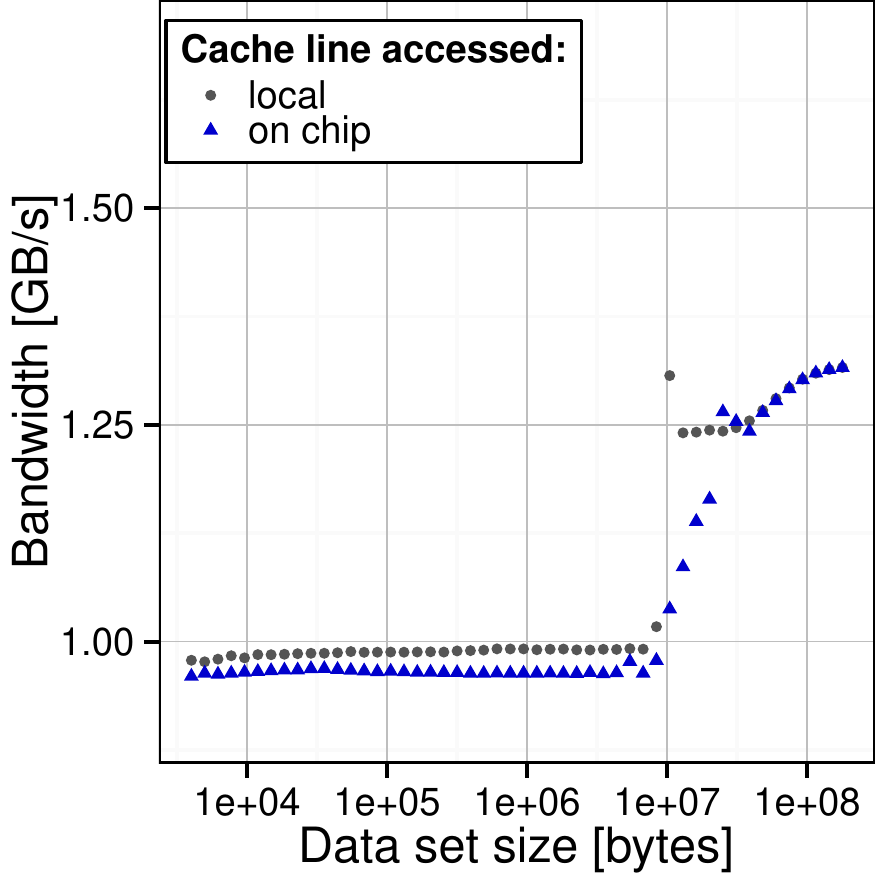}
 }
  \subfloat[\textsf{SWP}, the Shared state]{
  \includegraphics[width=0.22\textwidth]{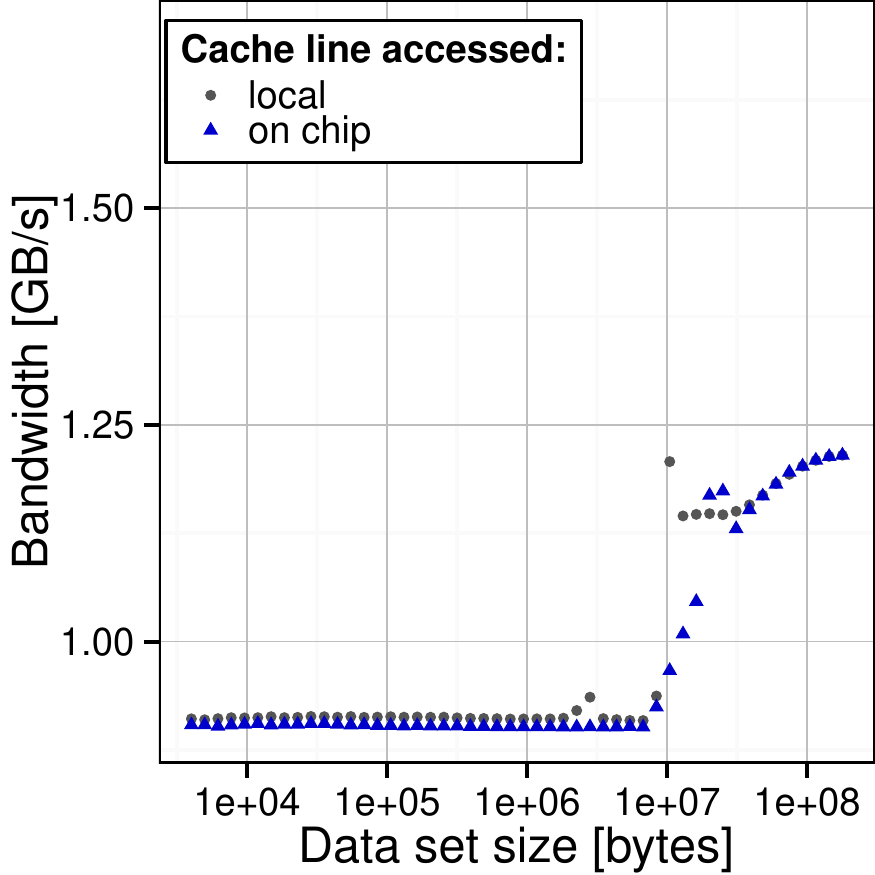}
 }
  \subfloat[\textsf{read}, the Shared state]{
  \includegraphics[width=0.22\textwidth]{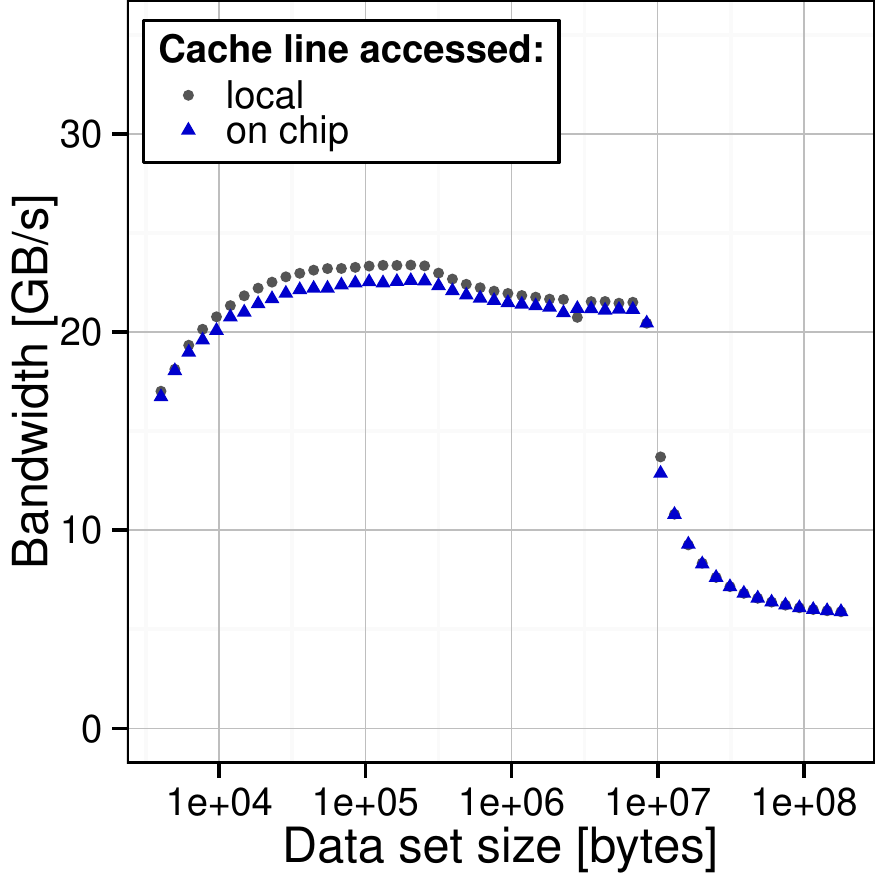}
 }
\caption{The comparison of the bandwidth of \textsf{CAS}, \textsf{FAA}, \textsf{SWP}, and \textsf{writes} on Haswell. The requesting core accesses its own cache lines (local) and cache lines of a different core from the same chip (on chip).}
\label{fig:BW_Has___APP}
\end{figure*}

In the Appendix, we provide full results in Figures~\ref{fig:Latency_results_Phi___APP}--\ref{fig:BW_Has___APP}.

\begin{figure*}[t]
\centering
  \subfloat[\textsf{CAS}, Exclusive state]{
  \includegraphics[width=0.32\textwidth]{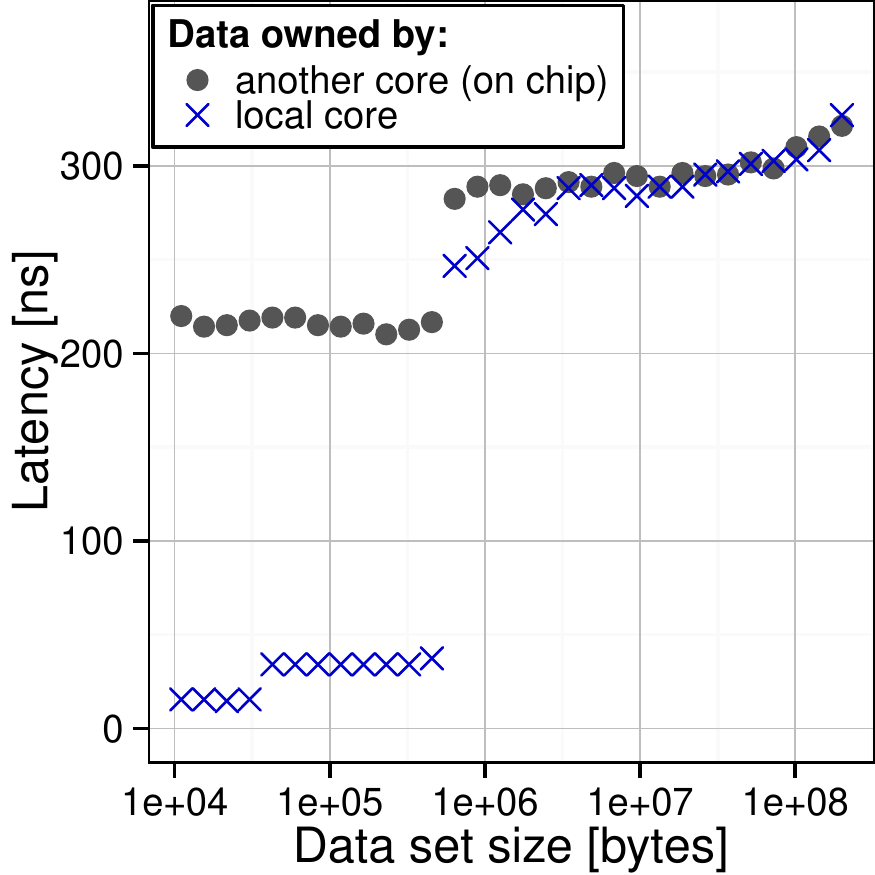}
 }
 \subfloat[\textsf{CAS}, Modified state]{
  \includegraphics[width=0.32\textwidth]{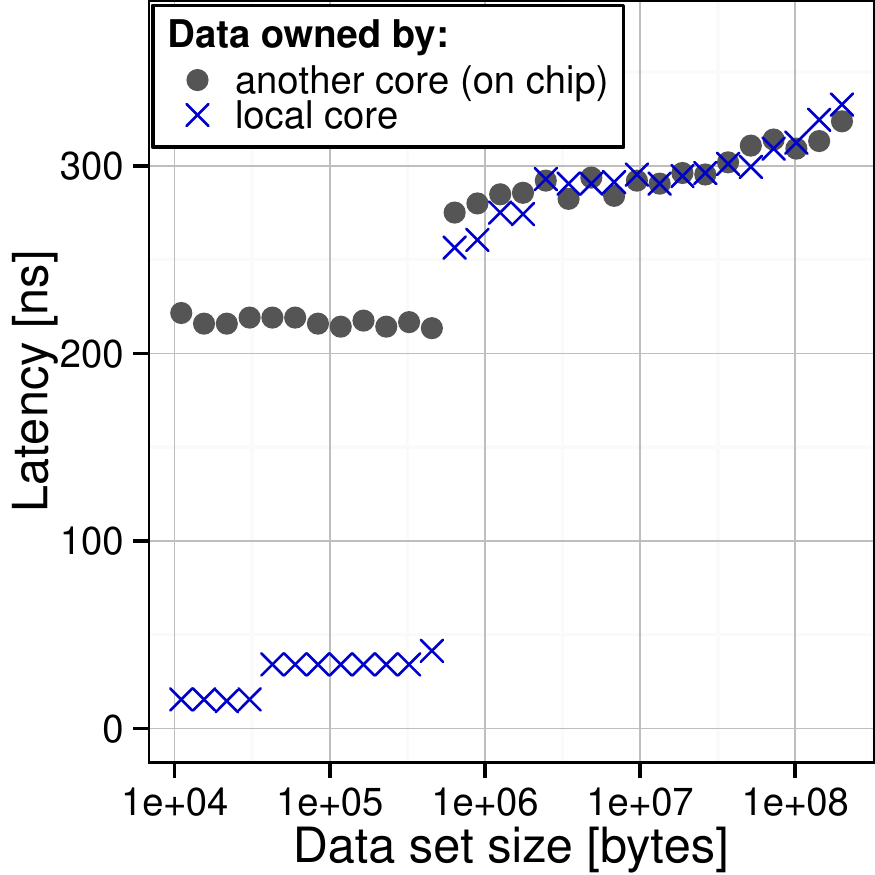}
 }
  \subfloat[\textsf{CAS}, Shared/Owned state]{
  \includegraphics[width=0.32\textwidth]{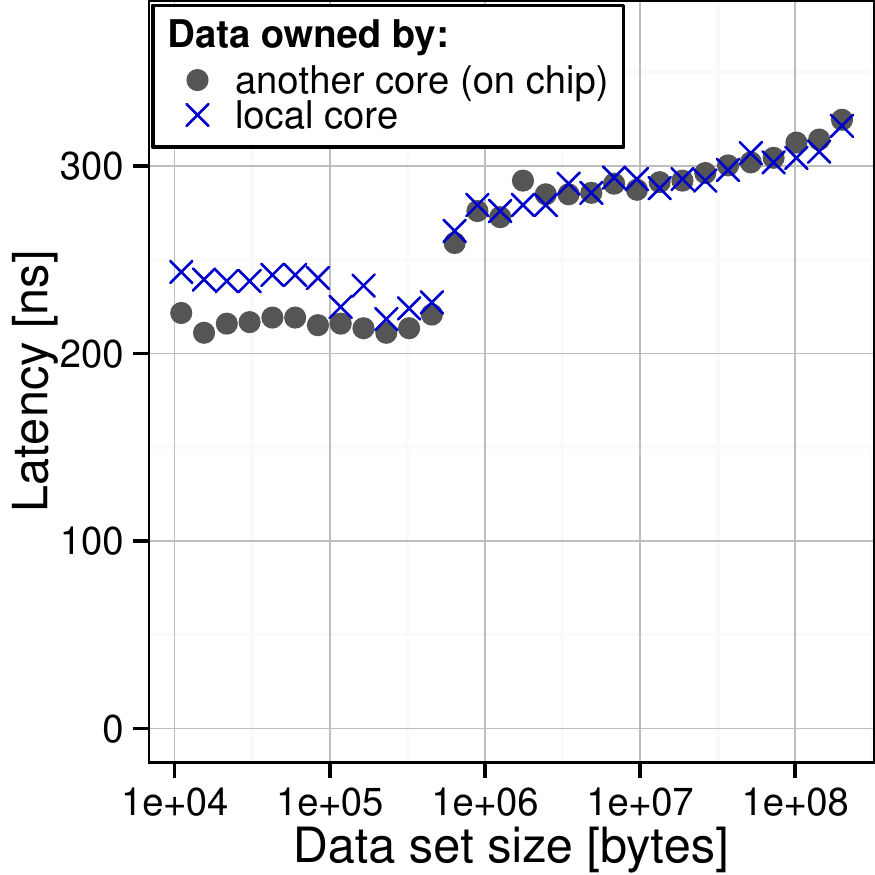}
 }\\
 \subfloat[\textsf{FAA}, Exclusive state]{
  \includegraphics[width=0.23\textwidth]{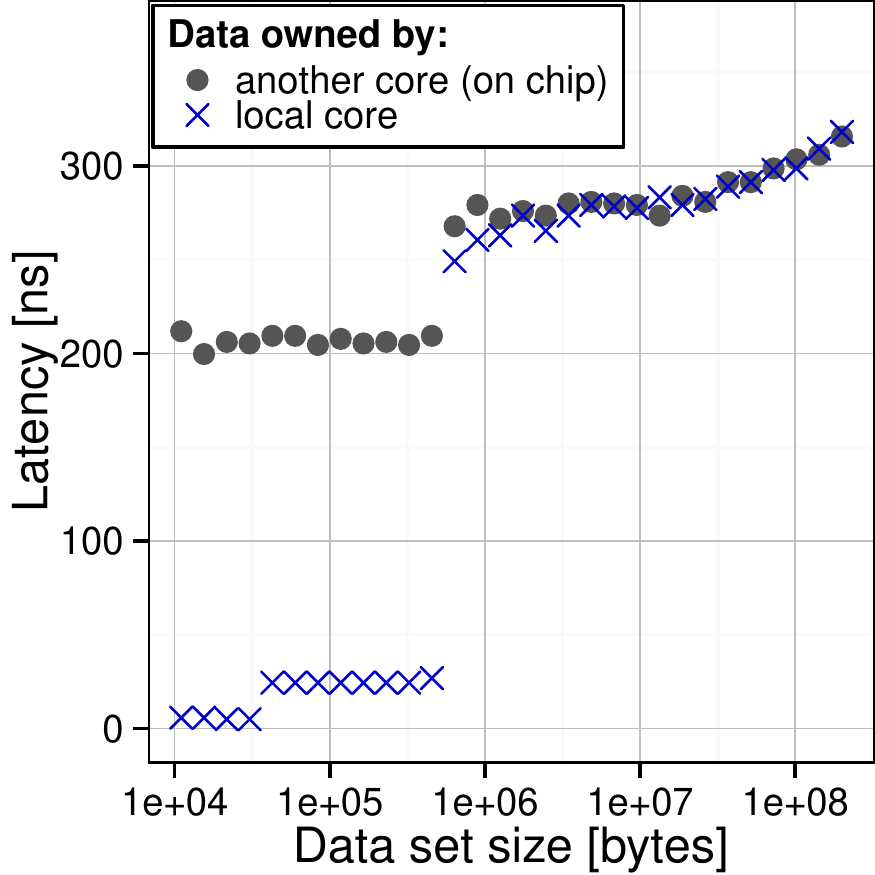}
 }
 \subfloat[\textsf{FAA}, Modified state]{
  \includegraphics[width=0.23\textwidth]{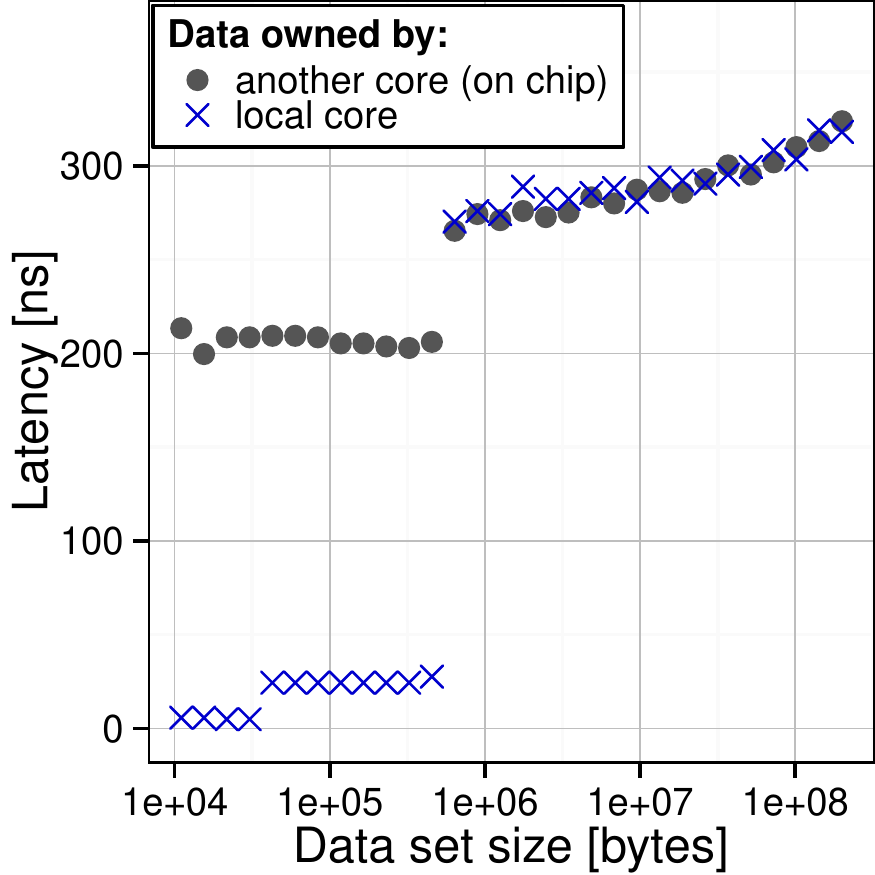}
 }
 \subfloat[\textsf{FAA}, Shared state]{
  \includegraphics[width=0.23\textwidth]{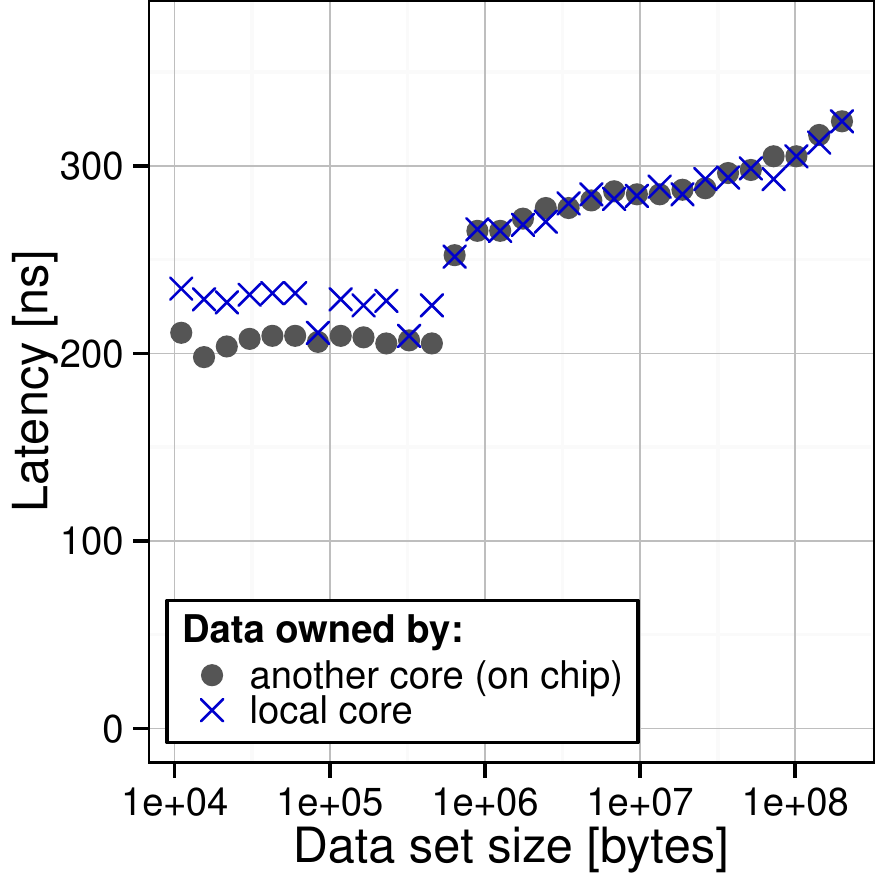}
 }
 \subfloat[\textsf{FAA}, Owned state]{
  \includegraphics[width=0.23\textwidth]{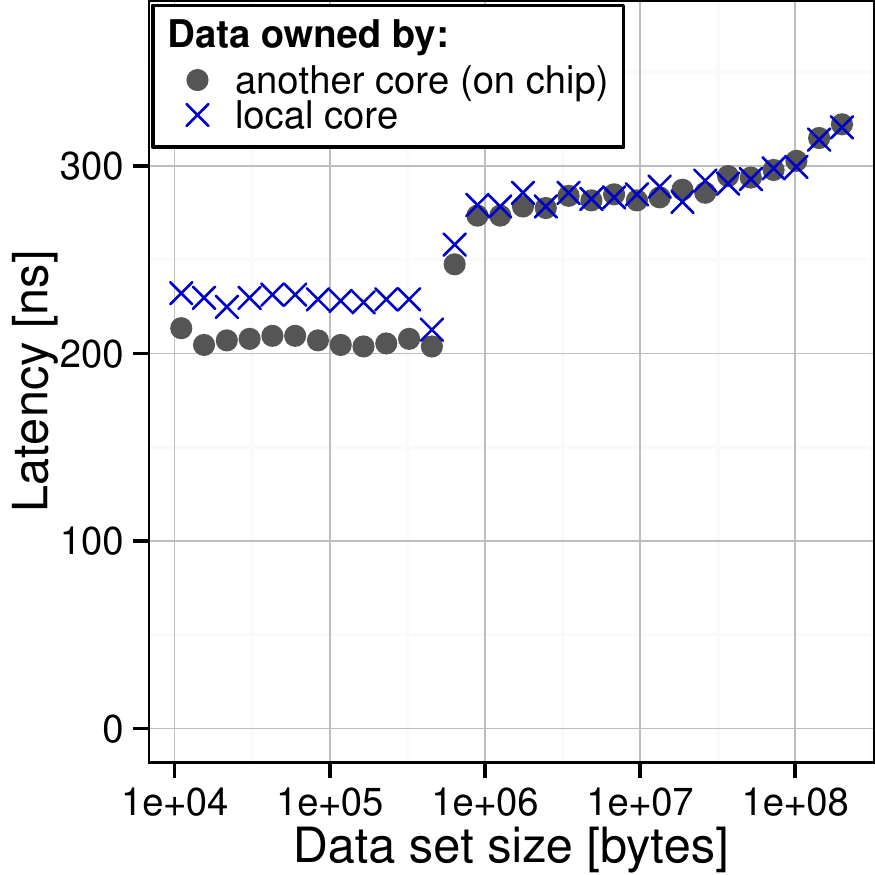}
 }\\
  \subfloat[\textsf{read}, Exclusive state]{
  \includegraphics[width=0.23\textwidth]{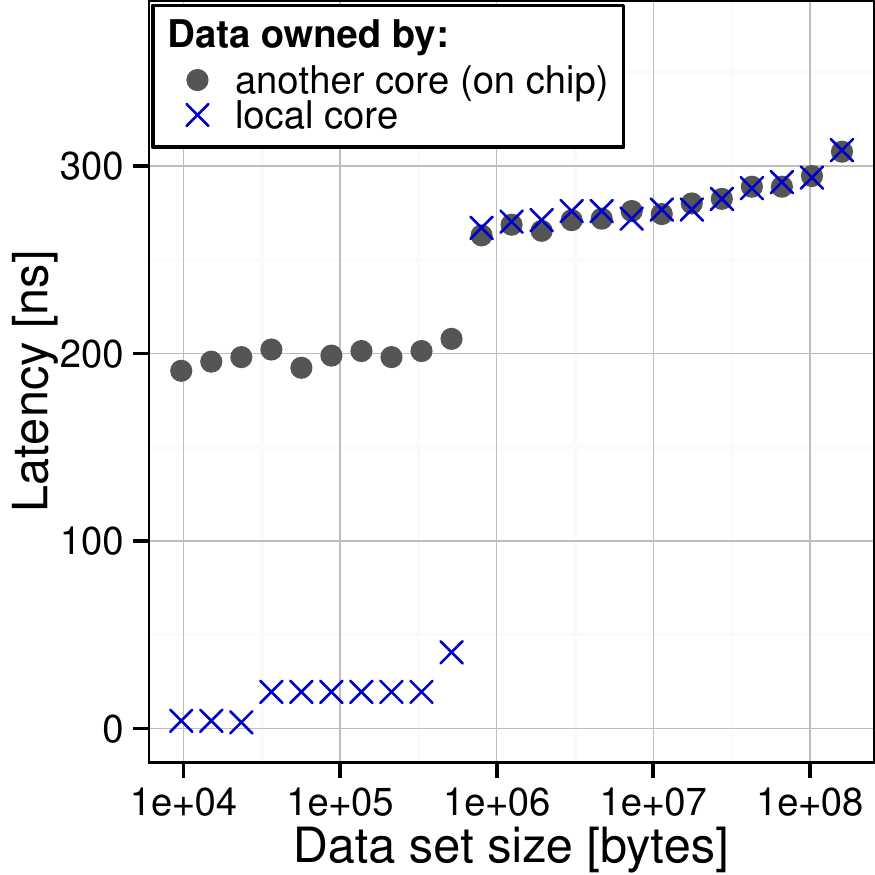}
 }
  \subfloat[\textsf{read}, Modified state]{
  \includegraphics[width=0.23\textwidth]{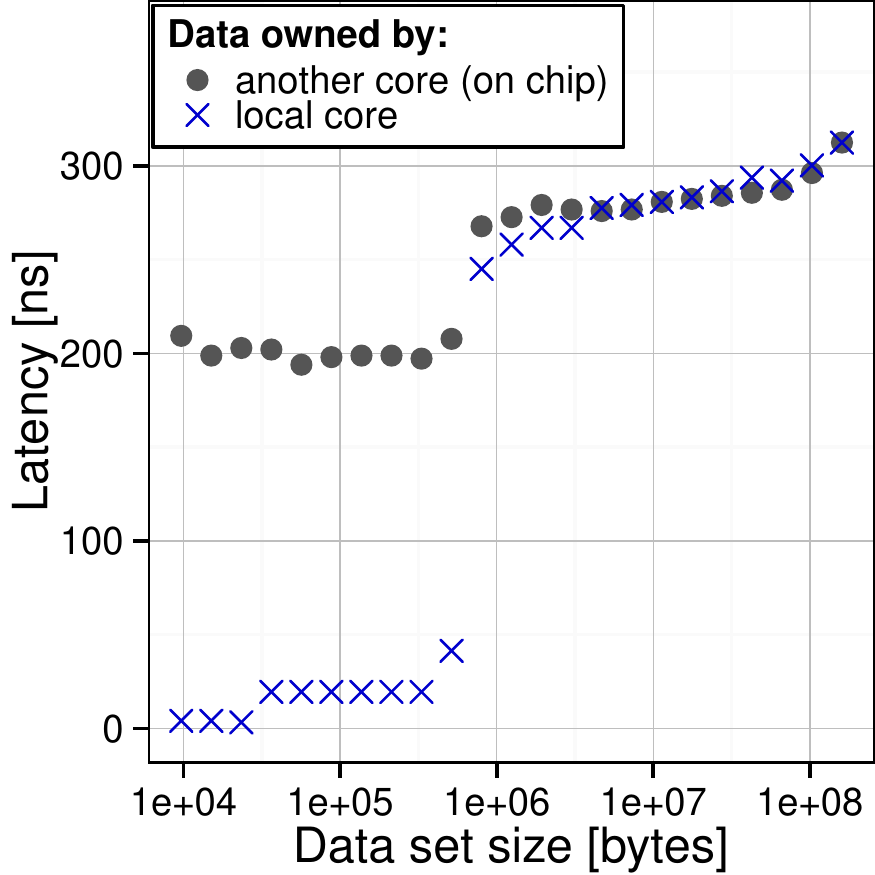}
 }
  \subfloat[\textsf{read}, Shared state]{
  \includegraphics[width=0.23\textwidth]{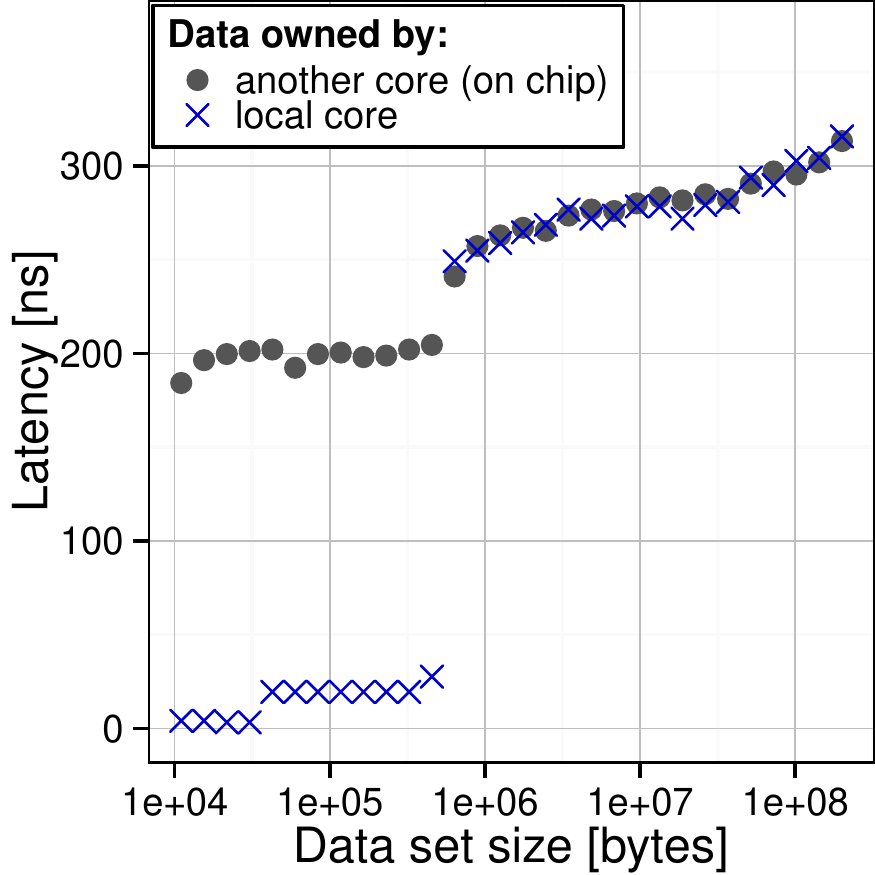}
 }
   \subfloat[\textsf{read}, Owned state]{
  \includegraphics[width=0.23\textwidth]{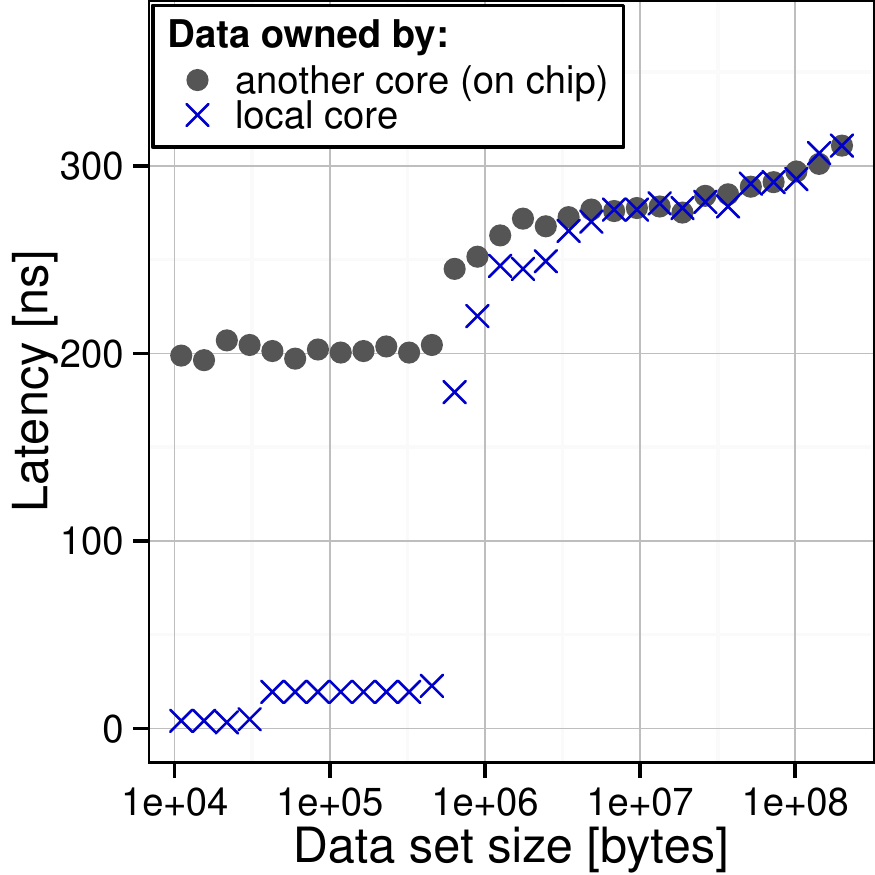}
 }
\caption{The comparison of the latency of \textsf{CAS}, \textsf{FAA}, and \textsf{read} on Xeon Phi. The requesting core accesses its own cache lines (local) and  cache lines of a different core from the same chip (on chip).}
\label{fig:Latency_results_Phi___APP}
\end{figure*}

\bibliographystyle{abbrv}
\bibliography{references}

\end{document}